\documentclass{jfm}
\usepackage{graphicx}
\usepackage{newtxtext}
\usepackage{newtxmath}
\usepackage{natbib}
\usepackage{hyperref}
\hypersetup{
    colorlinks = true,
    urlcolor   = blue,
    citecolor  = black,
}

\newcommand{\RomanNumeralCaps}[1]
\linenumbers
\usepackage{epstopdf, epsfig,color}
\usepackage{subcaption}
\usepackage{amsmath}
\usepackage{amsfonts}
\usepackage{amssymb}
\usepackage{xcolor}

\shorttitle{Instabilities in strongly shear-thinning viscoelastic flows}

\title{Instabilities in strongly shear-thinning viscoelastic flows through channels and tubes}

\author{Ramkarn Patne\aff{1,2}, Shraddha Mandloi\aff{2},
V. Shankar\aff{2}\corresp{\email{vshankar@iitk.ac.in}}
\and Ganesh Subramanian\aff{3}\corresp{\email{sganesh@jncasr.ac.in}}}
\affiliation{\aff{1}Department of Chemical Engineering, Indian Institute of Technology Hyderabad, Kandi, Sangareddy, Telangana 502285, India.
\aff{2}Department of Chemical Engineering, Indian Institute of Technology Kanpur, 208016, India.
\aff{3}{Engineering Mechanics Unit, Jawaharlal Nehru Center for Advanced Scientific Research, Bangalore 560064, India.
}}
\begin{document}

\maketitle
\begin{abstract}

The linear stability of a shear-thinning, viscoelastic fluid undergoing any of the canonical rectilinear shear flows, viz., plane Couette flow and pressure-driven flow through a channel or a tube is analyzed in the creeping-flow limit using the White--Metzner model with a power-law variation of the viscosity with shear rate. While two-dimensional disturbances are considered for plane Couette and channel flows, axisymmetric disturbances are considered for pressure-driven flow in a tube.
For all these flows, when the shear-thinning exponent is less than $0.3$, there exists an identical instability at wavelengths much smaller than the relevant geometric length scale (gap between the plates or tube radius).  There is also a finite-wavelength instability in these configurations governed by the details of the geometry and boundary conditions at the centerline of the channel or tube.  The most unstable mode could be either of the short-wave or finite-wavelength instabilities depending on model parameters. For pressure-driven channel flow, it is possible to have sinuous  or varicose unstable modes depending on the symmetry of the normal velocity eigenfunction about the channel centerline. This difference in symmetry is relevant only for the finite wavelength instability, in which case sinuous modes turn out to be more unstable, in accordance with experimental observations. 
In all the three configurations, the short wavelength unstable modes are localized near the walls, and are insensitive to symmetry conditions at the centerline. It is argued that this instability should be a generic feature in any wall-bounded shear flow of strongly shear-thinning viscoelastic fluids. 
Our predictions for the finite-wavelength instability in pressure-driven channel and pipe flows are in good agreement with experimental observations for the flow of concentrated polymer solutions in these geometries.
\end{abstract}

\section{Introduction}

The understanding of instabilities in rectilinear shearing flows of polymer solutions has attracted renewed interest in part due to the  discovery of `elasto-inertial turbulence' \citep{samanta-et-al-2013,choueiri-et-al-2018,choueiri2021experimental} in pipe flow of relatively dilute polymer solutions (of concentrations around the overlap value) at Reynolds numbers ($Re \sim 800$) much lower than the typical values ($\approx 2000$) at which the Newtonian transition is typically observed \citep[also see][]{chandra-et-al-2018,chandra_shankar_das_2020}. The observed transition at $Re \sim 100$ and higher has been attributed to an elastoinertial linear instability in the pressure-driven flow of an Oldroyd-B fluid that requires the simultaneous presence of inertia, elasticity and solvent-induced viscous effects \citep{piyush-et-al-2018,chaudharyetal_2021,khalid2020centermode}.
 There have also been several reports of instabilities in flows of highly elastic, dilute polymer solutions in the low Reynolds number regime,  by Arratia and coworkers for flow in a square channel \citep{Pan_2012_PRL,Arratia_PRL_2017,Arratia_PRL_2019} and by Steinberg and coworkers for flow in channels of high aspect ratios \citep{jha2020universal,shnapp2021elastic}.  The instabilities observed in the low Reynolds number regime have been traditionally proposed to occur via a subcritical, nonlinear mechanism \citep{bertola_saarloos2003,meulenbroek_sarloos2004,morozov_saarloos2005,morozov_saarloos2007,morozov_saarloos2019}. There have also been efforts \citep{hoda_jovanovic_kumar_2008,hoda_jovanovic_kumar_2009,jovanovic_kumar_2010} that have explored the importance of nonmodal (linear) growth for inertialess viscoelastic flows.
However, and in contrast, \cite{khalid_creepingflow_2021} recently reported a linear (modal) instability in the inertialess regime for viscoelastic channel flows, using the Oldoyrd-B model, albeit in the limit of highly elastic (with Weissenberg number $W \sim O(1000)$), ultra-dilute (with solvent to solution viscosity ratio $\approx 0.99$)  polymer solutions,  one that continues to the elastoinertial regime at higher $Re$. This instability has subsequently been predicted to exist at relatively moderate $W \sim 100$ using the FENE-P model \citep{buza2021weakly}.


While the aforementioned experimental efforts often employed polymer solutions of concentrations less than or around the overlap value, there have been several reports of instabilities in the flow of highly concentrated polymer solutions in tubes and channels, and the $Re$ at which this transition is observed is $O(1)$ or lower \citep{bodiguel-et-al-2015,poole-2016,picaut-et-al-2017,wen-et-al-2017,chandra-et-al-2019}. For flows involving concentrated polymer solutions, shear thinning is likely to play a dominant role on the transition, in contrast to the linear instability predicted  using the Oldroyd-B model which does not exhibit shear thinning \citep{piyush-et-al-2018,khalid_creepingflow_2021}. Although the FENE-P model \citep{birdvol2} goes beyond the Oldroyd-B model by accounting for finite extensibility of the polymer, and predicts shear thinning of both viscosity and first normal stress coefficient, the degree of shear thinning is fixed: the polymer viscosity in the FENE-P model scales as $W^{-2/3}$ for $W \gg 1$. The prediction of instabilities observed in shear thinning (concentrated) polymer solutions appears, however, to require a much stronger shear thinning exponent than the one predicted by the FENE-P model.
Consequently, the instability of these flows have been analyzed using the phenomenelogical \cite{white-metzner-1963} model by Wilson and co-workers \citep{wilson-rallison-1999,wilson-loridan-2015,castillo-wilson-2017}, wherein the degree of shear thinning is a model parameter that can be independently specified. These efforts have implicated both shear thinning and elasticity to underlie the observed instability. Importantly, the instability is predicted even in the absence of fluid inertial and solvent viscous effects, again in direct contrast to the linear instabilities predicted using the Oldroyd-B model \citep{piyush-et-al-2018,khalid_creepingflow_2021}. In a somewhat different context, the simultaneous importance of shear thinning and elasticity has also been shown to be relevant to the emergence of flow asymmetries in viscoelastic flow past a cylinder \citep{Haward_JNNFM_2020}, although the flow in this case has both extensional and shear components.

The present work augments the earlier theoretical efforts of Wilson and coworkers by analyzing the linear stability of a shear-thinning White-Metzner (WM) fluid in all the canonical rectilinear shear flows, viz., plane Couette flow (PCF), pressure-driven channel flow (henceforth referred to simply as `channel flow'), and pressure-driven tube flow (henceforth referred to simply as `tube flow'). The aim of the present work is to demonstrate the universal nature of the instability in the flow of shear-thinning viscoelastic fluids undergoing any wall-bounded shear flow.  Similar to the earlier efforts of Wilson and co-workers, we restrict our attention to creeping-flow limit in this study.

It is useful at the outset to distinguish between  hydrodynamic instabilities which can occur even when the constitutive relation (for stress vs. shear rate) is monotonic, and constitutive instabilities which occur only when the constitutive relation is non-monotonic \citep{yerushalmi-1970}. 
In the present study, we restrict ourselves to regimes where the constitutive relation is monotonic. 
 The linear stability of PCF of an upper-convected Maxwell (UCM) fluid, with a shear-rate independent viscosity, was first studied by \cite{gorodtsov-leonov-1967} who found two stable discrete modes (referred hereafter as the `GL' modes) in the creeping-flow limit.
 \cite{wilson-et-al-1999} extended this analysis to the case of inertialess plane-Couette and channel flows of an Oldroyd-B fluid, again finding both the flows to be stable for $W \sim O(1)$.
  \cite{renardy86} studied plane Couette flow, but for non-zero Reynolds number, and concluded that the flow is linearly stable for arbitrary $Re$. Similarly, \cite{lee-finlayson-1986,sureshkumar92} concluded that channel flow of a UCM fluid is linearly stable at low $Re$. Planar shearing flows (PCF and channel flow) of an inelastic shear-thinning (power-law) fluid are also linearly stable at low $Re$ \citep{nouar-et-al-2007}.  During the course of the present study, we found that tube flows of such fluids are also linearly stable in the creeping-flow limit. Thus, canonical internal shear flows of inelastic shear-thinning or elastic non-shear thinning fluids are linearly stable in the absence of inertia, with pressure-driven channel flow being the only exception owing to the recent discovery of a linear instability in the limit of highly elastic ultra dilute polymer solutions \citep{khalid_creepingflow_2021,buza2021weakly}. The prediction of a stable flow in the creeping-flow limit, especially when the solution is not dilute, is at odds with the aforementioned experiments \citep{bodiguel-et-al-2015,poole-2016,picaut-et-al-2017}, 
 which reveal an instability in the flow of concentrated polymer solutions at considerably low $Re \sim O(1)$, suggesting that both shear-thinning and elastic effects are necessary to predict the observed instability.


To this end, 
\cite{wilson-rallison-1999} used the WM model to study the linear stability of a pressure-driven channel flow using the power-law model for the viscosity variation with shear-rate, while assuming a constant relaxation modulus. They  
predicted the existence of a `shear-thinning elastic' instability for power-law index $n$ less than $0.3$ in the creeping-flow limit. 
In a subsequent study, which included a shear-rate-dependent elastic modulus (characterized by a second index), \cite{wilson-loridan-2015} predicted a destabilizing effect of strain softening on the instability predicted in the previous study \citep{wilson-rallison-1999}. Recently, \cite{castillo-wilson-2017} showed that solvent stresses have a stabilizing effect on the predicted instability. 
Channel flow allows for both varicose and sinuous modes depending on the symmetry (odd vs. even; see Sec.~\ref{sec:pf-problem-formulation}) of the  normal velocity eigenfunction about the centerline. However, Wilson and coworkers have, in the above efforts, only analyzed varicose modes
 due to the presence of an apparent singularity at the channel-center for sinuous modes. We show in Sec.~\ref{sec:pf-problem-formulation} that the singularity in the WM equations can be removed provided that relaxation modulus is independent of the shear rate. 
 
 
While the studies reviewed above have employed relatively simple viscoelastic models in the context of hydrodynamic stability, there have also been many studies, discussed below, that used more complex models which are nonlinear in the stresses.
\cite{grillet-et-al-2002} studied the linear stability of a shear-thinning viscoelastic fluid using the \cite{giesekus-1982} and  \cite{thien-tanner-1977} (PTT) models in the creeping-flow limit. Their analysis predicted that PCF of a Giesekus fluid is linearly stable, while channel flow of the same fluid is linearly unstable. However, both PCF and channel flow of a PTT fluid were found to be linearly unstable. 
The instability of PCF of PTT fluid, however, requires a highly shear-thinning viscoelastic fluid and high Weissenberg number. In contrast, 
\cite{arora-et-al-2004} used the pom-pom model and predicted PCF to be linearly stable in the creeping-flow limit. During the course of the present study, we found that PCF of the \cite{johnson-segalman-1977} model is stable in the constitutively stable regime. 
There have been some studies \citep{chokshi-kumaran-2009,cromer-et-al-2013,cromer-et-al-2014}
that invoked a non-local constitutive relation via a coupling between the stress and polymer concentration, the latter being governed by an additional convection-diffusion equation, and predicted
 a linear instability of plane Couette flow beyond a critical $W$ in the absence of inertia. Thus, plane Couette flow of polymer solutions (with monotonic constitutive flow curves) appears to exhibit an instability, even in the creeping-flow limit, due to the stress-concentration coupling. This coupling, however, is not considered in the present study.


Thus, barring the lone study of \cite{grillet-et-al-2002}, PCF has been found to be stable in the creeping-flow limit in the absence of stress-concentration coupling. In contrast, recently, \cite{barlow-et-al-2019} showed that (pressure-driven) channel flow exhibits an instability that is relatively model independent by using the Rolie-Poly, Johnson-Segalman, and WM models, provided that the fluid is sufficiently shear-thinning. Specifically, for the WM model, channel flow becomes unstable for $n < 0.3$, a conclusion that is broadly in agreement with the aforementioned works of Wilson and co-workers.   The mechanism of instability for Rolie-Poly, Johnson-Segalman \citep{barlow-et-al-2019} and PTT \citep{grillet-et-al-2002} models was attributed to the destabilization of the quasi-interface formed due to strong viscosity stratification in the base state, because of the varying shear rates across the channel. However, for the WM model, there was no quasi interface present in the base flow \citep{barlow-et-al-2019} yet it exhibits an instability; the reason behind this result is not fully understood yet.

The experimental observations of \cite{bodiguel-et-al-2015}  suggest that sinuous modes could be more unstable than varicose modes for channel flow, implying that no restrictions should be made vis-a-vis the symmetry of the perturbations. The experiments of \cite{poole-2016,picaut-et-al-2017} strongly suggest that, similar to pressure-driven channel flow, a shear-thinning elastic instability is also operative in tube flows of concentrated polymer solutions. However, thus far, there has been no attempt in the literature to predict the instability in a tube flow from a theoretical standpoint.  The earlier work of \cite{wilson-rallison-1999,wilson-loridan-2015,castillo-wilson-2017} for pressure-driven channel flow suggests that the disturbances corresponding to shear-thinning elastic instability are confined near the wall. This suggests that an identical instability could be present in any wall-bounded flow of a WM fluid, including in all the three canonical unidirectional shear flows, viz., plane Couette flow and pressure-driven flow in channels and tubes.

In this study, we demonstrate that an identical short-wave instability is present in all the three canonical rectilinear shearing flows,
 by carrying out a systematic stability analysis using the WM model.  While the recent work of \cite{barlow-et-al-2019} focussed on demonstrating possible similarities across different constitutive models for a specific flow (viz., channel flow), in the present work, our goal is to explore the possibility of an identical instability across canonical shear flows, but for a given constitutive (i.e., WM) model. Thus, the present effort, combined with the previous work of \cite{barlow-et-al-2019}, should provide a comprehensive picture of possible instabilities in rectilinear flows of shear thinning fluids across different constitutive models and flow geometries. Another important way in which the present work significantly goes beyond the existing literature is by presenting a detailed account of the origin of these unstable modes in the viscoelastic spectra as $n$ is decreased, an aspect not addressed in the prior efforts in this area.
We also show that our theoretical predictions are broadly consistent with the instabilities observed in both channels and tubes for the flow of concentrated polymer solutions.

The rest of the paper is organized as follows. In Sec.~\ref{sec:problem-formulation}, we derive the base state quantities and linearised perturbation equations for all three canonical shearing flows.  Sections~\ref{sec:pcf-results}, \ref{sec:ppf-results} and \ref{sec:hpf-results} respectively discuss the results for PCF, channel and tube flows. These sections focus on how the unstable modes emerge in the eigenspectrum with varying shear thinning index, $n$, the dispersion curves in the wavespeed-wavenumber plane, and neutral stability diagrams in the relevant parameter space for all three flows. The universal nature of the instability at sufficiently small wavelengths is demonstrated in Sec.~\ref{sec:equivalence}. Section~\ref{sec:comparison-with-experiments} provides a comparison of our theoretical predictions with the experimental results of \cite{bodiguel-et-al-2015}, \cite{poole-2016}
and \cite{picaut-et-al-2017} for channel and tube flows.
The salient results and implications of the present work are summarized in Sec.~\ref{sec:conclusions}.

\section{Problem formulation} \label{sec:problem-formulation}

We consider an incompressible, shear-thinning, viscoelastic fluid of  density $\rho$ and constant relaxation modulus $G$ flowing through a rigid channel or tube with impermeable walls. The dimensional fluid viscosity ($\eta^*_p$) and relaxation time ($\lambda^*_p$) are assumed to be functions of the second invariant ($\dot \gamma^*$) of the strain-rate tensor ($\boldsymbol{\dot \gamma^*}$; Eq.~\ref{eq:shear-rate}). For simplicity, we use the power-law model for the polymer viscosity  \citep{birdvol1, wilson-rallison-1999},
\begin{eqnarray}
\eta^*_p(\dot \gamma^*) = K \dot \gamma^{*(n-1)}, \label{eq:power-law-viscosity-dimensional}\\
\dot \gamma^*=\sqrt{\frac{1}{2} \dot \gamma_{ij}^* \dot \gamma_{ji}^*}.
\end{eqnarray}
\noindent
Here, $K$ denotes the consistency parameter in the power-law model and $n$ is the power-law index. The superscript $``*"$ in the above equation and in the subsequent discussion denotes dimensional quantities while subscript $``p"$ denotes a polymer quantity. By using the above definition of viscosity, the relaxation time ($\lambda^*_p$) for the WM fluid becomes
\begin{eqnarray}
\lambda^*_p(\dot \gamma^*) =\frac{\eta^*_p(\dot \gamma^*)}{G}= \frac{K}{G} \dot \gamma^{*(n-1)}\, . \label{eq:power-law-relaxation-time-dimensional}
\end{eqnarray} 
Before proceeding further, we present the non-dimensionalisation scheme for the problem under consideration. Because the power-law model does not have a natural viscosity scale, we define the viscosity scale as $\eta^*_{sc}=K \left( \frac{V_m}{R} \right)^{(n-1)}$ where $V_m$ and $R$ are respectively characteristic velocity and length scales of the system. For PCF, $V_m$ and $R$ are respectively the moving plate velocity and channel height. For channel flow, $V_m$ and $R$ are respectively the center-line velocity and half channel height and for tube flow, $V_m$ and $R$ are respectively the center-line velocity and tube radius. We non-dimensionalize lengths, velocities, viscosities, pressure and stresses respectively by $R$, $V_m$, $\eta^*_t=\eta^*_s+\eta^*_{sc}$ and $\eta^*_t V_m/R$ where $\eta^*_t$ and $\eta^*_s$ are the total (solution) and solvent viscosities. To relate the  equations governing the perturbations with their Oldroyd-B counterparts for a non-shear-thinning viscoelastic fluid, the viscosity of the polymer is non-dimensionlized by $\eta^*_{sc}$. Hence the dimensionless viscosity for the power-law model (\ref{eq:power-law-viscosity-dimensional}) becomes
\begin{eqnarray}
\eta_p(\dot \gamma)=  \dot \gamma^{(n-1)}. \label{eq:power-law-viscosity-non-dimensional}
\end{eqnarray}
\noindent
The dimensionless continuity and Cauchy momentum equations are
\begin{eqnarray}
\nabla \cdot \mathbf{v}=0, \label{eq:continuity-equation}\\
Re \left(\frac{\partial \mathbf{v}}{\partial t} + (\mathbf{v} \cdot \nabla) \mathbf{v}  \right) = -\nabla p + \nabla \cdot \boldsymbol{\tau }, \\
\boldsymbol{\tau }=\boldsymbol{\tau^s }+\boldsymbol{\tau^p},
\end{eqnarray}
\noindent
where, 
$\mathbf{v}=(v_1,v_2, v_3)$ is the velocity field, $v_i$ is the component of the velocity in the $i^{th}$ direction,
Reynolds number $Re=\rho V_m R/\eta^*_t$, $p$ is the pressure, $ \boldsymbol{\tau }, \boldsymbol{\tau^s }, \boldsymbol{\tau^p }$ are respectively total, solvent and polymer stress tensors and $\nabla$ is the gradient operator.  We use the following constitutive relations for the solvent and polymer stresses:
\begin{eqnarray}
\boldsymbol{\tau^s }=\beta \boldsymbol{\dot \gamma}, \\
\boldsymbol{\tau^p}+W \lambda(\dot \gamma) \left( \frac{\partial \boldsymbol{\tau^p}}{\partial t} + \mathbf{v} \cdot \nabla \boldsymbol{\tau^p}- (\nabla\mathbf{v})^T \cdot\boldsymbol{\tau^p}-\boldsymbol{\tau^p} \cdot (\nabla\mathbf{v}) \right)=(1-\beta) \eta_p(\dot \gamma) \boldsymbol{\dot \gamma}\, . \label{eq:WM-equation}
\end{eqnarray}
The strain rate tensor $\boldsymbol{\dot \gamma}$ is given by
\begin{equation}
\boldsymbol{\dot \gamma}=(\nabla\mathbf{v})+(\nabla\mathbf{v})^T\, .\label{eq:shear-rate}
\end{equation}
\noindent
Here, $\beta=\eta^*_s/\eta^*_t$, and the shear-rate dependent dimensionless relaxation time  $\lambda(\dot \gamma)=\frac{\eta_p(\dot \gamma)}{G}$  is non-dimensionalised by using the scale $\lambda^*_{sc}=\eta^*_{sc}/G$ and $W=\lambda^*_{sc} V_m/ R$ is the Weissenberg number.

\subsection{Planar shear flows} \label{sec:pf-problem-formulation}

In this section, we derive the base state and the governing equations describing the perturbed state for planar flows. For planar flows of an Oldroyd-B fluid, Squire's theorem is applicable \citep{bistagnino-et-al-2007}, i.e., two-dimensional disturbances are more unstable than the corresponding three-dimensional ones. Although a similar demonstration is not possible
for the WM fluid, we nonetheless restrict our attention to two-dimensional perturbations in the interests of simplicity. The two-dimensional velocity field is therefore represented as  $\mathbf{v}=(v_x,v_y)$ for planar flows. By using the governing equations in the preceding section, the fully-developed, steady-state base state velocity profile for PCF is
\begin{eqnarray}
\bar v_x=y.
\end{eqnarray}
\noindent
 In the above equation and henceforth, an overbar signifies a base-state quantity. Here, we have considered PCF with only one plate moving to aid comparison with the other two geometries subsequently discussed in this study.  However, whenever necessary, we will also refer to the configuration with both plates moving, or the configuration with a lower moving plate and stationary upper plate. The problem formulation for all three configurations of PCF remains the same except for some minor changes as explained in the relevant sections. 
In this study, we explore in detail the role of solvent viscous stresses ($\beta \neq 0$) only for PCF.

For channel flow (and tube flow below), we  restrict our analysis to $\beta=0$, because only in this limit is there an analytical solution to the laminar velocity profile. 
We nevertheless expect the qualitative trends obtained concerning the role of solvent viscous stresses for PCF to carry over for channel and tube flows as well.
The steady-state, fully-developed base state velocity profile for channel flow is given by
\begin{eqnarray}
\bar v_x=1-y^{\frac{1+n}{n}}.
\label{eq:ppf-velocity-profile}
\end{eqnarray}
\noindent
By using the above velocity profiles and constitutive equations, the base state stresses for the channel flow configuration  are given by
\begin{eqnarray}
\bar \tau^p_{xy}=(1-\beta)\bar \eta_p(\bar{\dot\gamma}) \frac{d \bar v_x}{dy}; \quad \bar \tau^p_{xx}=2(1-\beta)W\bar \lambda(\bar{\dot\gamma}) \bar \tau^p_{xy} \frac{d \bar v_x}{dy}; \quad \bar \tau^p_{yy}=0.
\end{eqnarray}
\noindent
Here, $\bar \eta_p(\bar{\dot\gamma})$ and $\bar \lambda(\bar{\dot\gamma})$ are the dimensionless shear rate dependent viscosity and relaxation time of the polymer in the base state.

The two-dimensional perturbations are expressed in the form of Fourier modes as
 \begin{eqnarray}
 f'(\mathbf{x},t)= \tilde f(y) \, e^{i\, k \,(x-ct)},
 \label{eq:normal-modes}
 \end{eqnarray}
\noindent
where $f'(\mathbf{x},t)$ is the perturbation to any of the field variables  and $\tilde f(y)$ is the corresponding eigenfunction. Here, $k$ is the (real-valued) wavenumber, while $c=c_r+i c_i$ is the complex wavespeed of the disturbances. The flow is linearly unstable if atleast one eigenvalue satisfies $c_i>0$. Substituting (\ref{eq:normal-modes}) in the linearised governing equations, we obtain the following   ordinary differential equations from the continuity and momentum equations:
\begin{eqnarray}
ik \tilde v_x+D\tilde v_y=0, \label{eq:ppf-ConEq}\\
Re[ik(\bar v_x-c)\tilde v_x+ D\bar v_x \tilde v_y]=-ik\tilde p+\beta(D^2-k^2)\tilde v_x+ik\tilde \tau^p_{xx}+D\tilde \tau^p_{xy}, \label{eq:ppf-x-mom}\\
ikRe(\bar v_x-c)\tilde v_y=-D\tilde p+\beta(D^2-k^2)\tilde v_y+ik\tilde \tau^p_{xy}+D\tilde \tau^p_{yy}, \label{eq:ppf-y-mom}
\end{eqnarray}
\noindent
where $D=d/dy$. The constitutive Eq.~(\ref{eq:WM-equation}) gives
\begin{eqnarray}
\nonumber
W\bar \lambda(\bar{\dot\gamma}) [ik(\bar v_x-c) \tilde \tau^p_{xx}+\tilde v_y D\bar \tau^p_{xx}-2ik \bar \tau^p_{xx} \tilde v_x-2 D\bar v_x \tilde \tau^p_{xy}-2 \bar \tau^p_{xy} D \tilde v_x]\\
-\underline{ 2 W \bar \tau^p_{xy} D \bar v_x \left(\frac{d \lambda}{d \dot \gamma}\right)_{\dot \gamma=\bar{\dot \gamma}} [ik\tilde v_y+D\tilde v_x]}+\tilde \tau^p_{xx}=2ik(1-\beta)\bar \eta_p(\bar{\dot\gamma})\tilde v_x, \label{eq:pf-Txx}\\
\nonumber
W\bar \lambda(\bar{\dot\gamma}) [ik(\bar v_x-c) \tilde \tau^p_{xy}+\tilde v_y D\bar \tau^p_{xy} -ik \bar \tau^p_{xy} \tilde v_x-D\bar v_x \tilde \tau^p_{yy}-ik \bar \tau^p_{xx} \tilde v_y-\bar \tau^p_{xy} D\tilde v_y]\\
+\tilde \tau^p_{xy}=(1-\beta) \underline{ \left[\bar \eta_p(\bar{\dot\gamma}) +\left(\frac{d \eta_p}{d \dot \gamma}\right)_{\dot \gamma=\bar{\dot \gamma}} \bar{\dot \gamma} \right] (ik\tilde v_y+D\tilde v_x)}, \label{eq:pf-Txy}\\
W\bar \lambda(\bar{\dot\gamma}) [ik(\bar v_x-c) \tilde \tau^p_{yy}-2ik \bar \tau^p_{xy} \tilde v_y]+\tilde \tau^p_{yy}=2(1-\beta)\bar \eta_p(\bar{\dot\gamma})D\tilde v_y, \label{eq:pf-Tyy}
\end{eqnarray}
\noindent
where $\left(\frac{d \lambda}{d \dot \gamma}\right)_{\dot \gamma=\bar{\dot \gamma}}$ and $\left(\frac{d \eta_p}{d \dot \gamma}\right)_{\dot \gamma=\bar{\dot \gamma}}$ are the derivatives of the polymer viscosity and relaxation time with respect to $\dot \gamma$ evaluated at the base state. The underlined terms in (\ref{eq:pf-Txx}-\ref{eq:pf-Txy}) are absent in the UCM/Oldroyd-B models and, as shown later, play a major role in determining the stability behaviour. Additionally, the WM model modifies all the terms of Oldroyd-B equations due to multiplication by $\bar \lambda(\bar{\dot\gamma})$ for the elastic terms and $\bar \eta_p(\bar{\dot\gamma})$ for viscous terms. The explicit expressions for $\bar \eta_p, \bar \lambda(\bar{\dot\gamma}),\left(\frac{d \eta_p}{d \dot \gamma}\right)_{\dot \gamma=\bar{\dot \gamma}}$ and $\left(\frac{d \lambda}{d \dot \gamma}\right)_{\dot \gamma=\bar{\dot \gamma}}$ for a WM fluid are
\begin{eqnarray}
\bar \lambda(\bar{\dot\gamma})=\bar \eta_p(\bar{\dot\gamma})=\bar{\dot \gamma}^{(n-1)}; \quad \left(\frac{d \lambda}{d \dot \gamma}\right)_{\dot \gamma=\bar{\dot \gamma}}=\left(\frac{d \eta_p}{d \dot \gamma}\right)_{\dot \gamma=\bar{\dot \gamma}}=(n-1)\bar{\dot \gamma}^{(n-2)}. \label{eq:power-law-base-etap}
\end{eqnarray}
\noindent
The above base-state terms are singular for shear-thinning fluids ($n<1$) in case of channel and tube flows for which $\bar{\dot \gamma}=0$ at the center of channel and tube, respectively.  For PCF, the above equations are then solved by using no-slip  and no-penetration boundary conditions at both plates 
\begin{eqnarray}
\tilde v_x=0; \quad \tilde v_y=0. \label{eq:ppf-bc1}
\end{eqnarray}

For channel flow, the no-slip conditions are applicable at $y=1$, while at the center of channel, the modes are classified as varicose and sinuous modes for which following boundary conditions are applicable  \citep{drazin-howard}
\begin{eqnarray}
\tilde v_{y}=0; \, D \tilde v_{x}=0,  \quad \text{varicose modes}, \label{eq:ppf-varicose}\\
\tilde v_{x}=0; \, D^2 \tilde v_{x}=0,  \quad \text{sinuous modes}.
\label{eq:ppf-sinuous}
\end{eqnarray}

\subsubsection{Regularity of the linearised constitutive equations}
\label{subsec:regularization}

The earlier analyses of \cite{wilson-rallison-1999,wilson-loridan-2015,castillo-wilson-2017} were restricted to varicose modes due to an apparent singularity at the center of the channel which arose due to the underlined terms of constitutive equations (\ref{eq:pf-Txx}-\ref{eq:pf-Txy}). \cite{wilson-rallison-1999} noted that these terms diverge at $y=0$ for $n<1$. This was circumvented by \cite{wilson-rallison-1999}  for varicose modes using the channel-center conditions (\ref{eq:ppf-varicose}) due to which the coefficient, $ik\tilde v_y+D\tilde v_x$, of these terms vanish, thereby making the differential equations regular. 

 For sinuous modes, however, by virtue of the channel-center conditions (\ref{eq:ppf-sinuous}), the coefficient, $ik\tilde v_y+D\tilde v_x$, of the underlined terms in the constitutive equations (\ref{eq:pf-Txx}-\ref{eq:pf-Tyy}) does not vanish. This led \cite{wilson-rallison-1999} to conclude that to regularize sinuous modes, a numerical patch would be required at the center. Furthermore, all the elastic and viscous terms of (\ref{eq:pf-Txx}-\ref{eq:pf-Tyy}) are respectively multiplied by $\bar \lambda(\bar{\dot\gamma})$ and $\bar \eta_p(\bar{\dot\gamma})$ which are singular at $y=0$ for both modes. This difficulty has been overcome in the present work by a rearrangement of terms as follows.

To obtain regularised equations, we divide both sides of equation (\ref{eq:pf-Txx}-\ref{eq:pf-Tyy}) by $\bar \lambda (\bar \dot{\gamma})$ and using (\ref{eq:power-law-base-etap}), $\bar{\dot \gamma}=\frac{d \bar v_x}{dy}$ such that the underlined terms of (\ref{eq:pf-Txx}-\ref{eq:pf-Txy}) become

\begin{eqnarray}
2 W \frac{\bar \tau^p_{xy} D \bar v_x}{\bar \lambda(\bar{\dot\gamma})} \left(\frac{d \lambda}{d \dot \gamma}\right)_{\dot \gamma=\bar{\dot \gamma}} [ik\tilde v_y+D\tilde v_x] =2 W (n-1)\left(\frac{n+1}{n}\right)^{n} y [ik\tilde v_y+D\tilde v_x],\\
\frac{1}{\bar \lambda(\bar{\dot\gamma})} \left(\bar \eta_p(\bar{\dot\gamma}) +\left(\frac{d \eta_p}{d \dot \gamma}\right)_{\dot \gamma=\bar{\dot \gamma}} \bar{\dot \gamma} \right) [ik\tilde v_y+D\tilde v_x] =n [ik\tilde v_y+D\tilde v_x].
\end{eqnarray}
\noindent
Thus, the purported singular terms in the constitutive equations are actually not singular for a WM fluid with constant modulus (for arbitrary values of $n$). 
 As an illustration, consider the equation for $\tilde \tau^p_{xy}$, (\ref{eq:pf-Txy}) which, upon following the above procedure, becomes 
\begin{eqnarray}
\nonumber
W [ik(\bar v_x-c) \tilde \tau^p_{xy}+\tilde v_y D\bar \tau^p_{xy} -ik \bar \tau^p_{xy} \tilde v_x-D\bar v_x \tilde \tau^p_{yy}-ik \bar \tau^p_{xx} \tilde v_y-\bar \tau^p_{xy} D\tilde v_y]\\
+\left(\frac{n+1}{n}\right)^{(1-n)} y^\frac{1-n}{n} \tilde \tau^p_{xy}=n(1-\beta) [ik\tilde v_y+D\tilde v_x]. \label{eq:pf-Txy-regularized}
\end{eqnarray}
\noindent
Thus, at $y=0$, all the terms are finite for a shear-thinning fluid ($n<1$). For a shear-thickening fluid, the original equations themselves are regular, thus demonstrating that the equations are regular for a WM fluid with a shear-rate independent modulus.  The regular nature  of the constitutive equations for both varicose and sinuous modes is further demonstrated later in Sec.~\ref{sec:ppf-results} with the aid of the eigenfunctions for the most unstable eigenvalue.

\subsection{Tube flow}
For tube flow, in the interests of simplicity, we restrict our analysis to axisymmetric modes. Following a procedure similar to channel flow, the fully-developed, steady-state, base state solution for tube flow is
\begin{eqnarray}
\bar v_z=1-r^{\frac{1+n}{n}}.
\end{eqnarray}
\noindent 
The base-state polymer stresses are
\begin{eqnarray}
\bar \tau^p_{rz}=\bar \eta_p(\bar{\dot\gamma}) \frac{d \bar v_z}{dr}; \quad \bar \tau^p_{zz}=2W\bar \lambda(\bar{\dot\gamma}) \bar \tau^p_{rz} \frac{d \bar v_z}{dr}; \quad \bar \tau^p_{rr}=\bar \tau^p_{\theta \theta}=0,
\end{eqnarray}
\noindent
Axisymmetric perturbations of the following form are imposed on the above base-state quantities, where $g'(\mathbf{x},t)$ is relevant field variable variable:
 \begin{eqnarray}
 g'(\mathbf{x},t)= \tilde g(r) \, e^{i\, k \,(z-ct)},
 \end{eqnarray}
\noindent
where, $\tilde g(r)$ is the corresponding eigenfunction. After substituting above perturbations in (\ref{eq:continuity-equation}-\ref{eq:WM-equation}), the continuity and momentum equations become
\begin{eqnarray}
ik \tilde v_z+D\tilde v_r=0, \label{eq:hpf-ConEq}\\
Re[ik(\bar v_z-c)\tilde v_z+ D\bar v_z \tilde v_r]=-ik\tilde p+ik\tilde \tau^p_{zz}+\frac{1}{r} \tilde  \tau^p_{rz}+D\tilde \tau^p_{rz}, \label{eq:hpf-z-mom}\\
ikRe(\bar v_z-c)\tilde v_r=-D\tilde p+ik\tilde \tau^p_{rz}+\frac{1}{r} (\tilde \tau^p_{rr}-\tilde \tau^p_{\theta \theta})+D\tilde \tau^p_{rr}, \label{eq:hpf-r-mom}
\end{eqnarray}
\noindent
The linearised constitutive equations are given by
\begin{eqnarray}
\nonumber
W\bar \lambda(\bar{\dot\gamma}) [ik(\bar v_z-c) \tilde \tau^p_{zz}+\tilde v_r D\bar \tau^p_{zz}-2ik \bar \tau^p_{zz} \tilde v_z-2 D\bar v_z \tilde \tau^p_{rz}-2 \bar \tau^p_{rz} D \tilde v_z]\\
-2 W \bar \tau^p_{rz} D \bar v_z \left(\frac{d \lambda}{d \dot \gamma}\right)_{\dot \gamma=\bar{\dot \gamma}} [ik\tilde v_r+D\tilde v_z]+\tilde \tau^p_{zz}=2ik\bar \eta_p(\bar{\dot\gamma}) \tilde v_z, \label{eq:hpf-Tauzz}\\
\nonumber
W\bar \lambda(\bar{\dot\gamma}) [ik(\bar v_z-c) \tilde \tau^p_{rz}+\tilde v_r D\bar \tau^p_{rz} -ik \bar \tau^p_{rz} \tilde v_x-D\bar v_x \tilde \tau^p_{rr}-ik \bar \tau^p_{zz} \tilde v_r-\bar \tau^p_{rz} D\tilde v_r]\\
+\tilde \tau^p_{rz}=\left[\bar \eta_p(\bar{\dot\gamma}) +\left(\frac{d \eta_p}{d \dot \gamma}\right)_{\dot \gamma=\bar{\dot \gamma}} \bar{\dot \gamma} \right] (ik\tilde v_r+D\tilde v_z), \label{eq:hpf-Taurz}\\
W\bar \lambda(\bar{\dot\gamma}) [ik(\bar v_z-c) \tilde \tau^p_{rr}-2ik \bar \tau^p_{rz} \tilde v_r]+\tilde \tau^p_{rr}=2\bar \eta_p(\bar{\dot\gamma}) D\tilde v_r, \label{eq:hpf-Taurr} \\
W\bar \lambda(\bar{\dot\gamma}) [ik(\bar v_z-c) \tilde \tau^p_{\theta \theta}]+\tilde \tau^p_{\theta \theta}=2\bar \eta_p(\bar{\dot\gamma}) \frac{\tilde v_r}{r}, \label{eq:hpf-Tautt}
\end{eqnarray}
\noindent
where, $D=d/dr$. The above equations have an apparent singularity at $r = 0$ which can be regularized by using a procedure similar to the one outlined for channel flows in Sec.~\ref{sec:pf-problem-formulation}, but by multiplying the constitutive equations by $r^2/\lambda(\bar{\dot\gamma})$. The extra $r^2$ term (compared to planar flows) is a consequence of the curvature of cylindrical geometry and is symptomatic of a centerline singularity in the Newtonian case. The expressions (\ref{eq:power-law-base-etap}) are applicable in the case of tube flow also. At $r=0$, symmetry conditions give \citep{batgill62}
\begin{equation}
  \tilde v_{r}=0, \quad D\tilde v_{z}=0.
  \label{eq:bc-r-zero}
\end{equation}
\noindent
At $r=1$, no-slip conditions are applicable 
\begin{eqnarray}
\tilde v_r=0, \quad \tilde v_z=0.
\end{eqnarray}
\noindent
The above linearised equations are then solved for the eigenvalue $c$ using the pseudospectral \citep{boyd,weideman-reddy} method as explained below.

\subsection{Numerical methodology}

 In the pseudo-spectral method, we expand the velocity and stress components in terms of the Chebyshev polynomials as $\tilde f(y)=\sum_{m=0}^{m=N} a_m T_m (y)$ where $\tilde f(y), m, N, a_m$ and $T_m(y)$ are respectively the relevant velocity or stress component, number of the Chebyshev polynomials, highest degree of the polynomial in the series expansion, $m^{th}$ coefficient in the expansion, and $m^{th}$ Chebyshev polynomial. Thus, we need to evaluate the series expansion at $N$ collocation points to evaluate the series coefficients $a_m$ and/or to obtain the eigenvalue $c$. The discretized equations then constitute a generalized eigenvalue problem of the following form
\begin{eqnarray}
 \mathbf{A}\mathbf{e}=c \mathbf{B}\mathbf{e}.
\end{eqnarray}
\noindent
where $\mathbf{A}$ and $\mathbf{B}$ are the discretized matrices and $\mathbf{e}$ is the eigenvector. To solve the above eigenvalue problem, we use the \emph{polyeig} MATLAB routine. To filter out the spurious modes in the numerically computed spectrum, the spectrum is obtained for $N$ and $N+2$ collocation points, and the eigenvalues are compared with a specified tolerance (e.g. $10^{-4}$). To confirm the genuine eigenvalues, we change the number of collocation points by $25$ and examine the variation of the predicted eigenvalues. If the eigenvalue does not change up to sixth significant figure, we then use same number of collocation points to predict the critical parameters. In the present work, we used $N=50$--$100$ in different flow configurations, as mentioned below in the relevant sections, to predict the most unstable discrete eigenvalue.

\section{Results and dicussion} \label{sec:results-and-analysis}

\subsection{PCF} \label{sec:pcf-results}

\begin{figure}
\centerline{\includegraphics[width=0.6\textwidth]{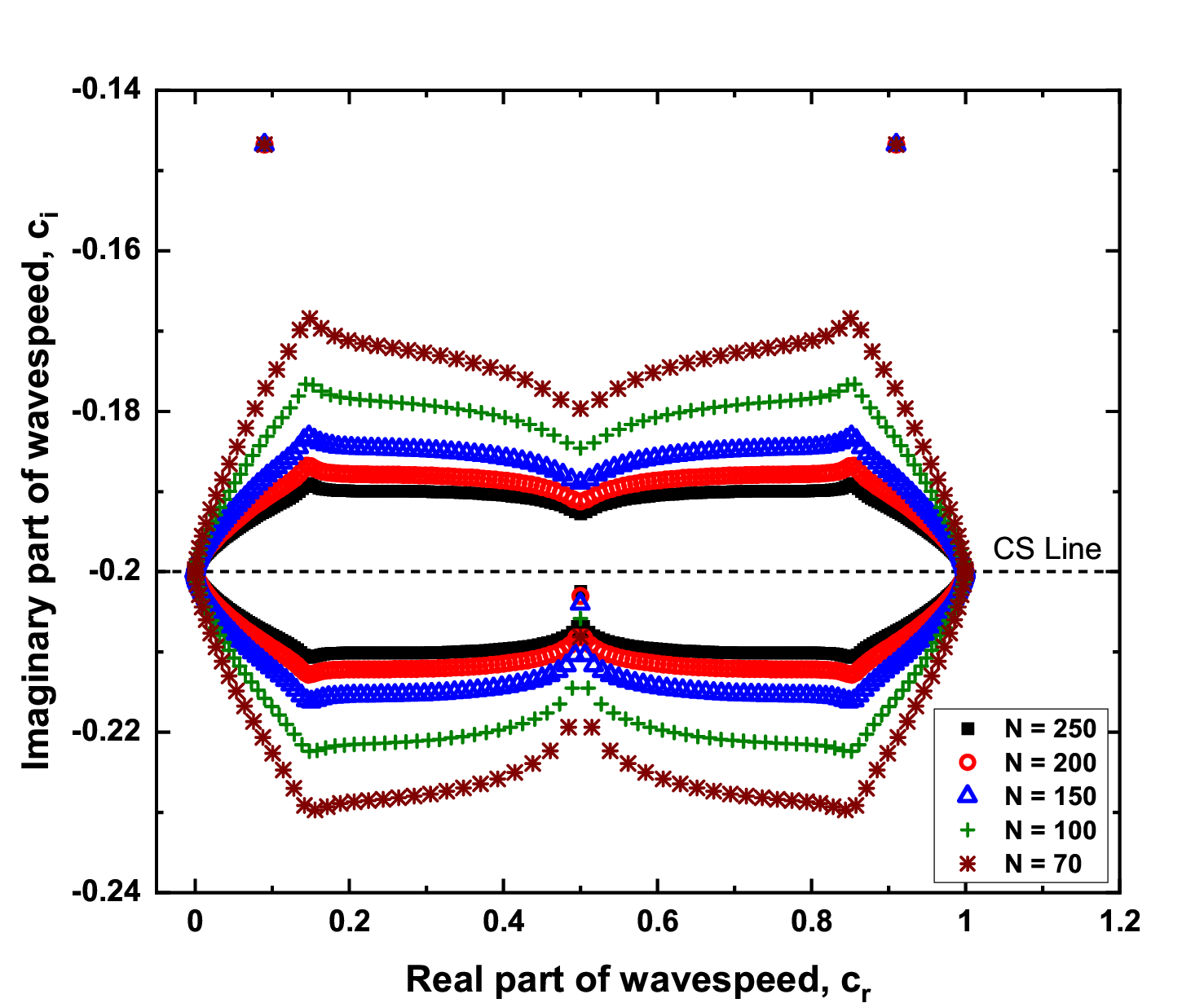}}
\caption{Spectrum for plane Couette flow of a UCM fluid, as obtained from our formulation for the White-Metzner fluid, by setting $\beta = 0$ and $n = 1$,  for $Re = 0$, $W = 2$ and $k = 2.5$. The spectra are shown for different values of $N$, the number of polynomials in the spectral method.}
\label{fig:variationofN}
\end{figure}
We begin this section by discussing the results obtained for plane Couette flow (PCF) of a White-Metzner fluid. Our numerical results for this flow are benchmarked with the results for PCF of inelastic power-law fluid \citep{liu-liu-2010} by setting $W = 0$, $n \neq 1$, and $\beta =0$, and UCM fluid \citep{gorodtsov-leonov-1967} by setting $W \neq 0$, $n = 1$, and $\beta =0$ in the present set of equations. We found excellent agreement between the results available in the existing literature and our results for plane Couette flow of the White-Metzner fluid in the respective parameter regimes. We later demonstrate in Sec.~\ref{sec:ppf-results} that our results for plane Poiseuille flow of White-Metzner fluid agree well with those of \cite{wilson-rallison-1999} for the same flow.\\
We next show, in Fig.~\ref{fig:variationofN}, the convergence of the discrete eigenvalues in the UCM limit by varying the number of collocation points. We notice that, the two discrete GL modes \citep{gorodtsov-leonov-1967} in the UCM limit (obtained by setting $W \neq 0$, $n = 1$, and $\beta =0$ for the PCF of WM fluid), converge for all values of $N > 70$, thereby demonstrating that these modes are physically genuine. 
We also observe the presence of a balloon-like structure of eigenvalues in the unfiltered eigenvalue spectra, that approximates the continuous spectrum and owes its origin to the presence of a singularity in the constitutive equations of the White-Metzner model, similar to the one present in the Oldroyd-B model \citep{graham-1998,wilson-et-al-1999}. To identify the location of the continuous spectrum, the coefficients of the stresses in Eqs. $(2.18)-(2.20)$ must vanish, which gives: 
\begin{equation}
c = y - \frac{i}{k W \bar{\eta}_{p}}.
\label{CSline}
\end{equation}
Since the above continuous spectrum is present even in the absence of a solvent, i.e. for $\beta = 0$, henceforth, we refer to it as the `polymer' continuous spectrum. For nonzero $\beta$, similar to the Oldroyd-B model \citep{wilson-et-al-1999}, a second continuous spectrum is also present in the White-Metzner fluid, and is termed as the `solvent' continuous spectrum. The eigenvalues belonging to the solvent continuous spectrum can be obtained by solving Eqs.~$(2.18)-(2.20)$ for $\tilde{\tau}_{x x}, \tilde{\tau}_{x y}$ and $\tilde \tau_{yy}$ and then substituting the resulting expressions in Eqs.~$(2.16)$ and $(2.17)$. The modified momentum balance equations~$(2.16)$ and $(2.17)$ along with Eq.~$(2.15)$ are then used to eliminate $\tilde{v_x}$ and $\tilde{p}$ to obtain a fourth-order equation in terms of the derivatives of $\tilde{v_y}$. For this differential equation to be singular, the coefficient of its highest derivative must vanish. This condition yields the expression for the eigenvalues belonging to the solvent continuous spectrum:
\begin{equation}
c = y-\frac{i\left[\beta+(1-\beta)\left(\bar{\eta}_{p}+\left(\frac{d \eta_{p}}{d \dot{\gamma}}\right)_{\dot{\gamma}=\bar{\gamma}}\right)\right]}{k \beta \bar{\eta}_{p} W}.
\label{solvent_CSline}
\end{equation}
The numerator of the second term on the right-hand side of the above equation can be shown to be related to the slope of the stress-shear rate curve in the base state. This term is always positive for a monotonic variation of the base state stress with shear rate, which requires $n > 0$ in the WM model. 
 The continuous spectra given by Eqs.~\ref{CSline} and \ref{solvent_CSline}, reduce to the well-known results for an Oldroyd-B fluid if we assume shear-rate independent viscosity. Thus, the two continuous spectra for a WM fluid are generalizations (for $n \neq 1$) of the \cite{gorodtsov-leonov-1967} and viscous continuous spectra \citep{wilson-et-al-1999} present in shear flows of Oldroyd-B fluids. While these equations give the theoretical location of the continuous spectra, they can only be resolved as spread out `balloon' of eigenvalues instead of a `line' (as described theoretically in Eqs. \ref{CSline}, \ref{solvent_CSline}) in the numerically computed spectra, as the spectral method used to calculate the eigenvalues uses a finite number of polynomials to approximate the governing dynamical variables. \textcolor{black}{It must be noted that the imaginary part of the equation \ref{solvent_CSline} is independent of the wall normal direction $y$, and thus the theoretically observed continuous spectrum appear as a line in the $c_r - c_i$ plane, for plane Couette flow. Whereas, for plane Poiseuille and pipe Poiseuille flows the theoretical continuous spectrum appear as a curve following equations \ref{eq:ppf-continuous-spectra-c}, \ref{eq:hpf-continuous-spectra-c} due to the presence of wall-normal directions in the respective equations.}\\
In figure \ref{fig:variationofN}, we observe that the unfiltered eigenvalues of the polymer continuous spectrum move slightly closer to the theoretical CS line as the number of collocation points ($N$) is increased. However, we find $N = 100$ to be sufficient to accurately capture the most unstable (or least stable) discrete eigenvalues. Thus, in the interests of clarity, unless required, we do not plot the CS balloons in our results for the spectra, and instead indicate the CS with dashed lines as specified by their analytical expressions.\\
In the following discussion, we focus on the effect of variation of the power-law index ($n$), and wavenumber ($k$) on the discrete GL modes and emergence of new modes as power-law index is decreased. In this regard, we find it convenient below to divide $n$ into different regimes where there is a qualitative change in the number and/or behaviour of the modes.
\textcolor{black}{\subsubsection{Origin of new wall and center modes}
 As the power law-index is decreased from $n =1$, i.e. the UCM limit, to lower values ($n = 0.4$), we see that the GL modes becomes slightly less stable. In addition, new discrete wall modes (Abbreviated as WM1 henceforth) emerge below the CS, that becomes more stable with decrease in the power-law index ($n$). A detailed discussion and effect of variation of wavenumber for these modes is available in appendix \ref{appA1}. As the power-law index is further decreased $n \in [0.33, 0.4]$, new center modes ($c_r =0.5$), emerge in the low-wavenumber regime ($k< 0.6$), both above and below the CS. The center modes (above the CS) becomes less stable with an further decrease in the power-law index $n$, as shown in appendix \ref{appA2}. Interestingly, for $n \in [0.15,0.33]$, new wall modes appear above (WM2), and below the CS ( WM3), in high-wavenumber regime ($k>0.6$). The wall modes above the CS (WM2) and the GL modes, becomes less stable with the decrease in the power-law index $n$. Whereas, in low-wavenumber regime, new center modes emerge from the polymer continuous spectra, and becomes less stable with decrease in $n$. Interestingly, as we vary the wavenumber for $n = 0.3$, we see that the wall modes which appear in the high $k$ regime (WM2), morph into the center modes at a critical $k$. A detailed discussion of this transformation and the effect of variation of wavenumber is provided in appendix \ref{appA3}.}
\subsubsection{Destabilization of GL, wall and center modes: $n \leq 0.15$}
\begin{figure*}
\begin{subfigure}[b]{0.5\textwidth}
        \includegraphics[width=\textwidth]{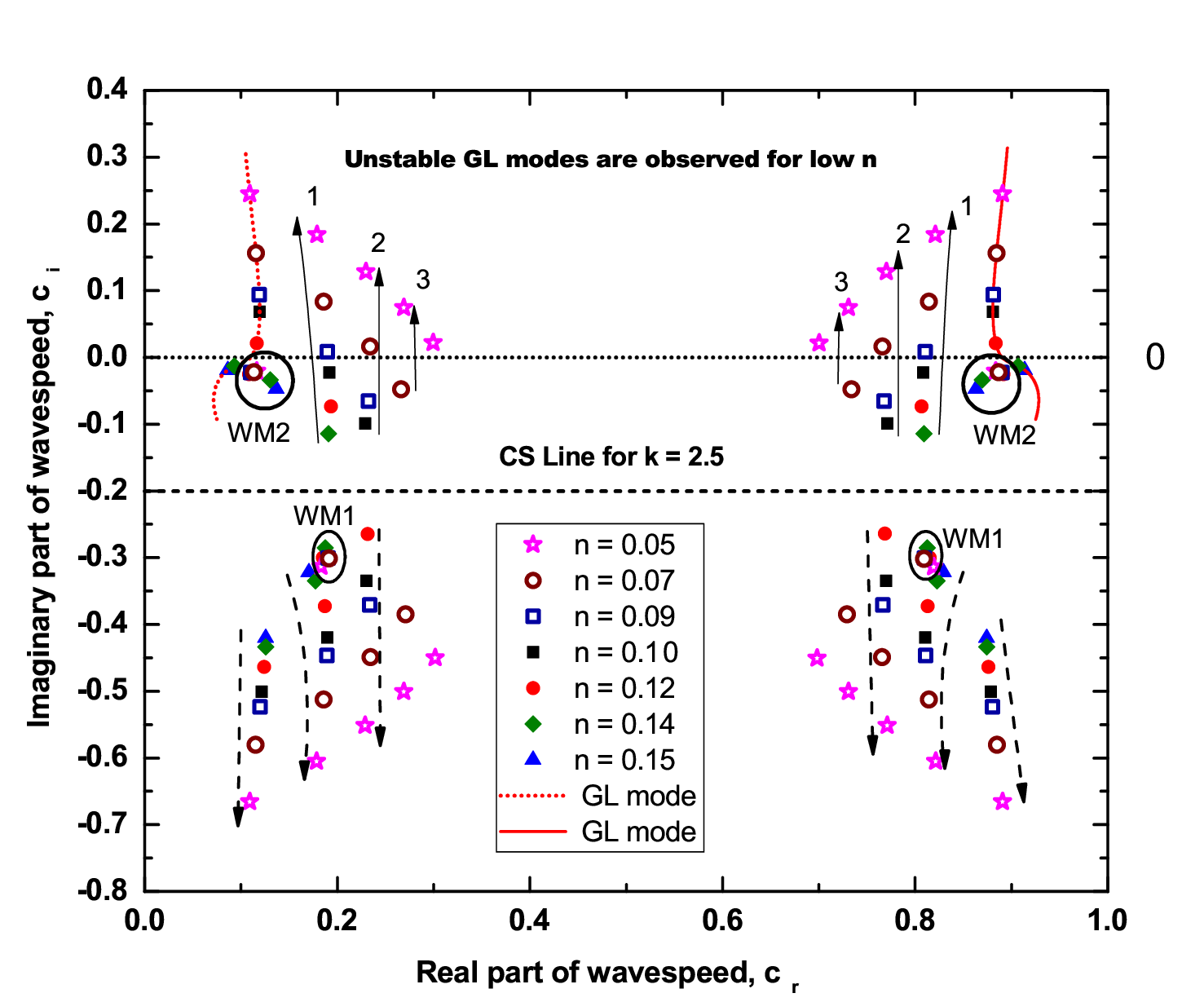}
\caption{$k = 2.5$}
\label{fig:n0p1to15_highk}
\end{subfigure}\hspace{1em}
\begin{subfigure}[b]{0.5\textwidth}
        \includegraphics[width=\textwidth]{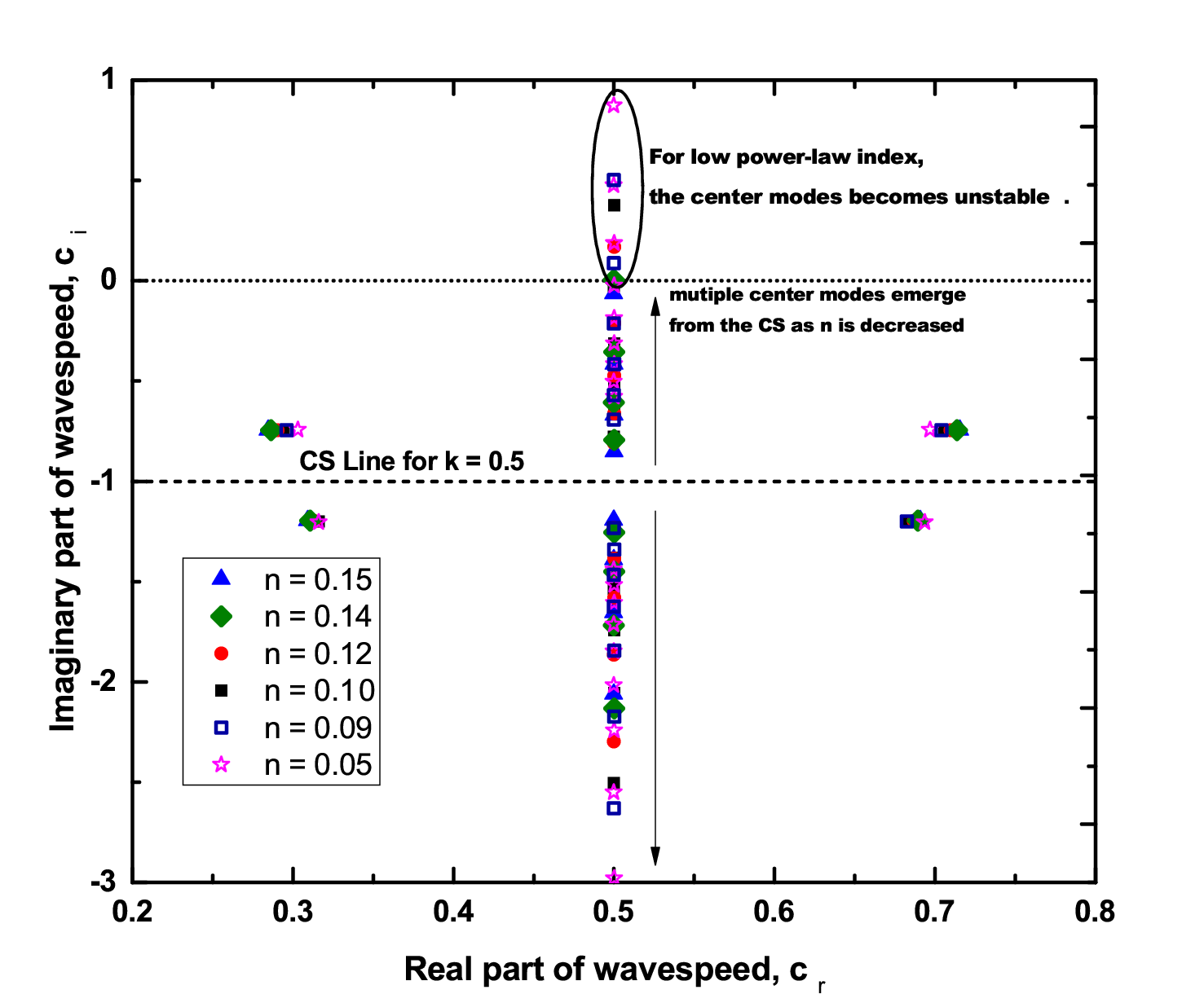}
\caption{$k = 0.5$}
\label{fig:k0p5_n0p15to05}
\end{subfigure}
\caption{Figure showing filtered eigenvalue spectrum for $W=2$, $Re=0$, $\beta=0$, $n \leq 0.15$, and $k$ as shown in figure.}
\label{fig:k_n0p3to15}
\end{figure*}

In this subsection, we discuss how the GL, wall, and center modes get destabilized in different wavenumber regimes for $n \leq 0.15$. While more pairs of the discrete wall and center modes emerge for $n \leq 0.15$ as well, the GL and center modes discussed above become unstable first as $n$ is decreased, as shown in figure \ref{fig:k_n0p3to15}.  For $k=2.5$, as displayed in figure \ref{fig:n0p1to15_highk}, we observe that the GL mode becomes unstable for $n \leq 0.12$ for fixed Weissenberg number. However, in accordance to the general behavior observed for $n \in [0.15, 0.30]$, we observe more pairs of discrete wall modes emerge from the polymer continuous spectra above the CS line. These discrete wall modes also eventually become unstable as the power-law index is decreased. Similarly, more pairs of discrete wall modes also emerge below the CS line but these modes further stabilize as the power-law index is decreased. It is noteworthy that the wall modes that emerged at higher $n$ above the CS line (labelled in figure \ref{fig:k_n0p3to15} as WM2) remain stable for lower values of power-law index as well. However, the wall modes that emerged from the polymer continuous spectrum for $n<0.15$ (marked by arrows $1,2$ and $3$) become unstable as the power-law index $n$ is decreased below $0.09$.

\begin{figure}
\centering
\includegraphics[width=0.6\textwidth]{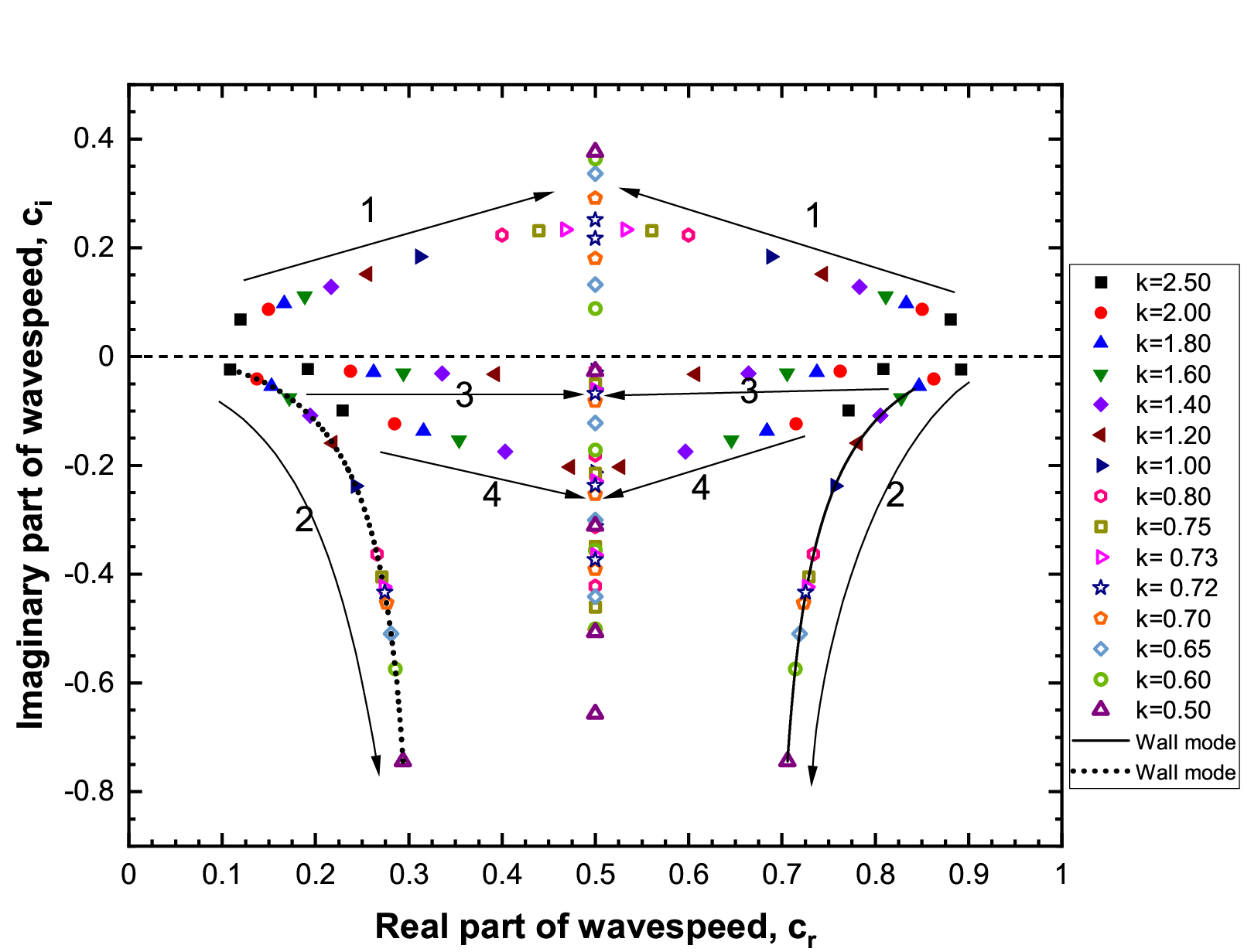}
\caption{Eigenvalue spectrum for $W=2$, $Re=0$, $n=0.1$, $\beta=0$ and a different $k$ to demonstrate the effect of the wavenumbmer.}
\label{fig:n0p1spectra}
\end{figure}

 We next show the low-wavenumber ($k = 0.5$) scenario of the different modes for $n \leq 0.15$ in figure \ref{fig:k0p5_n0p15to05}. The center modes in this wavenumber regime become unstable for  $n \leq 0.14$, while multiple center modes emerge from the polymer continuous spectra.
In figure \ref{fig:n0p1spectra}, we show the filtered spectrum focusing on the modes above the CS line for fixed power-law index ($n = 0.1$) and wavenumbers ranging from $0.1$ to $2.5$. The dashed line at $c_i = 0$ is marked as a visual aid to locate the movement of unstable modes with $k$. For $k =2.5$ and $n =0.1$, we observe four pairs of discrete modes (represented by black square symbols). As $k$ is decreased from $2.5$ to lower values, the two unstable GL modes (at $k = 2.5$) transform into two unstable center modes (marked as path $1$) at $ k \sim 0.73$. This transition can be seen for $k = 0.73$ and $k = 0.72$, where two GL modes (pink open triangles) with different $c_r$ but same $c_{i}$ transform into two center modes with same $c_{r}$ and different $c_{i}$ (marked by open star symbols). Similarly, the other set of wall modes that emerge from the continuous spectrum for $n < 0.15$, marked as path $3$ and $4$ convert into the center modes as $k$ is decreased from high to low values. However, the class of wall modes labeled WM2 (shown by arrows marked as $2$) smoothly continues to the lower-$k$ values as $k$ is decreased from high to low values, without morphing into center modes.

\begin{figure}
\centering
\includegraphics[width=0.6\textwidth]{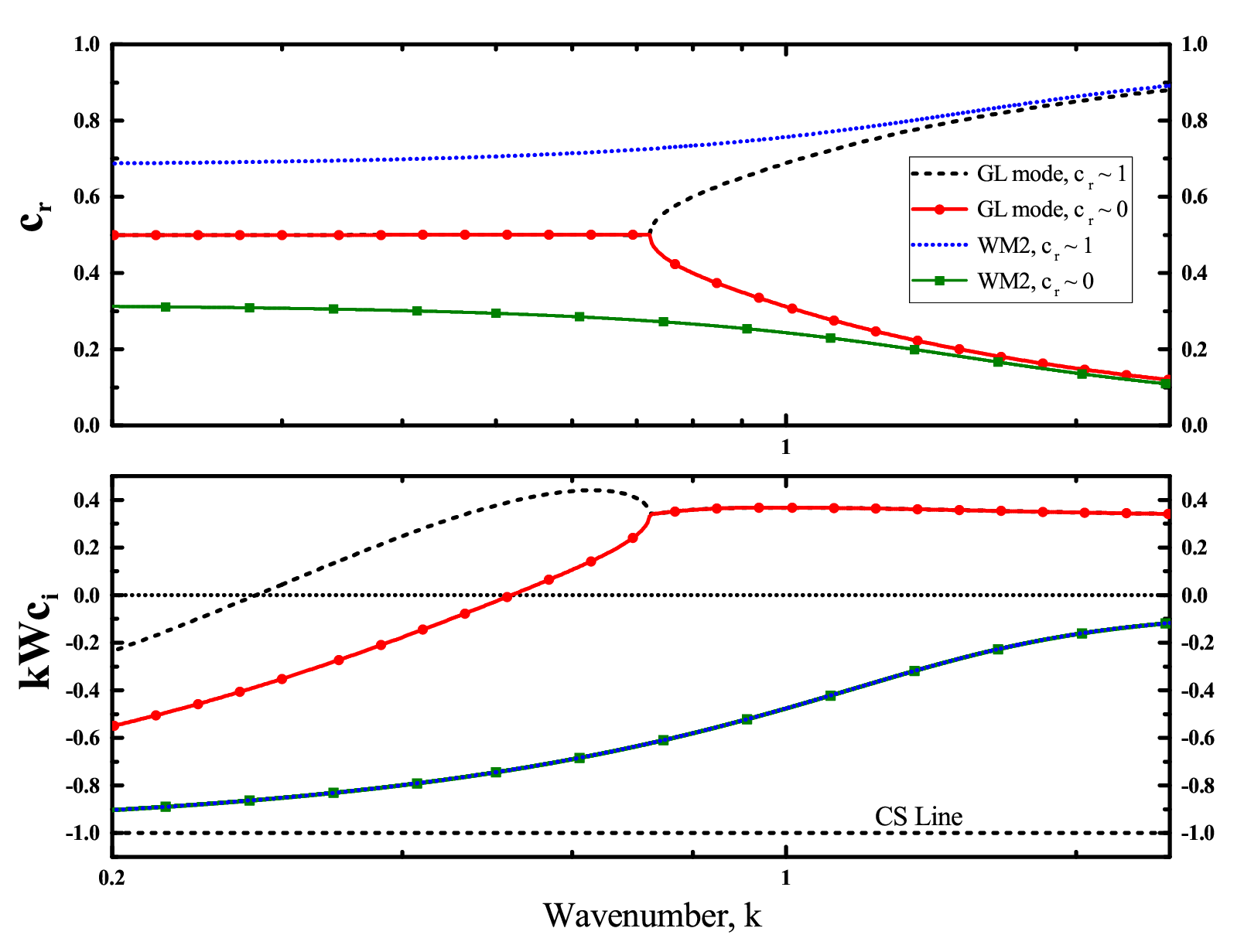}
\caption{The evolution of real part of wavespeed ($c_r$) and scaled growth rate $kWc_i$ with wavenumber, $k$. Data for $W=2$, $Re=0$, $n=0.1$ and $\beta=0$.}
\label{fig:n0p1}
\end{figure}

This picture becomes clearer in figure \ref{fig:n0p1} where we show the variation of scaled growth rate $kWc_{i}$ with the wavenumber $k$ for GL modes and wall modes (WM2) at $n =0.1$. The two unstable GL modes (different $c_r$ same growth rate) in the high wavenumber regime (figure \ref{fig:n0p1to15_highk}) convert into two unstable center modes (same $c_r$ different growth rate) at $k \approx 0.73$. Whereas, the wall modes (WM2) that were earlier converting to the center modes (for $n > 0.15$) smoothly vary with the wavenumber for $n =0.1$. Interestingly the growth rates for the GL and wall modes plateau off in the high-wavenumber regime, a phenomenon referred to as the `weak Hadamard instability' \citep{joseph-saut-1986}. The literature, however, suggests that the `Hadamard instability' is generally a consequence of neglecting a dissipative or stabilizing effect such as solvent viscous effects ($\beta$) \citep{joseph-saut-1986}, interfacial tension \citep{patne-shankar-2018}, or stress diffusion \citep{fielding2005}. In the present case, the role of solvent viscous contribution on  Hadamard instability is discussed in the subsequent subsection.\\
Note that the GL modes in the high-$k$ regime and the continuation of GL modes to the center modes in lower-$k$ regime remain the most unstable modes in the highly shear-thinning regime (with $n < 0.15$)
and hence dictate the stability of the system in this regime.\\
\textcolor{black}{Interestingly, the emergence of the GL modes can also be tracked by increasing the power-law index $n$, as discussed in appendix \ref{appA4}. The GL modes observed in the UCM limit \cite{gorodtsov-leonov-1967}, also emerge from the polymer continuous spectrum for $n > 1$.}
\subsubsection{Neutral Stability Curves}

In this subsection, we present the neutral stability curves in the $W$-$k$ plane, obtained by plotting variation of smallest Weissenberg number for the destabilization of the most unstable mode for a given wavenumber. The minimum of this curve, $W_{c}$, is the critical Weissenberg number for a given $\beta$ and $n$.
\begin{figure}
\centering
\includegraphics[width=0.6\textwidth]{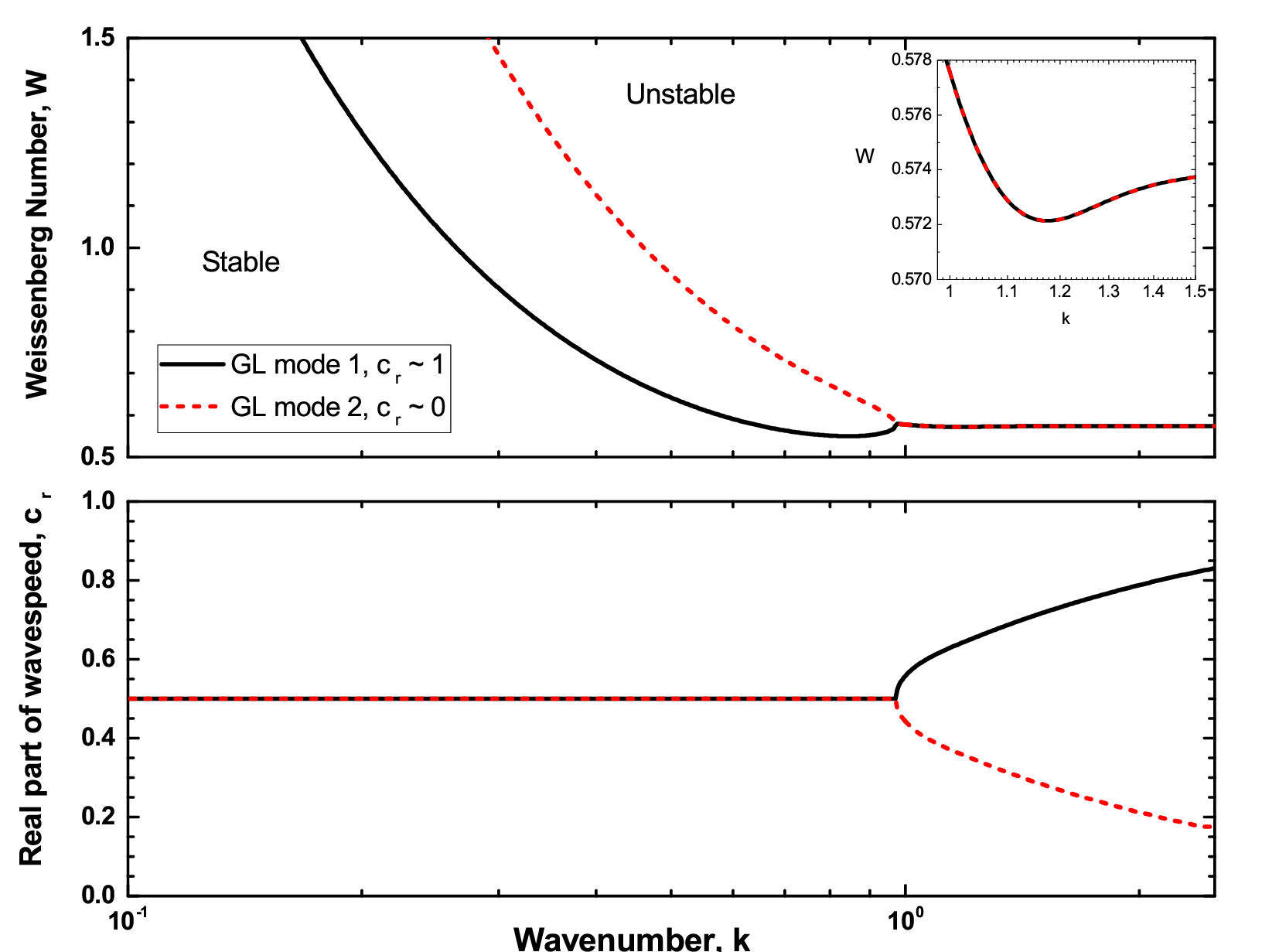}
\caption{Neutral stability curves: Panel~(a) Figure showing Neutral stability curves in $W$-$k$ plane for $Re=0$, $n =0.05$, and $\beta=0$. The corresponding $c_r$ values for the neutral stability curves in $c_r$-$k$ plane is displayed in panel~(b).}
\label{fig:n0p05_neutral_curve}
\end{figure}
 In figure \ref{fig:n0p05_neutral_curve}, panel $(a)$, for $k < 0.97$, we observe two distinct neutral stability curves corresponding to the two unstable center modes ($c_r = 0.5$) for $n = 0.05$. However, one center mode has a smaller critical Weissenberg number ($W_c$) than the other. Henceforth, we only consider the neutral curve corresponding to lower $W_c$. However, for $k > 0.97$, the two neutral stability curves overlap to represent two GL modes (with different $c_r$, but same $c_i = 0$, see figure \ref{fig:n0p05_neutral_curve}, panel $(b)$) which then extends to arbitrarily high-$k$. This crossover of the center to GL modes appears as a kink in the neutral stability curves, since we plot only the envelope for the most unstable mode. The presence of two different modes of instability is reflected in the neutral stability curves as two distinct minima, one for center mode and one for wall modes (shown in inset) in figure\ref{fig:n0p05_neutral_curve} for PCF. 

\begin{figure}
\begin{subfigure}[b]{0.5\textwidth}
        \includegraphics[width=\textwidth]{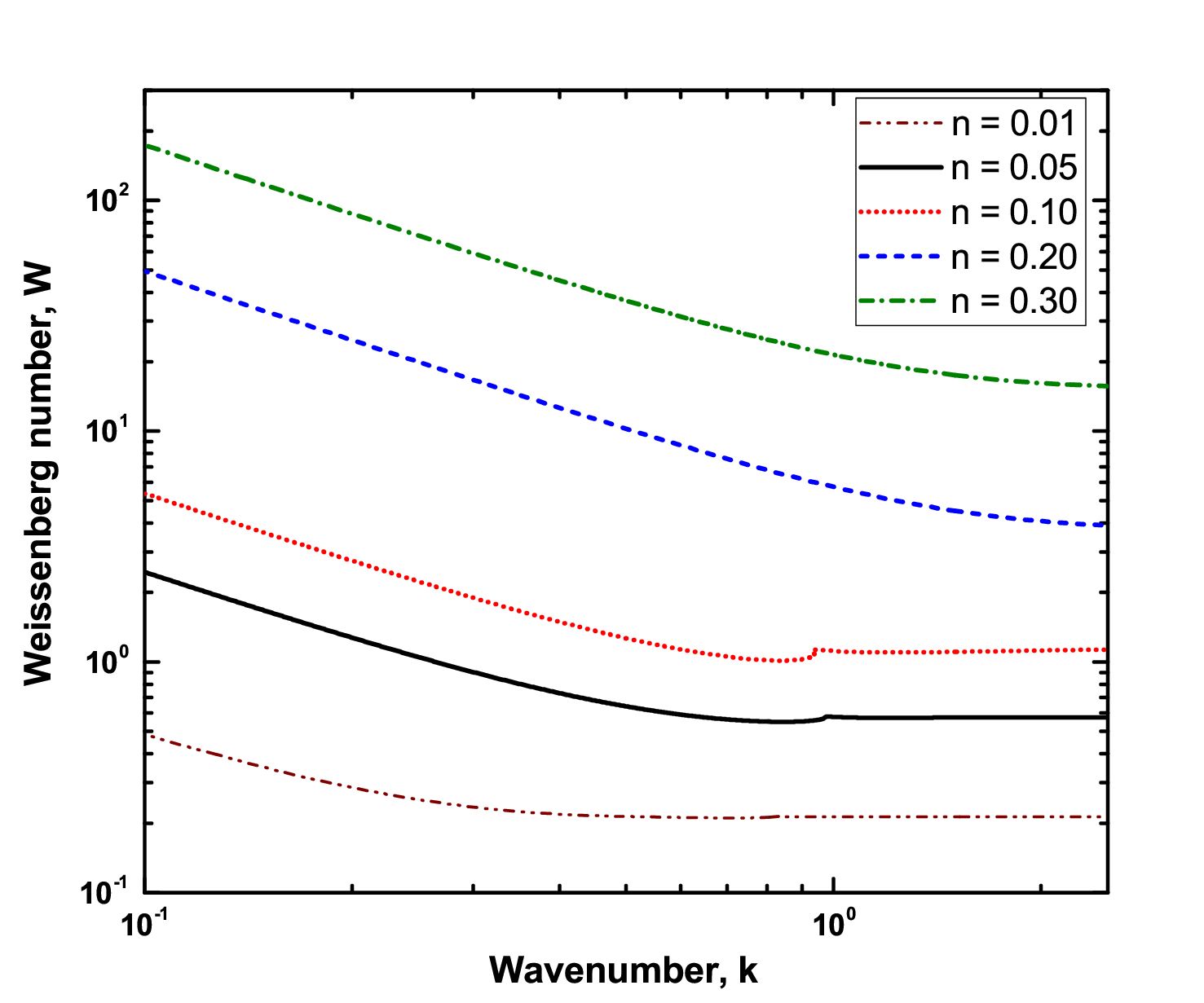}
   \caption{Variation of power-law index}
   \label{fig:variationofn_neutral_curves}
\end{subfigure}\hspace{1em}
 \begin{subfigure}[b]{0.5\textwidth}
        \includegraphics[width=\textwidth]{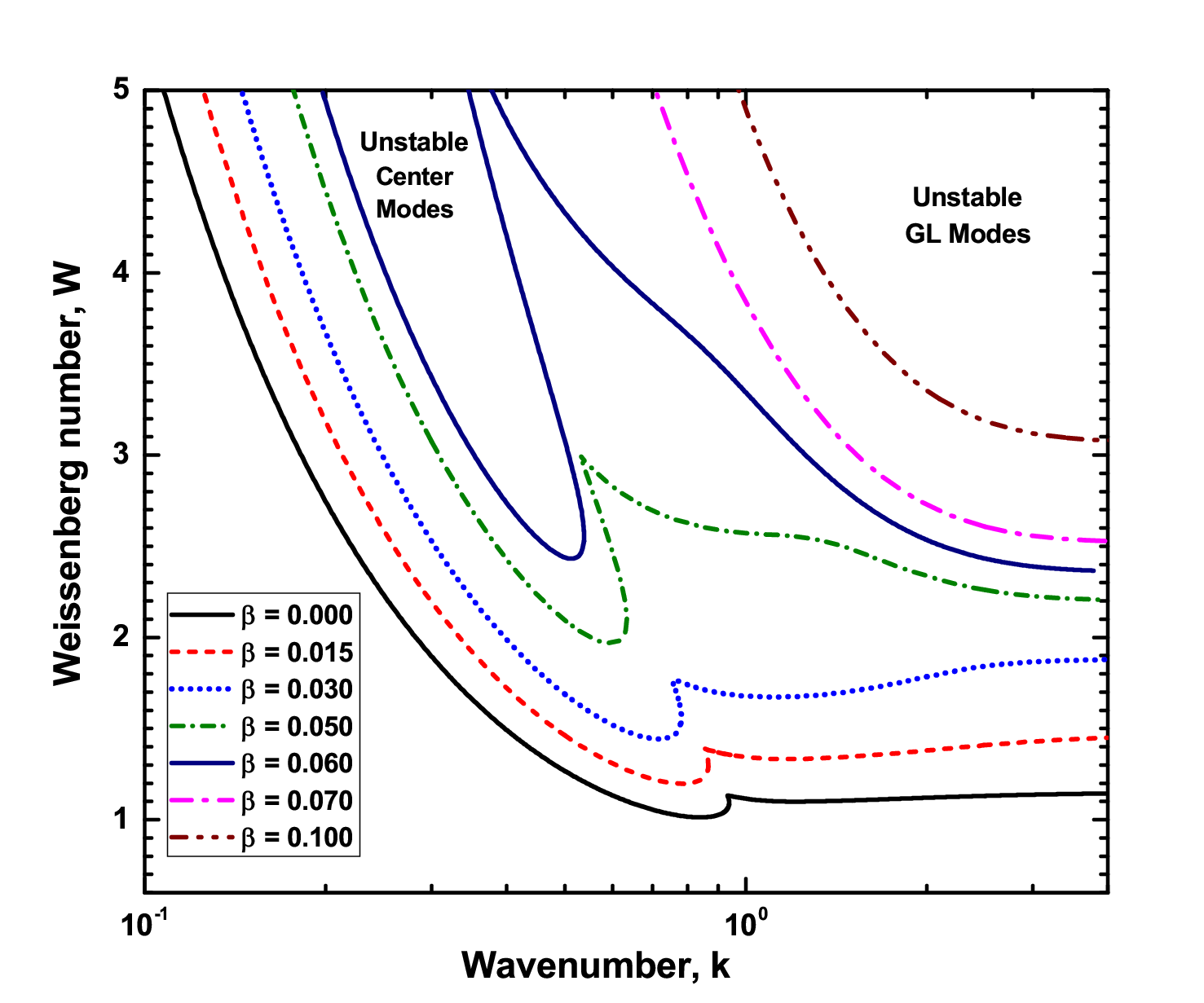}
   \caption{Variation of $\beta$}
    \label{fig:n0p1differentb}
\end{subfigure}
\caption{Neutral stability curves in the $W$--$k$ plane demonstrating the role of $n$ and $\beta$. Panel~(a) shows the neutral curves for $\beta=0$ and  for different values of $n$, while panel~(b) shows the neutral curves for $n = 0.1$ and for different $\beta$.}
\label{fig:n0p01_neutral_curves}
\end{figure}
We next show the effect of variation of power-law index $n$ and $\beta$ on the stability of the system in Fig.\,\ref{fig:n0p01_neutral_curves}. As the power-law index $n$ is increased, the neutral stability curves shift to higher $W$'s; the unstable center mode is absent for $n > 0.2$, which is characterized by the absence of the kink in the neutral envelopes in the $W$--$k$ plane as shown in subfigure \ref{fig:variationofn_neutral_curves}. Thus, for $n > 0.2$, the GL mode is the critical mode since the center mode is destabilized at
higher $W$. However, the center mode is most unstable mode for $n < 0.2$. In Fig.\,\ref{fig:n0p1differentb}, we show that, for $n = 0.1$, as the solvent viscosity parameter $\beta$ is increased, both center and GL modes are stabilized.  While the region of existence of the center modes shrinks rapidly with increase in $\beta$ and vanishes for $\beta > 0.07$, the local minimum for the GL mode instability shifts towards the high-$k$ region as $\beta$ is increased. In the high-$k$ regime, we notice that the neutral stability curves shift to higher $W$'s, but  plateau off at high $k$'s, indicating that while the solvent contribution alleviates the intensity of the Hadamard instability, it does not fully remove the weak-Hadamard instability from the system, by rendering arbitrarily high $k$ modes stable. For $\beta >0.3$, the elastic shear-thinning instability is found to be absent in all wavenumber regimes.

\begin{figure*}
\centering
    \begin{subfigure}[b]{0.48\textwidth}
        \includegraphics[width=\textwidth]{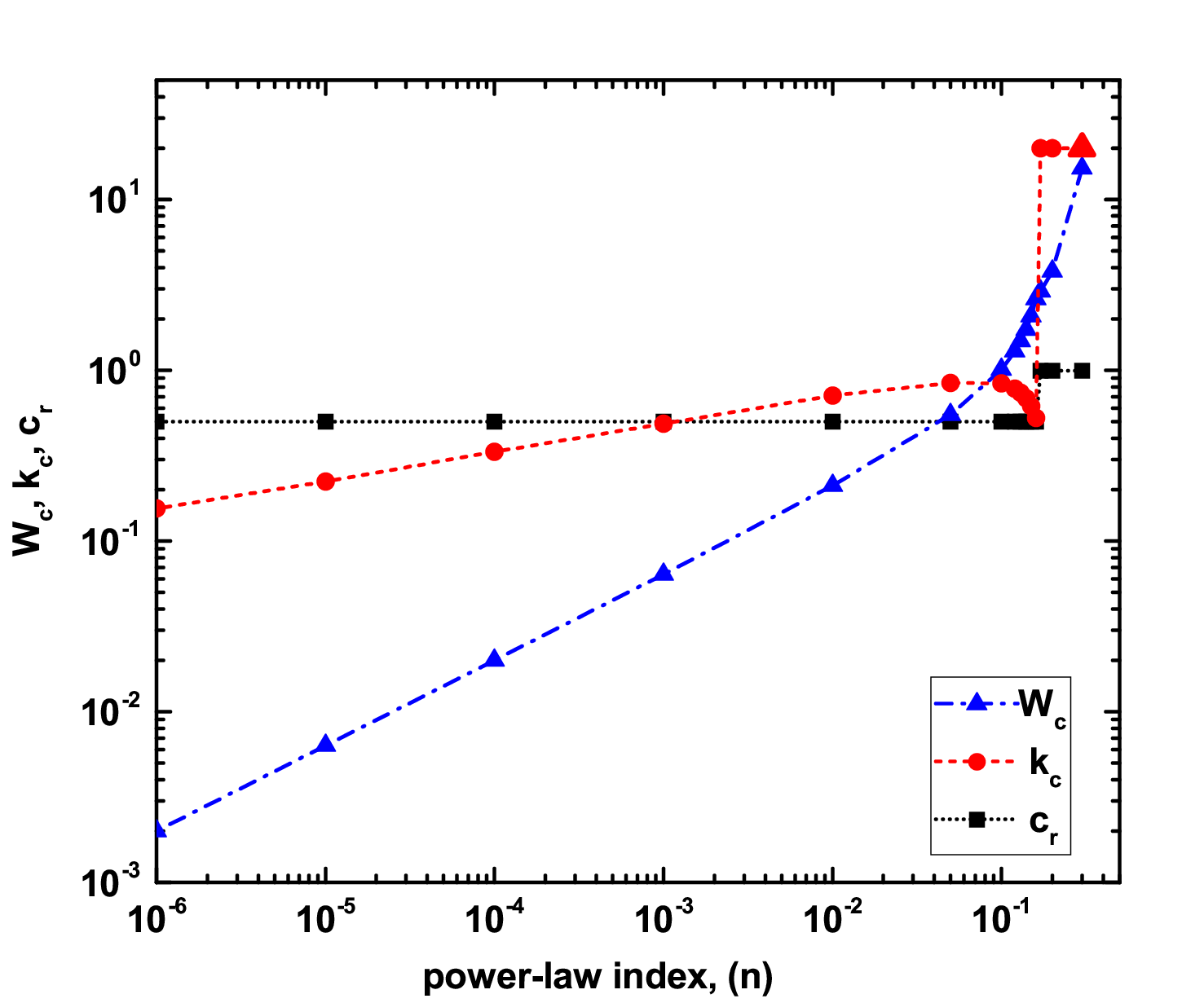}
   \caption{$\beta = 0$}
   \label{fig:critical_beta0}
\end{subfigure}\hspace{1em}
 \centering
    \begin{subfigure}[b]{0.48\textwidth}
        \includegraphics[width=\textwidth]{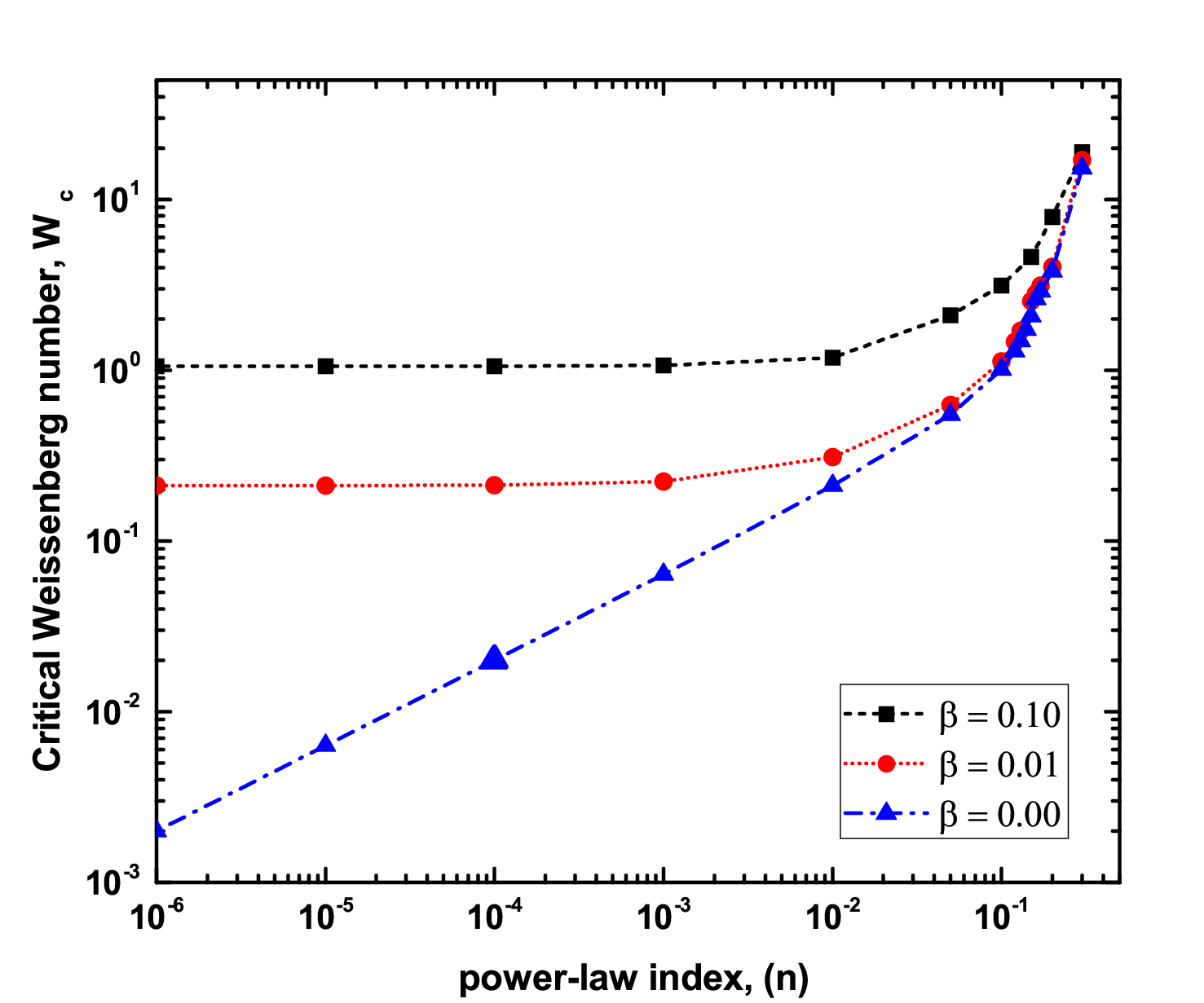}
   \caption{Effect of variation of $\beta$}
    \label{fig:critical_beta}
\end{subfigure}
\caption{Figure showing the variation of $W_c$, $k_c$ and $c_r$ with $n$: Panel~(a) shows $W_c \sim \sqrt{n}$ as $n  \rightarrow 0$ for $\beta = 0$ and panel~(b) shows the stabilizing effect of increase in the solvent viscosity in creeping flow limit.}
\label{fig:critical_neutral_curves}
\end{figure*}

Figure~\ref{fig:critical_neutral_curves} shows the critical parameters ($W_c, k_c$ and $c_r$) for the most unstable mode as a function of $n$. We observe that for $\beta =0$ (Fig.\,\ref{fig:critical_beta0}), the center modes are the most unstable modes for $n \leq 0.16$. However, a jump in the critical parameters $k_c$ and $c_r$ is observed at $n = 0.17$ as the GL modes become the most unstable modes for $n \geq 0.17$. The variation of the critical Weissenberg number $W_c$ with $n$ for $\beta = 0$ shows the scaling $W_c \sim \sqrt{n}$ as $n \rightarrow 0$ while the critical wavenumber $k_c$ decreases with the power law index with the scaling $ k_c \sim n^{0.16}$. However, a similar scaling is not observed for $\beta \neq 0$, as shown in figure \ref{fig:critical_beta}, where, instead, the critical Weissenberg number $W_c$ saturates to a constant value as $n \rightarrow 0$ due to the stabilizing effect of the solvent viscous effects ($\beta \neq 0$). The scaling $W_c \propto \sqrt{n}$ for $n \ll1$ (for $\beta = 0$) suggests that for $n = 0$, the flow should be unstable regardless of $W$. For $n < 0$, the constitutive curve is nonmonotonic and hence the homogeneous flow in such cases is known to be inherently unstable (dubbed the `constitutive' instability) \citep{yerushalmi-1970}. Thus, hydrodynamic stability predicted here for strongly shear thinning fluids with monotonic constitutive curves is likely to smoothly cross over to the constitutive instability for $n < 0$.

\subsection{Channel flow} \label{sec:ppf-results}

For channel flow of WM fluids, the earlier efforts of \cite{wilson-rallison-1999,wilson-loridan-2015,castillo-wilson-2017} were restricted to  varicose modes due to an apparent singularity at the center of channel. In stark contrast, the experiments of \cite{bodiguel-et-al-2015} observed sinuous modes
in channel flow of concentrated polymer solutions, suggesting that these could be more unstable than the varicose modes.  As discussed in Sec.~\ref{sec:pf-problem-formulation}, the apparent singularity can be removed. In this section, we analyze both sinuous and varicose modes based on the regularized constitutive equations discussed in Sec.~\ref{sec:pf-problem-formulation}.

\begin{figure*}
    \centering
    \begin{subfigure}[b]{0.5\textwidth}
        \centering
        \includegraphics[width=\textwidth]{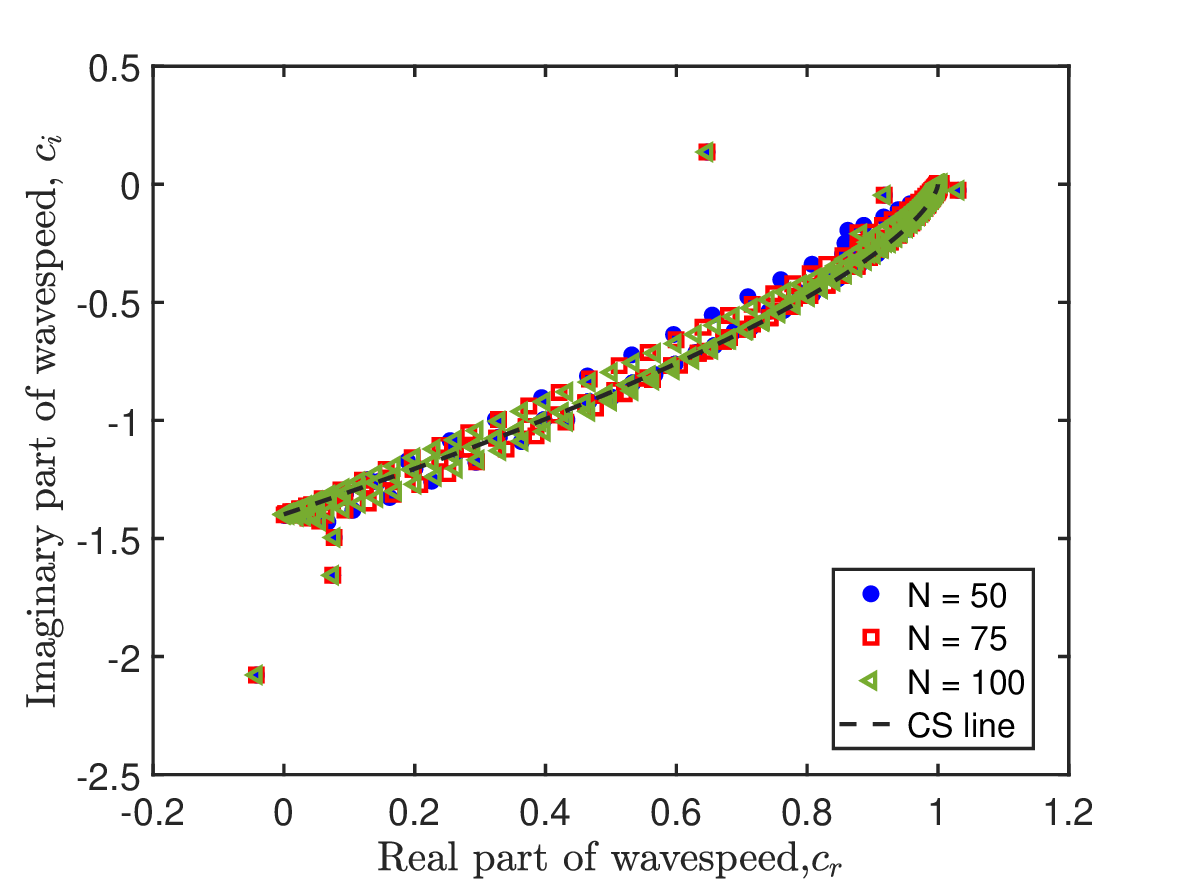}
        \caption{Varicose mode}
        \label{fig:ci_vs_cr_W_3_k_1_n_0p2_Var}
    \end{subfigure}%
    ~ 
    \begin{subfigure}[b]{0.5\textwidth}
        \centering\includegraphics[width=\textwidth]{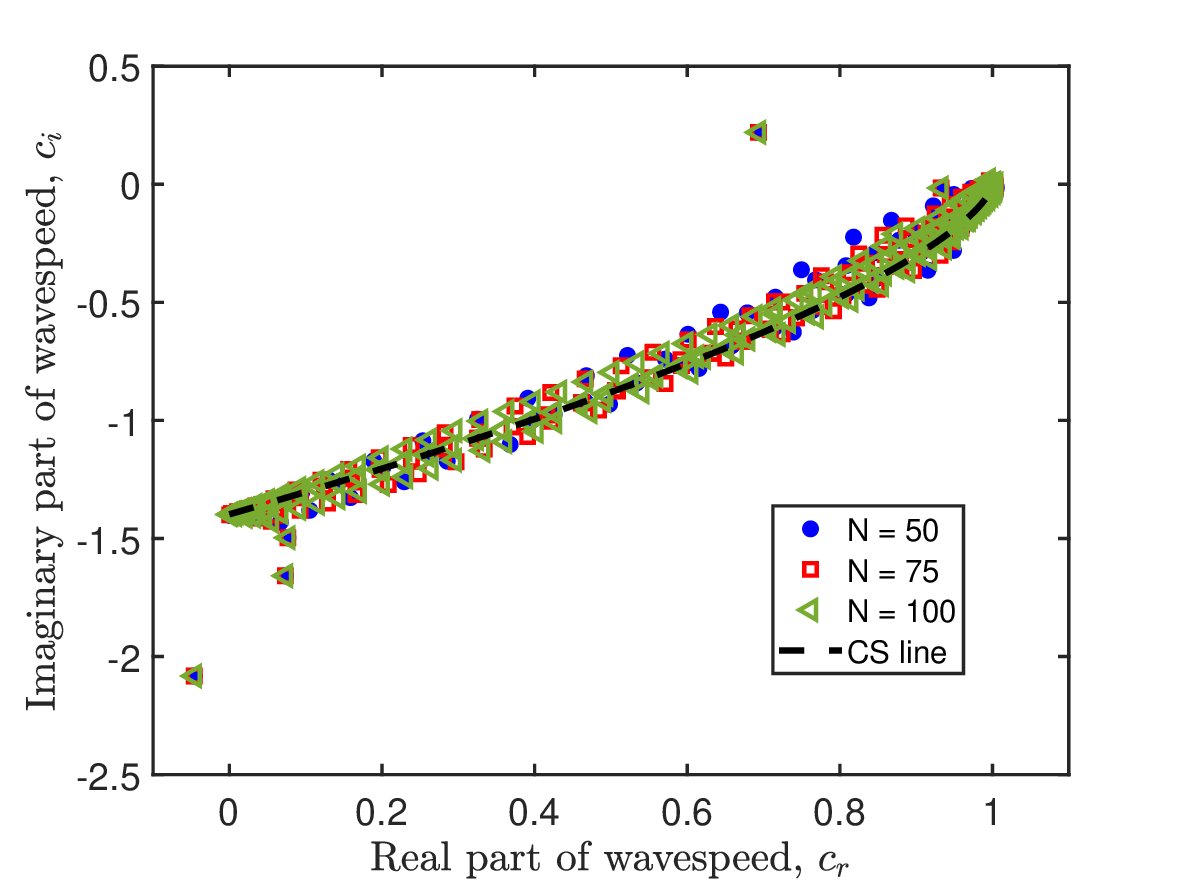}
        \caption{Sinuous mode}
          \label{fig:ci_vs_cr_W_3_k_1_n_0p2_Sin}
    \end{subfigure}
    \caption{\small The discrete and continuous spectra for the varicose (panel (a)) and sinuous (panel (b)) modes. The other parameters are $k=1, n=0.2, W=3$ and $\beta=0$. The most unstable converged eigenvalues are $c=0.6468244 + 0.1367089i$ and $c=0.6934075 +     0.22011645i$ respectively for the varicose and sinuous modes. The CS line appear as a curved line, following eq. \ref{eq:ppf-continuous-spectra-c}}
\end{figure*}

The convergence of the spectra for varicose and sinuous modes is shown in figures~\ref{fig:ci_vs_cr_W_3_k_1_n_0p2_Var} and \ref{fig:ci_vs_cr_W_3_k_1_n_0p2_Sin}, respectively. The eigenvalues belonging to the polymer continuous  spectrum for channel flow can be obtained by setting the coefficient of the stresses in (\ref{eq:pf-Txx}-\ref{eq:pf-Tyy}) to zero to obtain
\begin{eqnarray}
c= 1-y^{\frac{1+n}{n}}-\frac{i}{kW} \left(\frac{n+1}{n} \right)^{1-n}y^{\frac{1-n}{n}} \, . \label{eq:ppf-continuous-spectra-c}
\end{eqnarray}
\noindent
The eigenvalues which are $N$-dependent in figures~\ref{fig:ci_vs_cr_W_3_k_1_n_0p2_Var} and \ref{fig:ci_vs_cr_W_3_k_1_n_0p2_Sin} correspond to the (poorly resolved) continuous spectra. 
Note that, unlike the CS for plane Couette flow (Eq.\,\ref{CSline} and Fig.~\ref{fig:variationofN}), the location of the eigenvalues belonging to the CS in channel flow do not lie on a straight line (Figs.~\ref{fig:ci_vs_cr_W_3_k_1_n_0p2_Var} and \ref{fig:ci_vs_cr_W_3_k_1_n_0p2_Sin}), owing to the spatial variation of base-state shear rate in channel flow. The  genuine discrete mode converges for $N=50$ as shown in figures~\ref{fig:ci_vs_cr_W_3_k_1_n_0p2_Var}-\ref{fig:ci_vs_cr_W_3_k_1_n_0p2_Sin}.  This mode is unstable for the parameter values chosen here, and the origin of this mode is discussed below in Figs.~\ref{fig:ci_vs_cr_W_3_k_1_Destabilization} and \ref{fig:ci_vs_cr_W_3_k_1_emergence}.

\begin{figure*}
\centering
    \begin{subfigure}[b]{0.5\textwidth}
        \centering
\includegraphics[width=\textwidth]{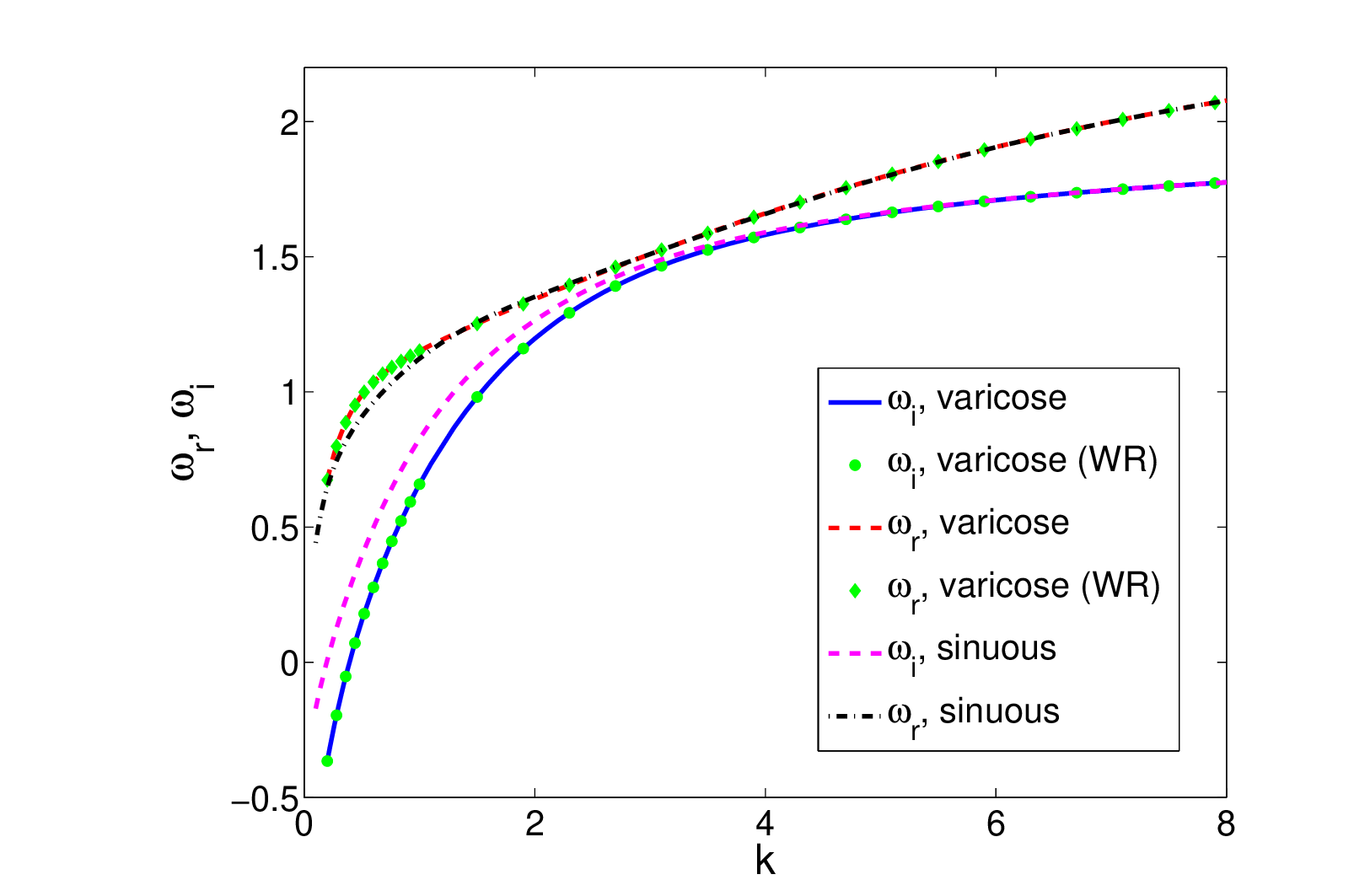}
\caption{$\omega$ vs $k$}
        \label{fig:w_vs_k_Re_0_W_2_n_0p1}
    \end{subfigure}%
\begin{subfigure}[b]{0.5\textwidth}
\centering
\includegraphics[width=\textwidth]{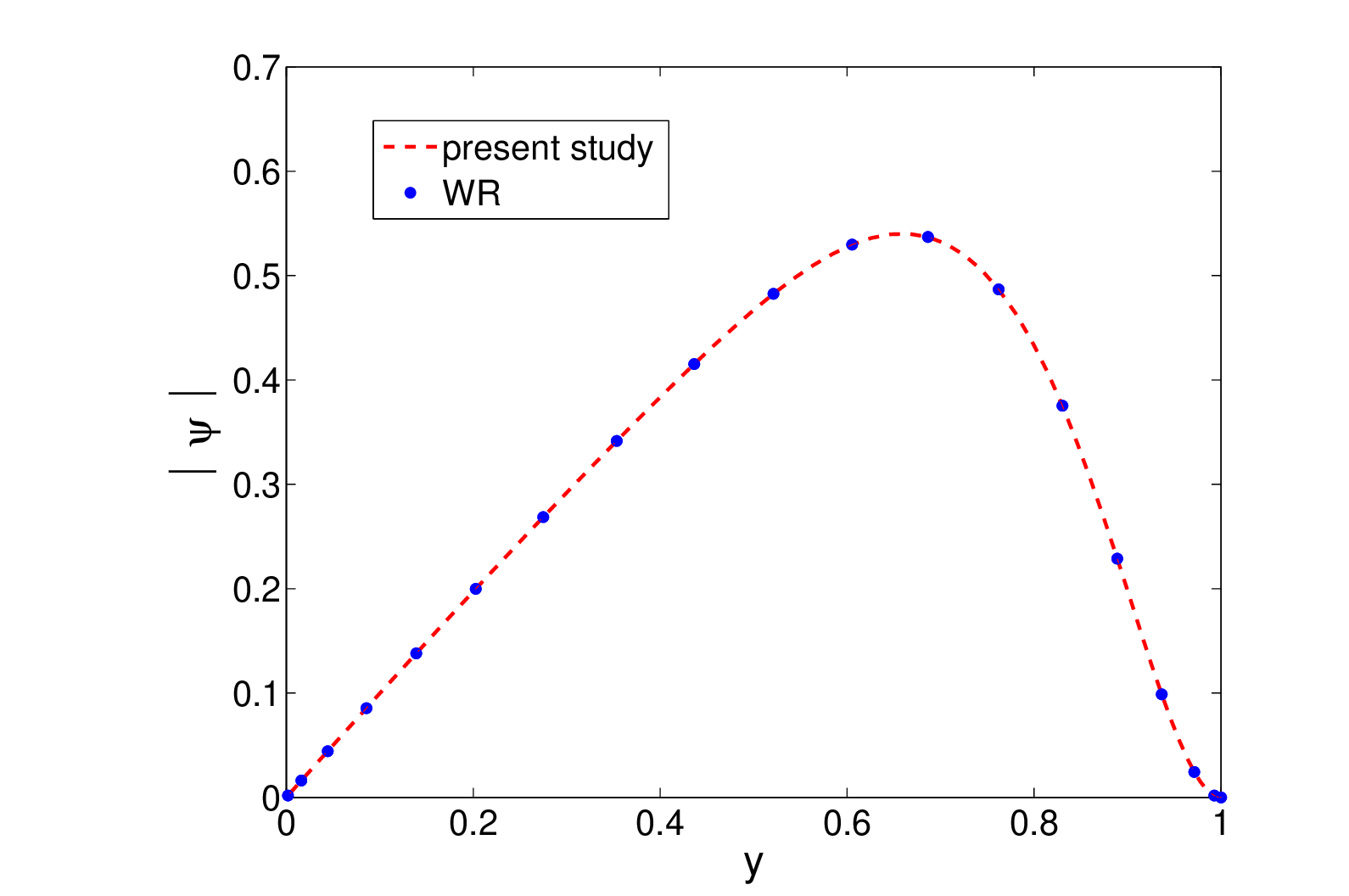}
\caption{Stream function}
\label{fig:si_vs_y_Re_0_W_2_k_1p77_n_0p2}
\end{subfigure}

\caption{\small Comparison of our channel flow results with those of \cite{wilson-rallison-1999} (abbreviated `WR' in the figure). Panel~(a) shows the variation of $\omega=kc$ for the most unstable varicose and sinuous modes at $W=3,n=0.1$ and $\beta=0$. Panel~(b) shows 
the variation of magnitude of the perturbation stream function, $\psi=-ik \tilde{v}_y$, with $y$ for the most unstable varicose mode $\omega=0.917638 +     0.093016i$ at $k=1.77, W=2, n=0.2$ and $\beta=0$.}

\end{figure*}

We first validate, in figure~\ref{fig:w_vs_k_Re_0_W_2_n_0p1}, our numerical results for the varicose mode with the earlier results of \cite{wilson-rallison-1999}; we find very good agreement between the two. Figure~\ref{fig:w_vs_k_Re_0_W_2_n_0p1} also shows that the growth rate for sinuous modes is greater than that for  varicose modes suggesting the possibility of the former being more unstable than the latter. As $k$ is increased, the eigenvalues corresponding to both the modes begin to collapse onto a  single curve showing that, at high $k$, disturbances are more localized near the wall, and center-line boundary conditions become increasingly irrelevant. The confinement of disturbances near the wall is discussed below in Sec.~\ref{sec:equivalence} in detail with the aid of eigenfuctions for all three shear flows. The perturbation eigenfunctions for the most unstable varicose mode also agree very well (figure~\ref{fig:si_vs_y_Re_0_W_2_k_1p77_n_0p2})  with those given in \cite{wilson-rallison-1999}.

The origin of the unstable mode shown in figure~\ref{fig:ci_vs_cr_W_3_k_1_n_0p2_Var} can be traced as follows. For channel flow of a UCM fluid ($n=1$), there are two discrete modes with one being varicose and the other sinuous. In figure~\ref{fig:ci_vs_cr_W_3_k_1_Destabilization}, we track the varicose mode for the UCM fluid with decreasing $n$ and find that the unstable mode for channel flow is, in fact, the continuation of the UCM discrete mode to lower $n$. Thus, we find that shear-thinning destabilizes discrete UCM modes in both PCF and channel flows. Similar to plane Couette flow, with decrease in $n$, discrete modes emerge from the continuous spectrum and become unstable with further decrease in $n$ as illustrated in figure~\ref{fig:ci_vs_cr_W_3_k_1_emergence}.

\begin{figure*}
    \centering
    \begin{subfigure}[b]{0.5\textwidth}
        \centering
        \includegraphics[width=\textwidth]{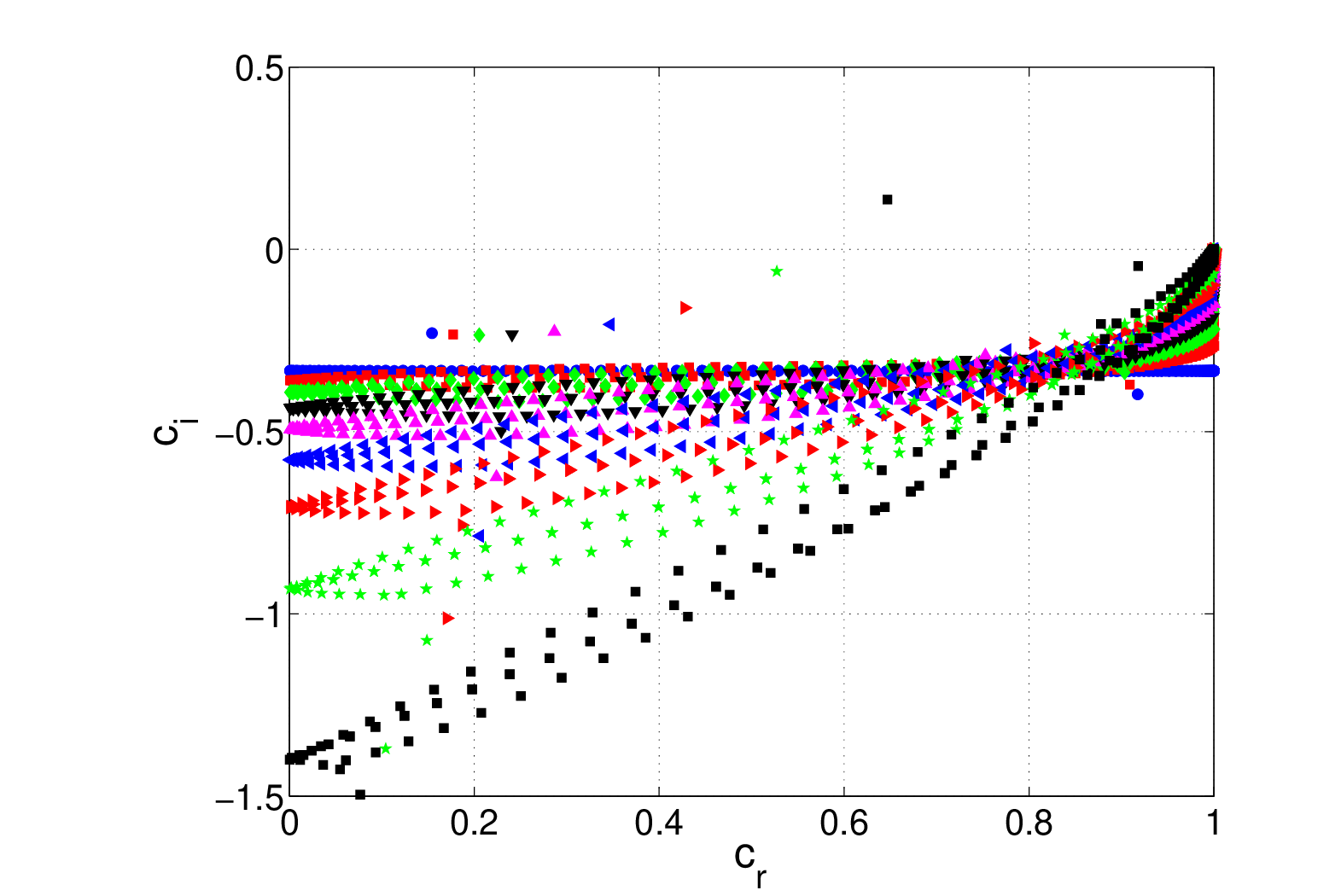}
        \caption{Destabilization of UCM modes}
        \label{fig:ci_vs_cr_W_3_k_1_Destabilization}
    \end{subfigure}%
    ~ 
    \begin{subfigure}[b]{0.5\textwidth}
        \centering
        \includegraphics[width=\textwidth]{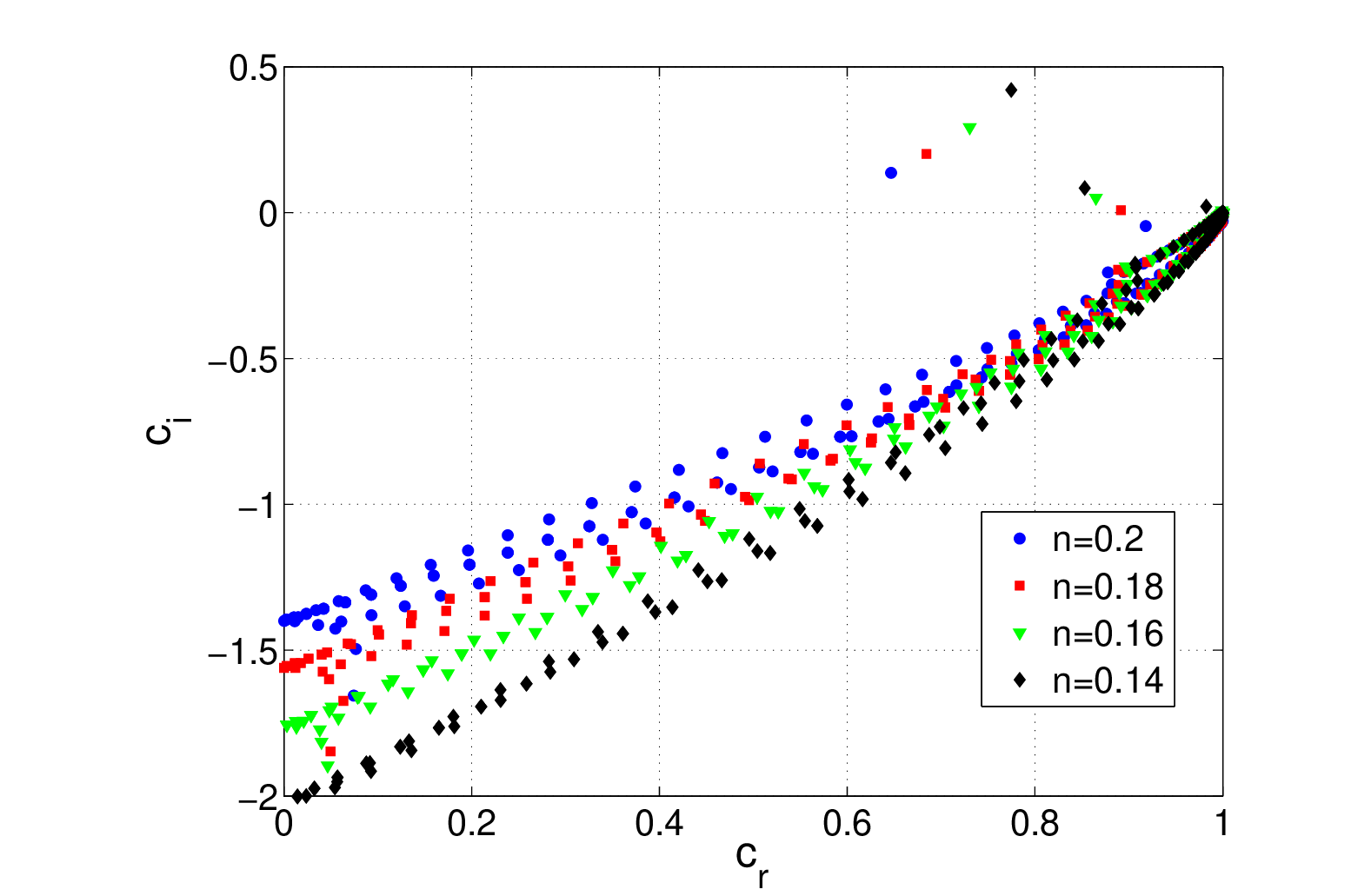}
        \caption{Emergence of a new mode}
          \label{fig:ci_vs_cr_W_3_k_1_emergence}
    \end{subfigure}
    \caption{\small The destabilization of UCM mode and emergence of of new discrete modes for the varicose modes of channel flow. Panel (a) shows destabilization of UCM mode ($n=1$) as $n$ decreases upto $0.2$ in steps of $n=0.1$. Panel (b) shows emergence of a new discrete mode with decrease in $n$. Other parameters are $k=1,W=3$ and $\beta=0$.  }
\end{figure*}

\begin{figure*}
    \centering
    \begin{subfigure}[b]{0.5\textwidth}
        \centering
        \includegraphics[width=\textwidth]{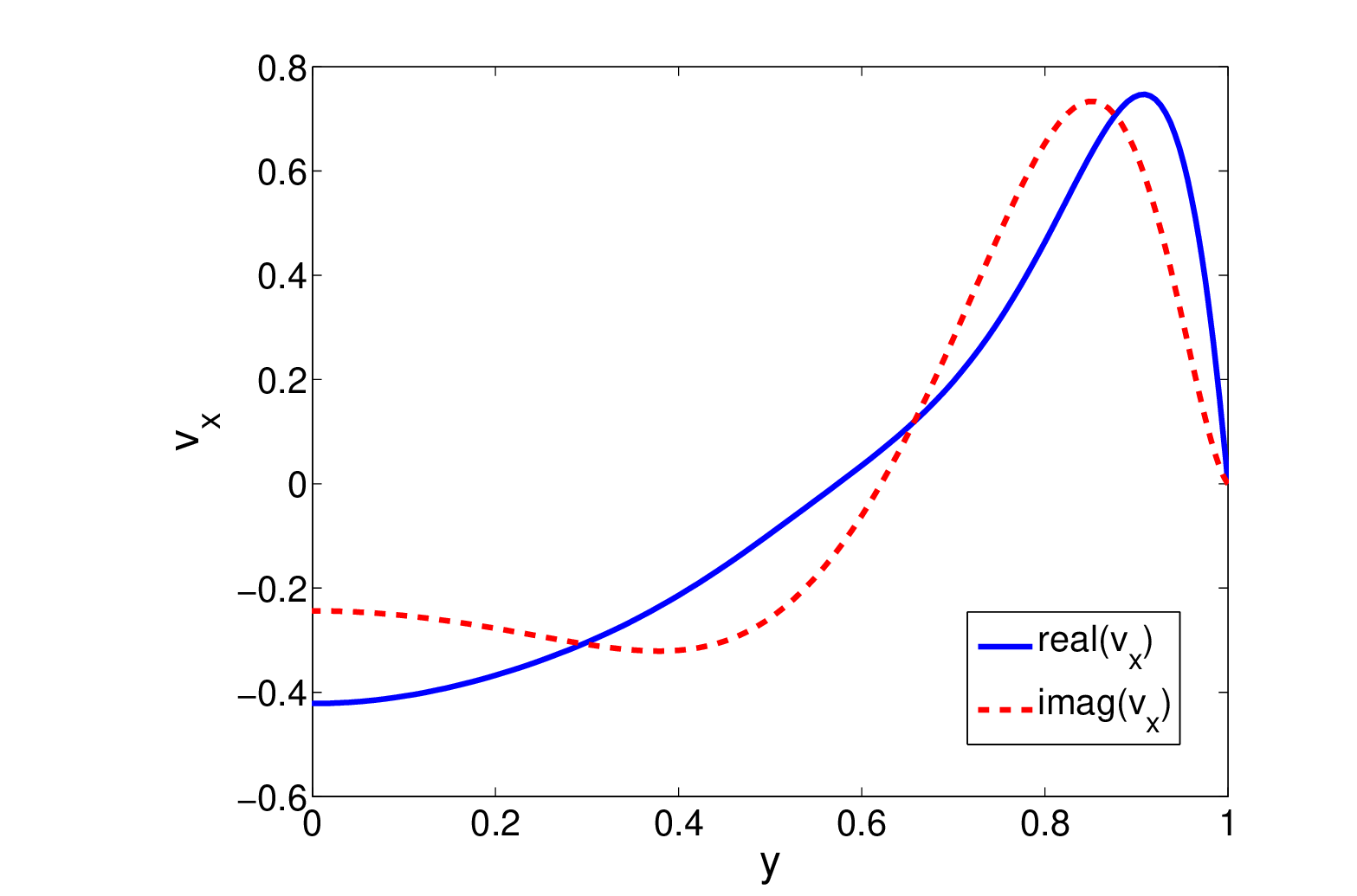}
        \caption{$v_x$-eigenfunction}
        \label{fig:vx_vs_y_W_3_k_1_n_0p2_Var}
    \end{subfigure}%
    ~ 
    \begin{subfigure}[b]{0.5\textwidth}
        \centering
        \includegraphics[width=\textwidth]{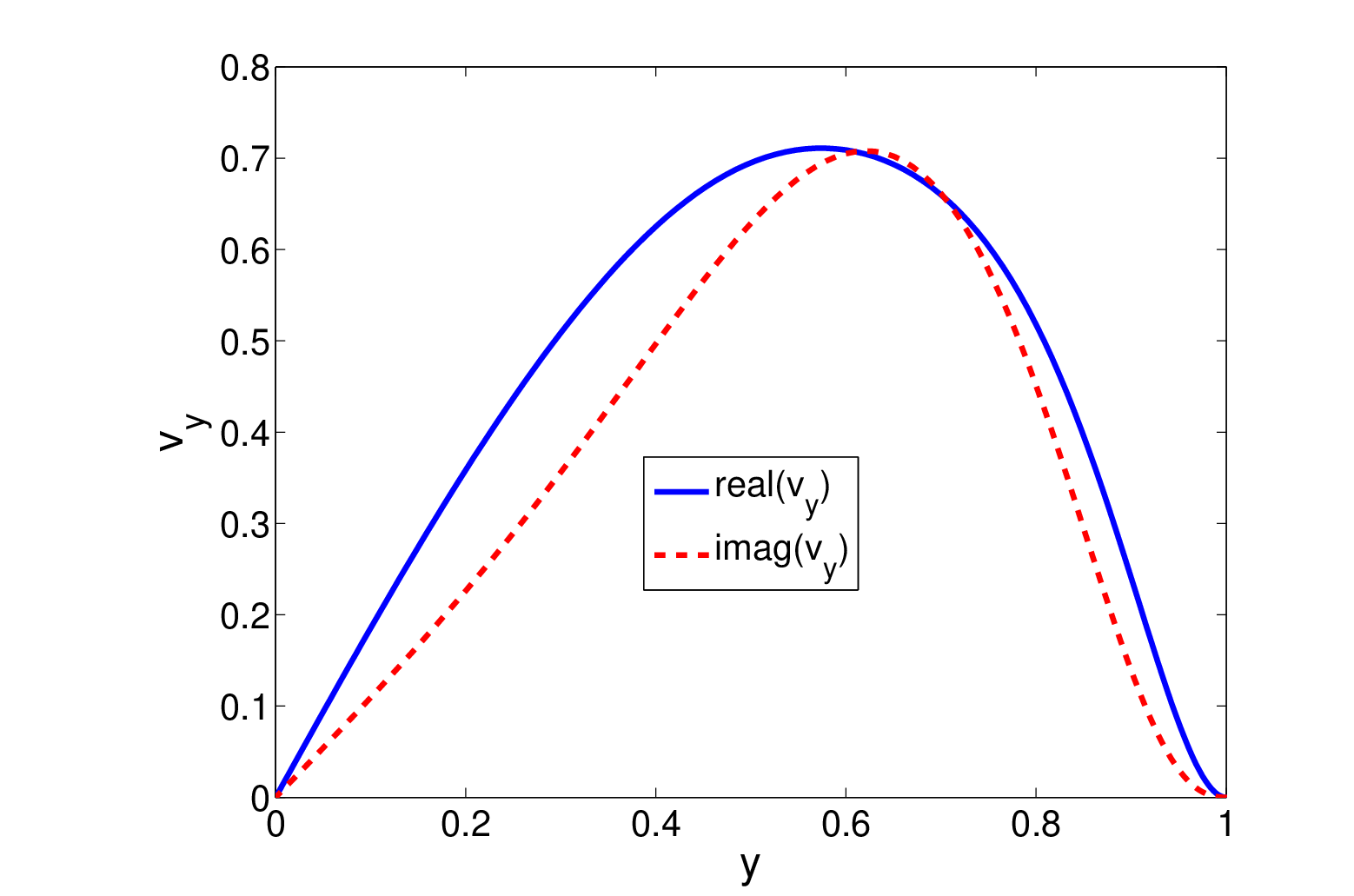}
        \caption{$v_y$-eigenfunction}
          \label{fig:vy_vs_y_W_3_k_1_n_0p2_Var}
    \end{subfigure}\\
        \centering
    \begin{subfigure}[b]{0.5\textwidth}
        \centering
        \includegraphics[width=\textwidth]{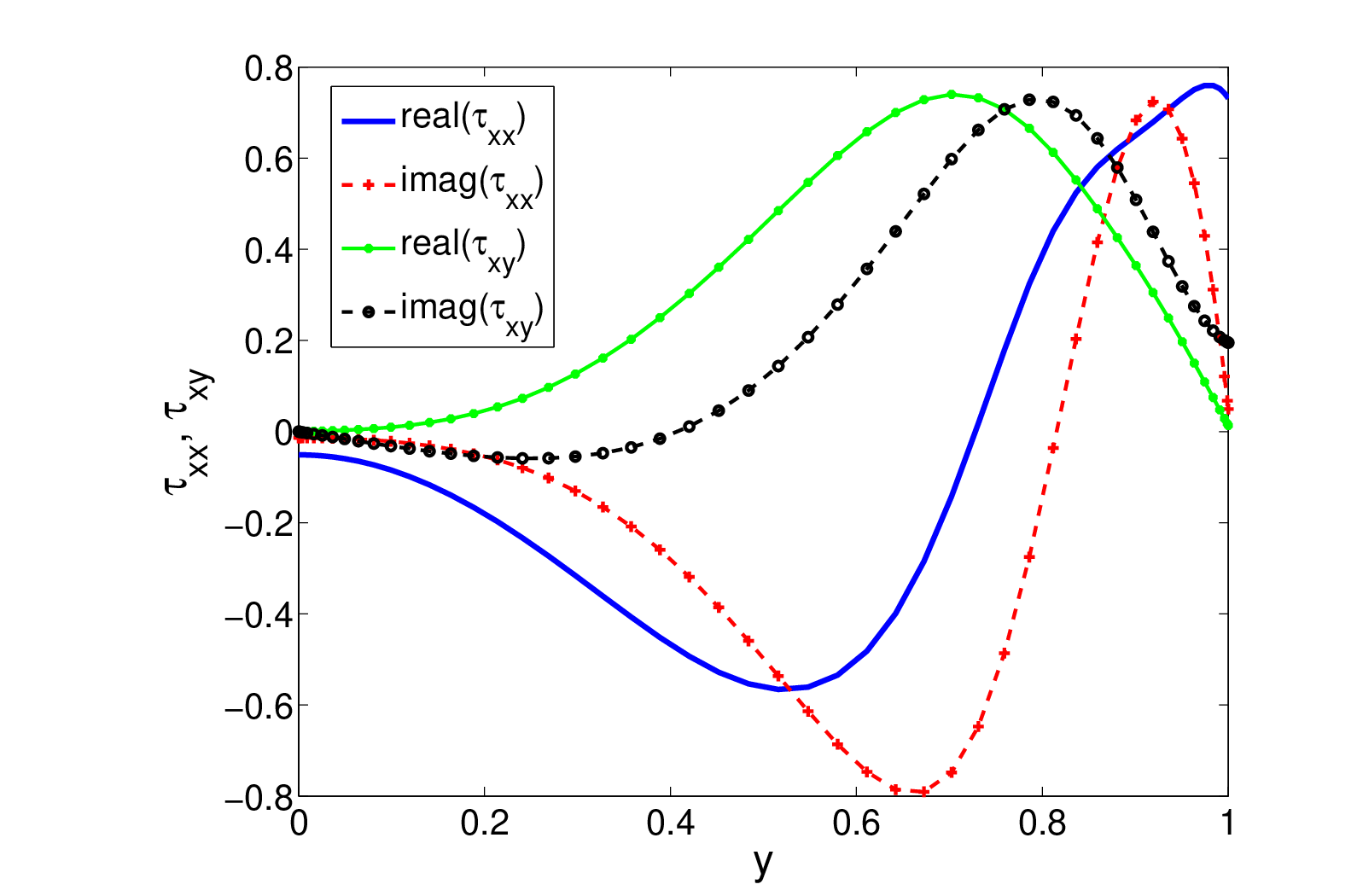}
        \caption{$\tau_{xx}$ and $\tau_{xy}$-eigenfunctions}
        \label{fig:Txx_Txy_vs_y_W_3_k_1_n_0p2_Var}
    \end{subfigure}%
    ~ 
    \begin{subfigure}[b]{0.5\textwidth}
        \centering
        \includegraphics[width=\textwidth]{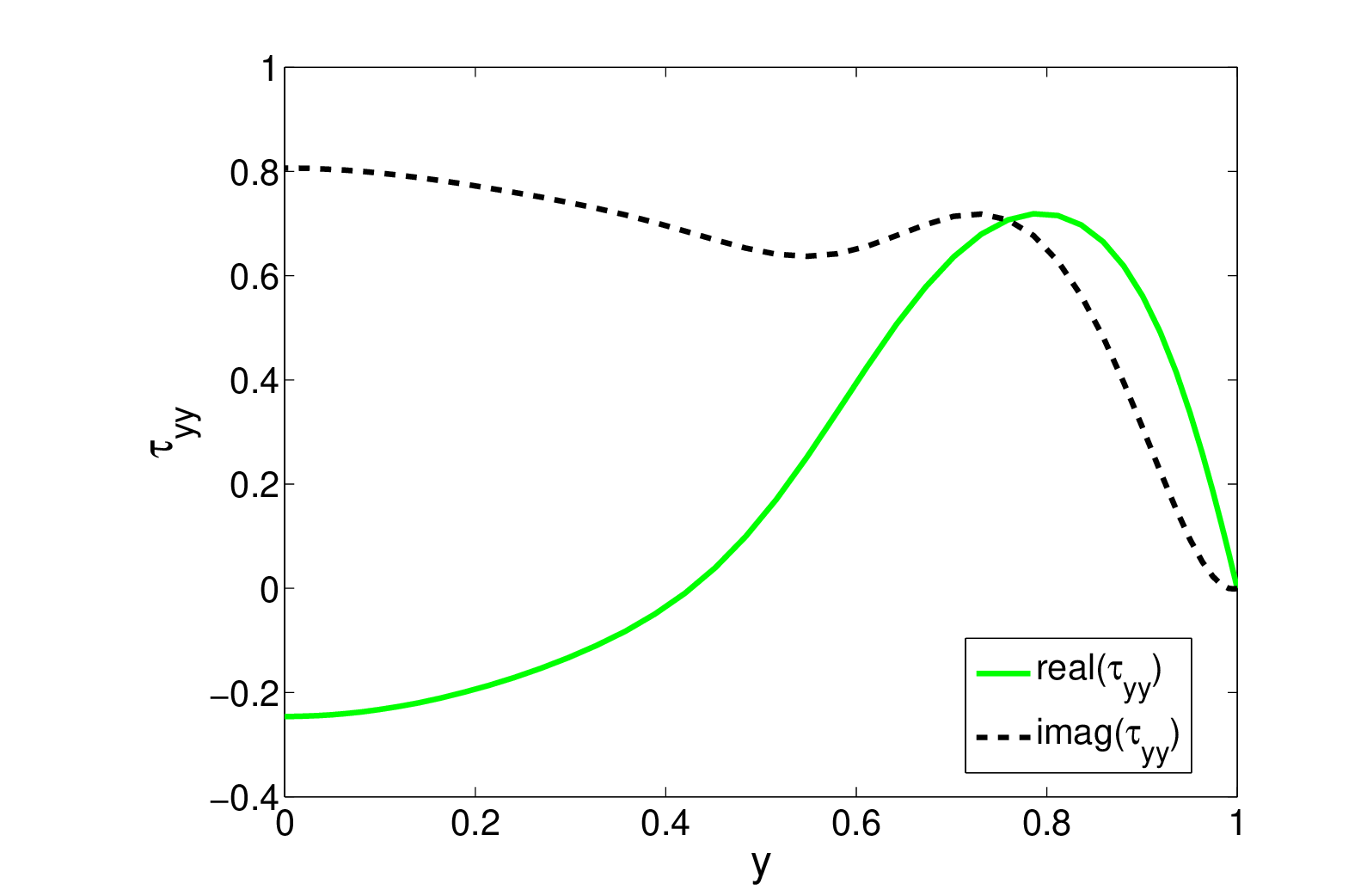}
        \caption{$\tau_{yy}$-eigenfunction}
          \label{fig:Tyy_vs_y_W_3_k_1_n_0p2_Var}
    \end{subfigure}
    \caption{\small The normalised eigenfunctions for the most unstable varicose mode, $c=0.6468244 + 0.1367089i$. Other parameters are $k=1, n=0.2, W=3$ and $\beta=0$.}
\end{figure*}

The tangential and normal velocity eigenfunctions corresponding to the most unstable varicose mode in figure~\ref{fig:ci_vs_cr_W_3_k_1_n_0p2_Var} are shown in figures~\ref{fig:vx_vs_y_W_3_k_1_n_0p2_Var} and \ref{fig:vy_vs_y_W_3_k_1_n_0p2_Var}.  In order to demonstrate the validity of regularization discussed in Sec.~\ref{sec:pf-problem-formulation}, we plot the eigenfunctions for the stresses obtained from (\ref{eq:pf-Txx}-\ref{eq:pf-Tyy}) in figures~\ref{fig:Txx_Txy_vs_y_W_3_k_1_n_0p2_Var} and \ref{fig:Tyy_vs_y_W_3_k_1_n_0p2_Var} for the converged eigenvalue of figure~\ref{fig:ci_vs_cr_W_3_k_1_n_0p2_Var}. The eigenfunctions in \ref{fig:Txx_Txy_vs_y_W_3_k_1_n_0p2_Var} and \ref{fig:Tyy_vs_y_W_3_k_1_n_0p2_Var} are continuous and satisfy boundary conditions (\ref{eq:ppf-bc1}-\ref{eq:ppf-sinuous}) thereby demonstrating  the validity of the regularization of governing equations implemented in Sec.~\ref{sec:pf-problem-formulation}.

\begin{figure*}
    \centering
    \begin{subfigure}[b]{0.5\textwidth}
        \centering
        \includegraphics[width=\textwidth]{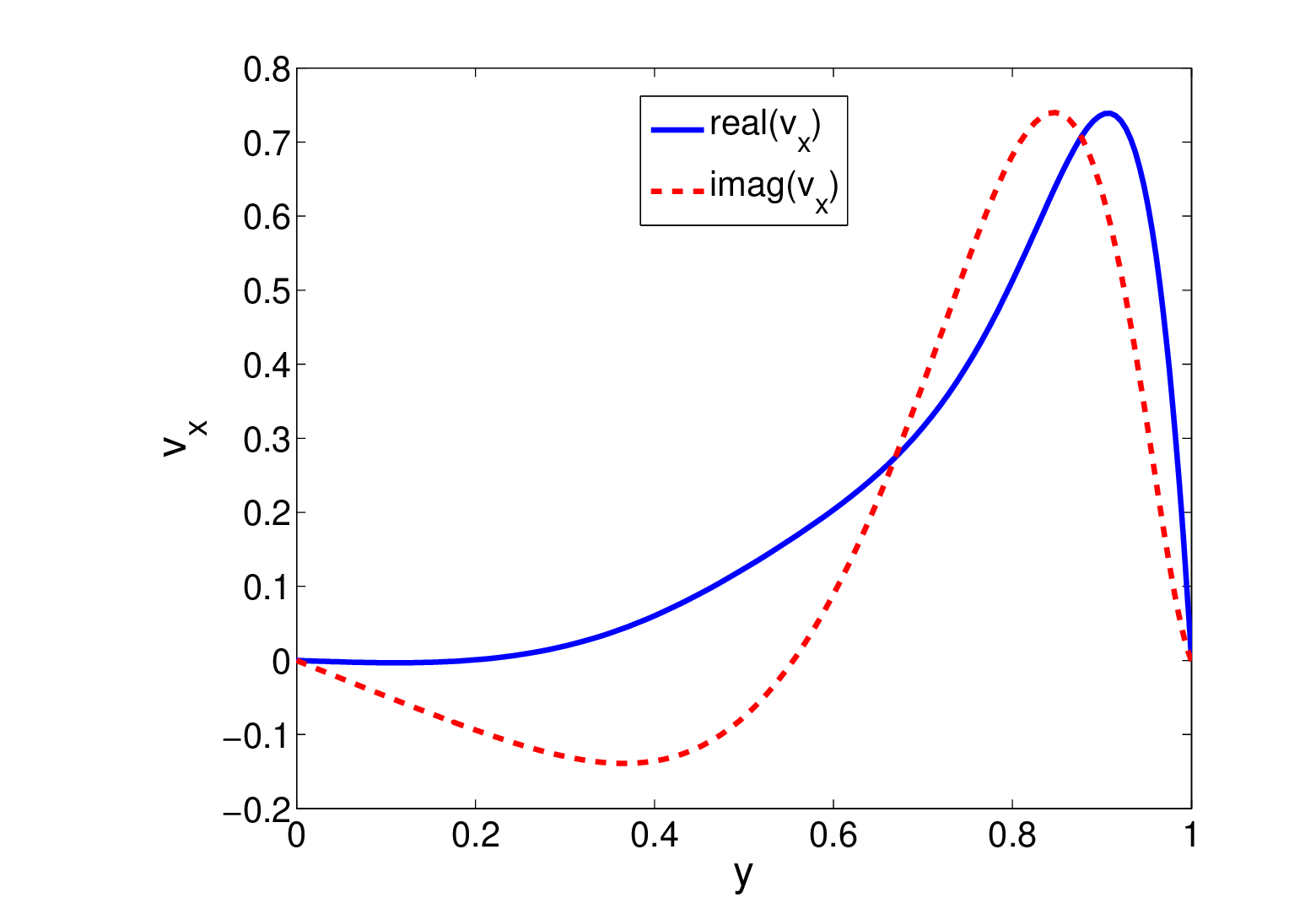}
        \caption{$v_x$-eigenfunction}
        \label{fig:vx_vs_y_W_3_k_1_n_0p2_Sin}
    \end{subfigure}%
    ~ 
    \begin{subfigure}[b]{0.5\textwidth}
        \centering
        \includegraphics[width=\textwidth]{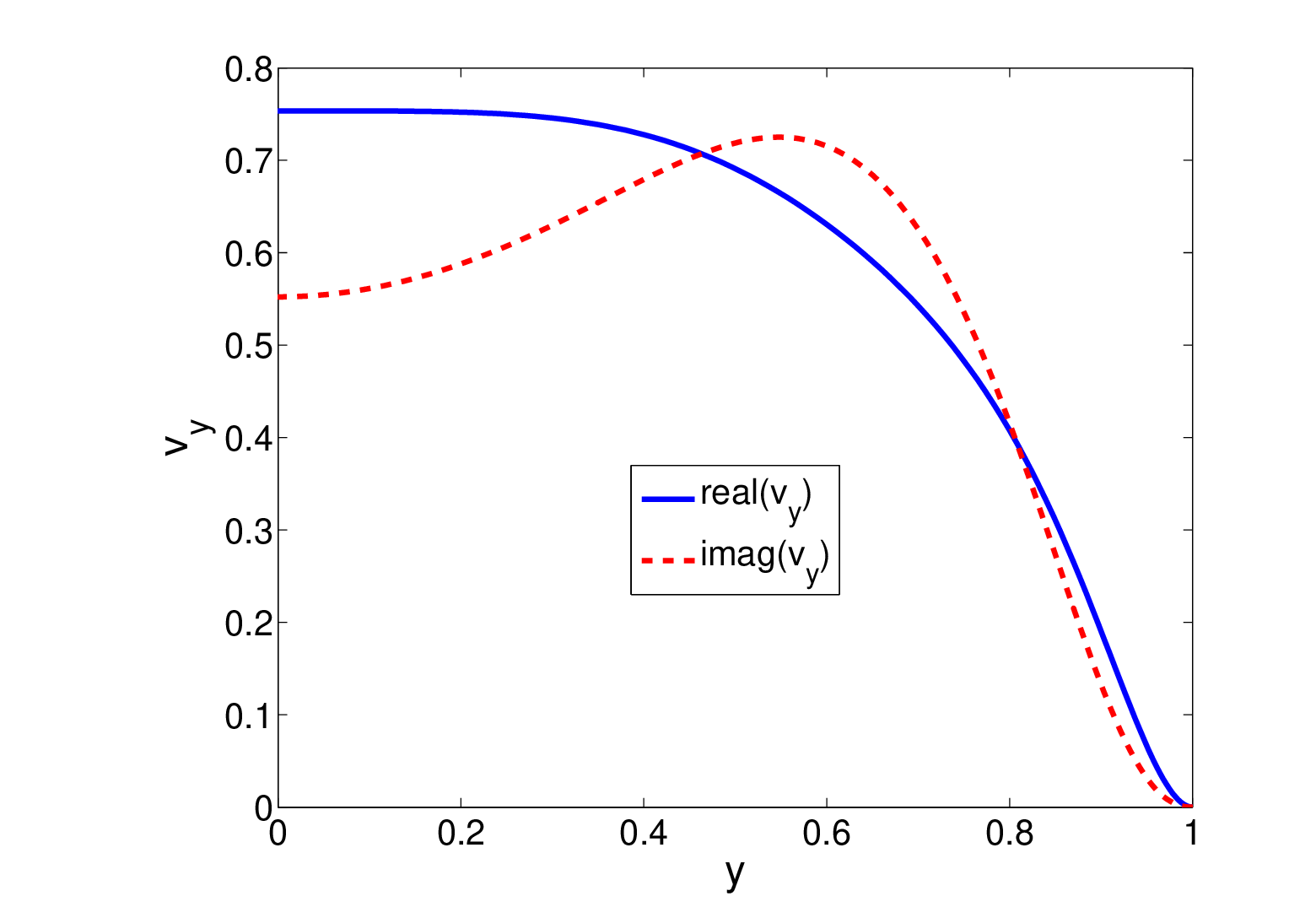}
        \caption{$v_y$-eigenfunction}
          \label{fig:vy_vs_y_W_3_k_1_n_0p2_Sin}
    \end{subfigure}\\
        \centering
    \begin{subfigure}[b]{0.5\textwidth}
        \centering
        \includegraphics[width=\textwidth]{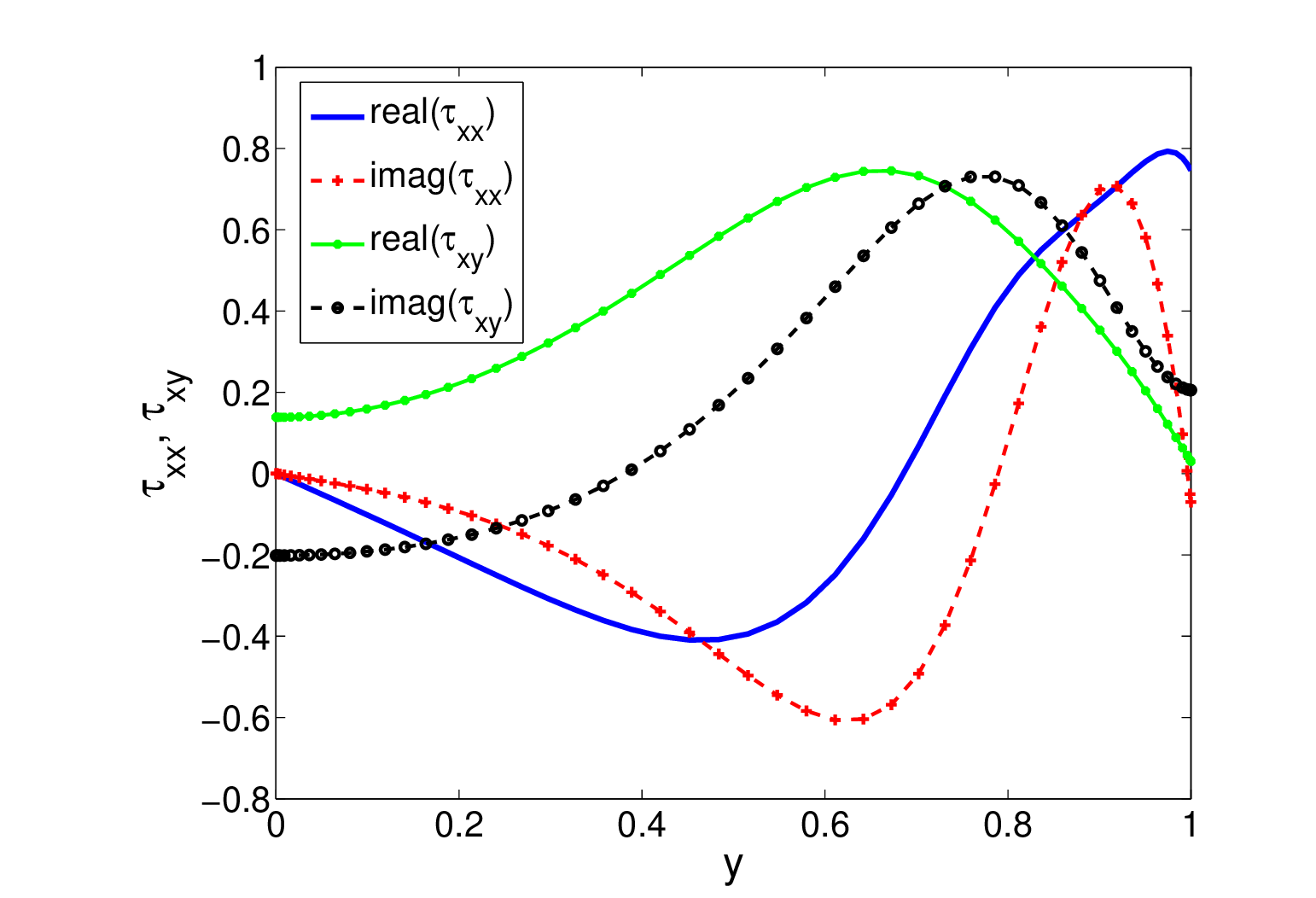}
        \caption{$\tau_{xx}$ and $\tau_{xy}$-eigenfunctions}
        \label{fig:Txx_Txy_vs_y_W_3_k_1_n_0p2_Sin}
    \end{subfigure}%
    ~ 
    \begin{subfigure}[b]{0.5\textwidth}
        \centering
        \includegraphics[width=\textwidth]{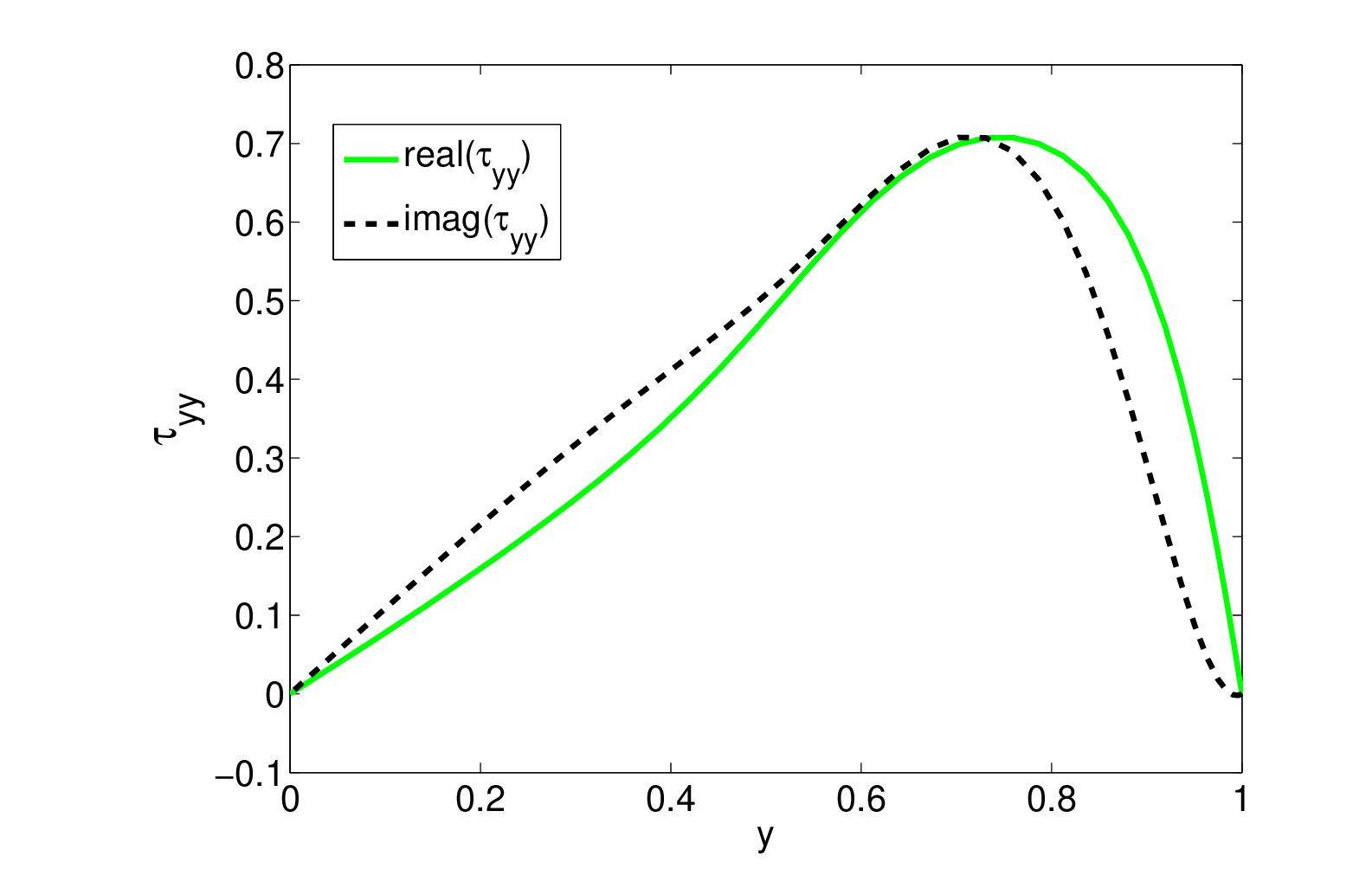}
        \caption{$\tau_{yy}$-eigenfunction}
          \label{fig:Tyy_vs_y_W_3_k_1_n_0p2_Sin}
    \end{subfigure}
    \caption{\small The normalised eigenfunction plots for the most unstable sinuous mode, $c=0.6934075 +     0.22011645i$ . Other parameters are $k=1, n=0.2, W=3$ and $\beta=0$. The smooth variation of the eigenfunctions indicates success of the numerical method. }
\end{figure*}

The tangential and normal velocity eigenfunctions for the most unstable sinuous mode discussed earlier in figure~\ref{fig:ci_vs_cr_W_3_k_1_n_0p2_Sin} show that the velocities are regular and are in agreement with the center-line conditions for sinuous modes (figures~\ref{fig:vx_vs_y_W_3_k_1_n_0p2_Sin} and \ref{fig:vy_vs_y_W_3_k_1_n_0p2_Sin}). Figures~\ref{fig:Txx_Txy_vs_y_W_3_k_1_n_0p2_Sin} and \ref{fig:Tyy_vs_y_W_3_k_1_n_0p2_Sin} demonstrate that the stress components too are regular at the channel center. The eigenfunctions shown in figures~\ref{fig:vx_vs_y_W_3_k_1_n_0p2_Var}-\ref{fig:Tyy_vs_y_W_3_k_1_n_0p2_Sin} exhibit a maximum closer to the wall indicating that the disturbances are confined near the wall. The latter feature is more pronounced at high $k$, and is common to all the three (viz., PCF, channel and tube) flows analyzed in this study. 

\begin{figure*}
    \centering
    \begin{subfigure}[b]{0.5\textwidth}
        \centering
        \includegraphics[width=\textwidth]{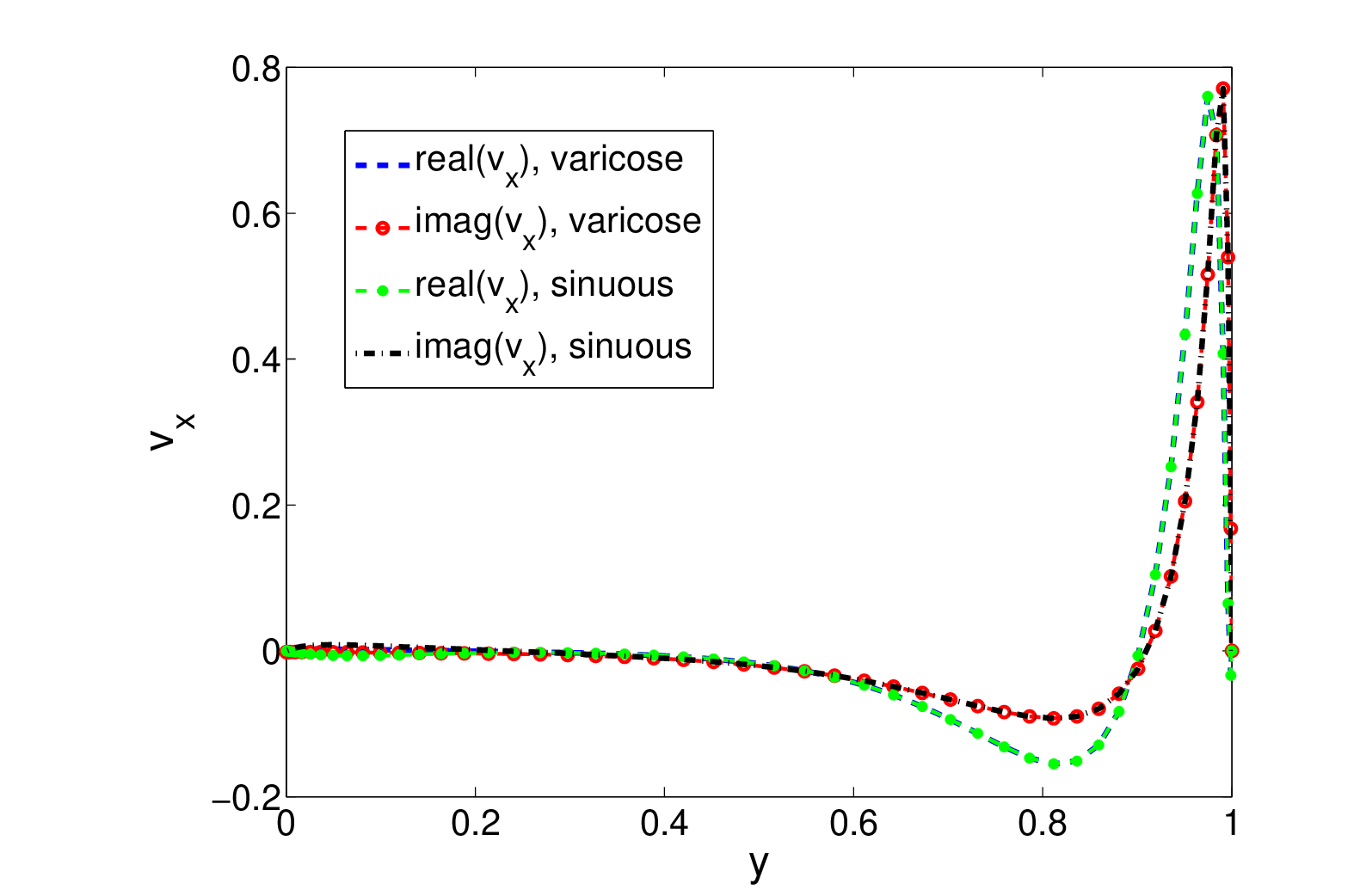}
        \caption{$v_x$-eigenfunction}
        \label{fig:vx_vs_y_W_5_k_10_n_0p2_ppf}
    \end{subfigure}%
    ~ 
    \begin{subfigure}[b]{0.5\textwidth}
        \centering
        \includegraphics[width=\textwidth]{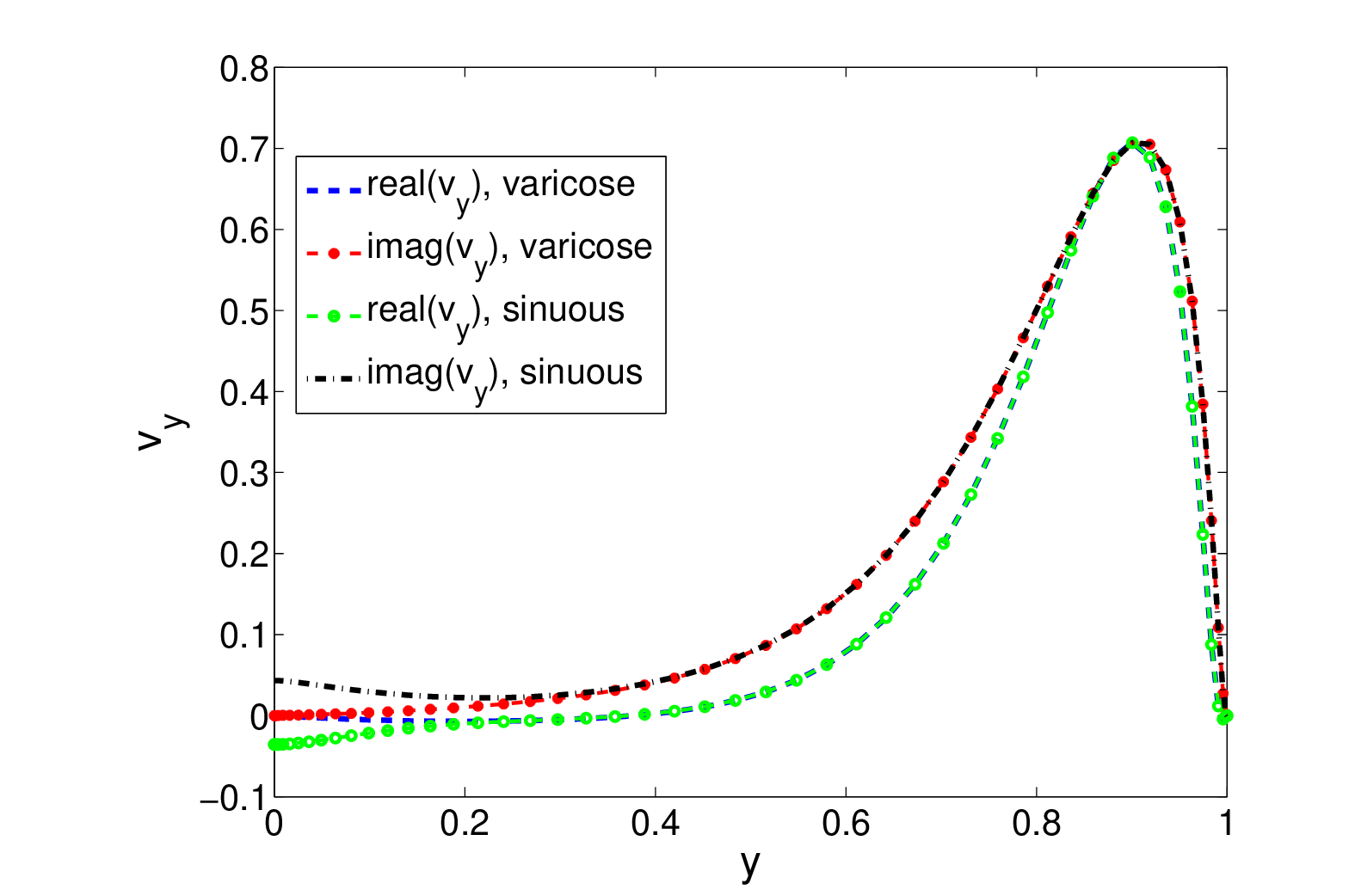}
        \caption{$v_y$-eigenfunction}
          \label{fig:vy_vs_y_W_5_k_10_n_0p2_ppf}
    \end{subfigure}
    \caption{\small The normalised eigenfunction plots for the most unstable varicose and sinuous modes, $c=0.04642377 +    0.01324292i$  for channel flow using WM-PL model. Other parameters are $k=10, n=0.2, W=5$ and $\beta=0$. The overlapping of the varicose and sinuous eigenfunctions for $y>0.3$ is due to the confinement of the disturbances to the wall region.}
\end{figure*}

A comparison of the eigenfunctions for varicose and sinuous modes at high $k$ shows that, close to the wall, the perturbation velocities and  stresses behave independently of the center-line boundary conditions, and both varicose and sinuous modes ought to converge to the same eigenvalue. This is illustrated in table~\ref{table:varicose-sinuous-comparison}, where, for $k>5$, both varicose and sinuous mode eigenvalues agree very well. The value of $k$ required for the agreement between varicose and sinuous modes depends on the values of the other parameters as well and, in particular, is a strong function of $n$. In order to show the behaviour of the disturbances at sufficiently high $k$, eigenfunctions corresponding to the last row of table~\ref{table:varicose-sinuous-comparison} are shown in figures~\ref{fig:vx_vs_y_W_5_k_10_n_0p2_ppf} and \ref{fig:vy_vs_y_W_5_k_10_n_0p2_ppf}. The overlap of the eigenfunctions for $y>0.3$ where maximum variation in the eigenfunction is observed, shows that the disturbances at high $k$ are confined to the wall. It must be noted that the eigenfunctions will not agree near $y=0$ i.e. the channel center-line, by virtue of the boundary conditions (\ref{eq:ppf-varicose}-\ref{eq:ppf-sinuous}) even as $k \gg 1$.  
The confinement of disturbances at sufficiently high $k$, in fact, has important consequences not only for channel flow but also for PCF (Sec.~\ref{sec:pcf-results}) and tube flow (Sec.~\ref{sec:hpf-results}), as a similar wall mode instability exists for all three shear flows. This suggests the possibility of universality of the wall mode instability in any wall-bounded shear flow.

 \begin{table}
  \begin{center}
    \def~{\hphantom{0}}
    \begin{tabular}{lcc}
      Parameters  & Varicose & Sinuous  \\[3pt]
      $ k=0.1$ &  $0.98441396 -     0.52407040i$ & $2.2070283 -    0.08556572i$  \\
      $ k=0.5$ &  $0.79270520 +     0.17861588i$ & $0.81523304 +     0.32999025i$  \\
      $ k=1$ &  $0.45611528 +     0.16271074i$ & $0.46503710 +     0.18762645i$  \\
      $ k=5$ &  $0.09335747 +    0.02928823i$ & $0.09336330 +    0.02929848i$  \\
      $ k=10$ &  $0.04642377 +    0.01324292i$ & $0.04642377 +    0.01324292i$  \\
    \end{tabular}
    \caption{\small Most unstable (or least stable) eigenvalues, $c$ (rounded off to eighth significant figure), for both modes. Other parameters are $n=0.2, W=5$ and $\beta=0$. Table shows the convergence of the varicose and sinuous modes to a single mode as $k$ is increased.}
    \label{table:varicose-sinuous-comparison}
  \end{center}
\end{table}

Similar to PCF, channel flow of a WM fluid exhibits two types of modes, viz., center and wall modes. For the wall modes discussed above, $c_r \to 0$ indicating their confinement near wall. However, for center modes $c_r \to 1$ and $k_c < 5$, thus from table~\ref{table:varicose-sinuous-comparison},  center-line conditions will affect the  predictions for the center mode. The earlier efforts of Wilson and coworkers \citep{wilson-rallison-1999,wilson-loridan-2015,castillo-wilson-2017} studied only the wall modes in detail. It must be noted that, for the center modes of PCF, $c_r=0.5$ and $k_c<1$. However, for the center modes of channel and tube flows $0.5<k_c<5$ and $c_r \sim 1$ at critical parameter values. This difference in $k_c$ and $c_r$ causes the stability of PCF and channel flows to be different. 

\begin{figure*}
    \centering
    \begin{subfigure}[b]{0.5\textwidth}
        \centering
        \includegraphics[width=\textwidth]{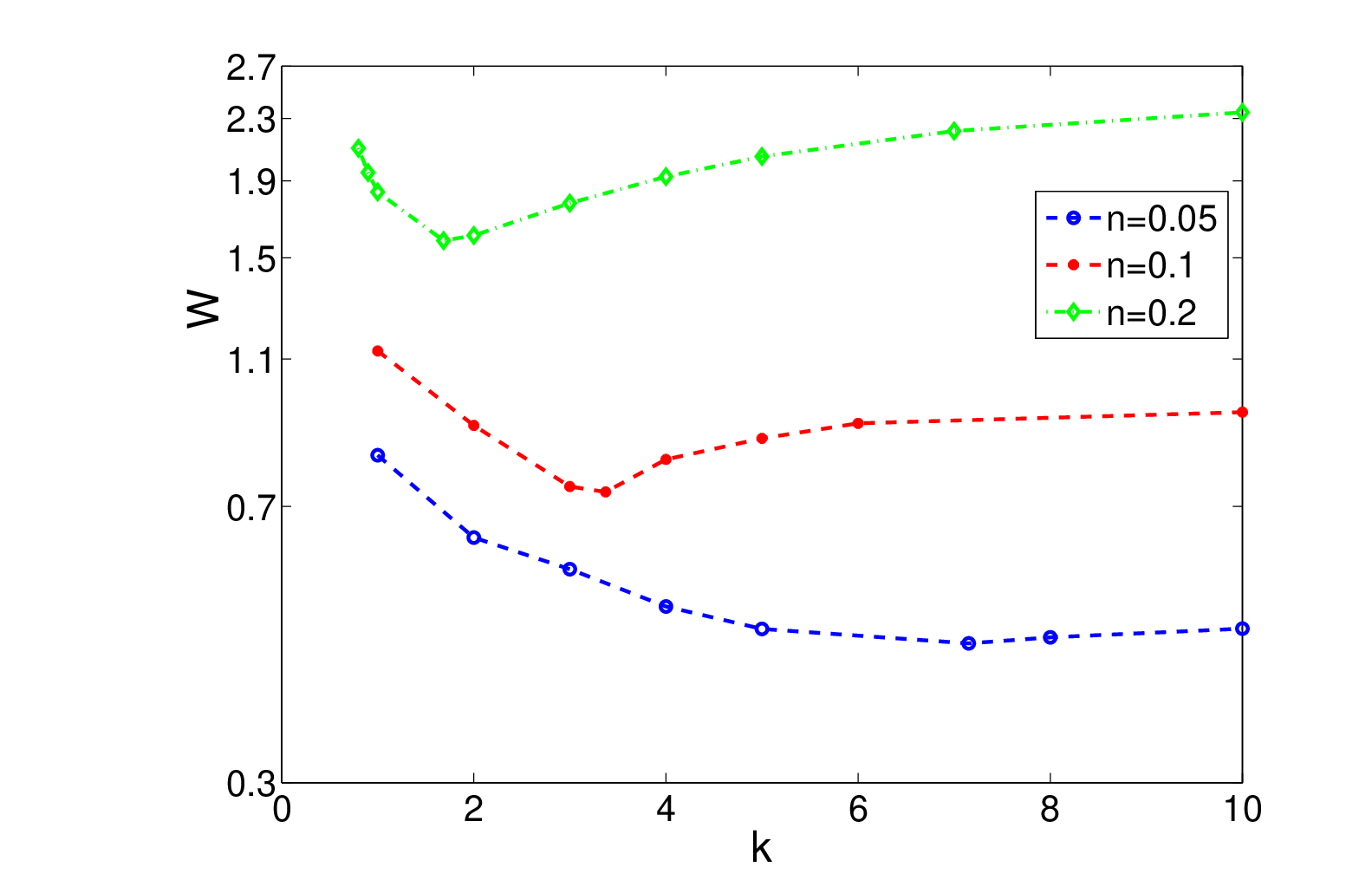}
        \caption{Varicose mode}
        \label{fig:W_vs_k_Var}
    \end{subfigure}%
    ~ 
    \begin{subfigure}[b]{0.5\textwidth}
        \centering
        \includegraphics[width=\textwidth]{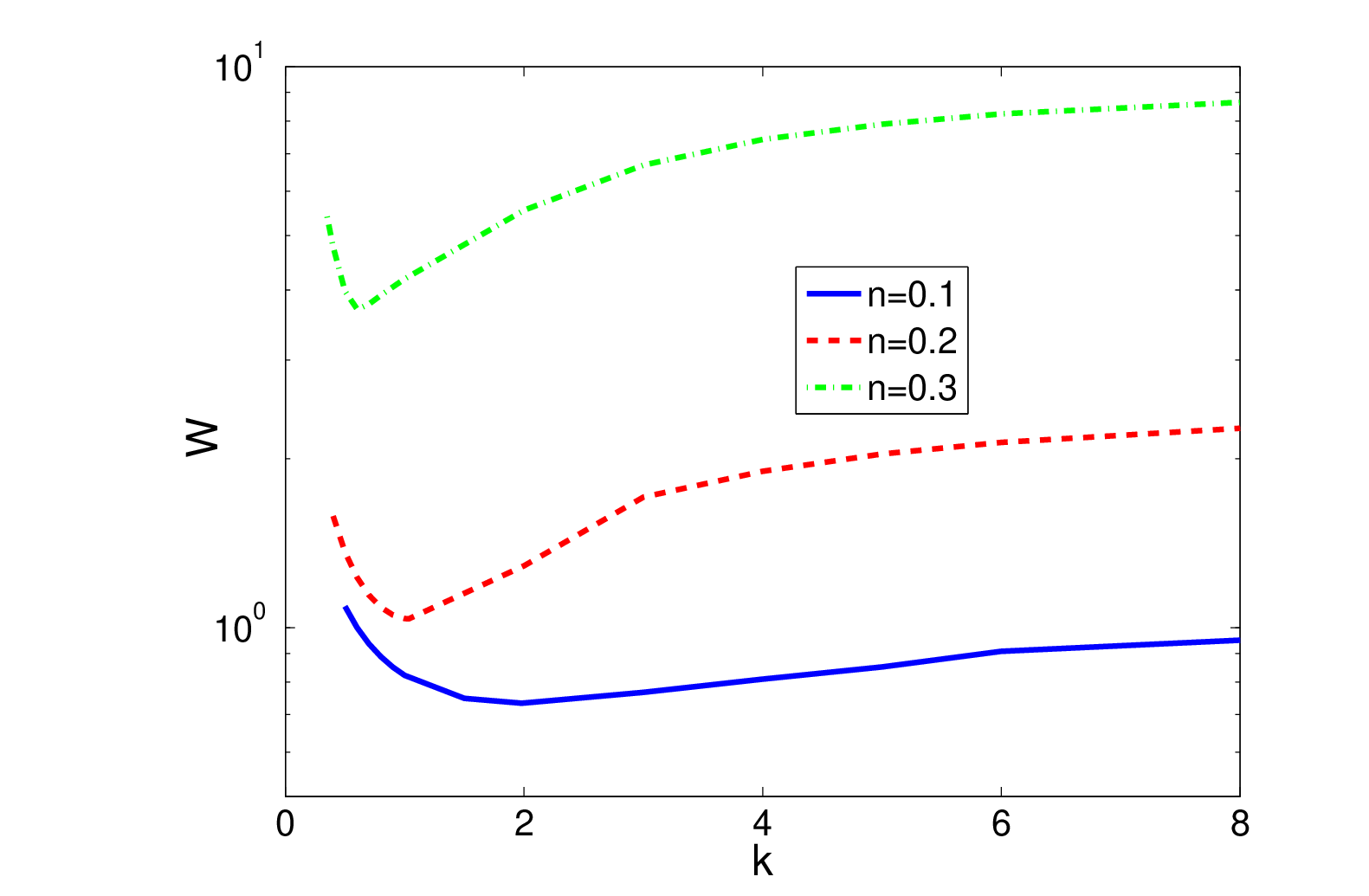}
        \caption{Sinuous mode}
          \label{fig:W_vs_k_Sin}
    \end{subfigure}
    \caption{\small Neutral stability curves for varicose and sinuous modes using $\beta=0$. The stabilising effect of increase in $n$ and the presence of the Hadamard instability is observed.}
\end{figure*}

To facilitate comparison with experiments, and to determine whether the sinuous mode is more unstable than the varicose mode, we computed $W_c, k_c$ and $c_r$ for the most unstable varicose and sinuous modes. To obtain the critical parameters, we plot the neutral stability curves for the most unstable varicose and sinuous modes respectively in figures~\ref{fig:W_vs_k_Var} and \ref{fig:W_vs_k_Sin}. The features of the neutral curves are similar to those for PCF, but for a more gradual crossover from the center to wall mode for channel flow. In the case of PCF, at lower $n$ values, the center mode is the critical mode (see figure~\ref{fig:n0p01_neutral_curves}) whereas for channel flow, wall mode is the critical mode.


%

\begin{figure}
\centering
\includegraphics[width=0.6\textwidth]{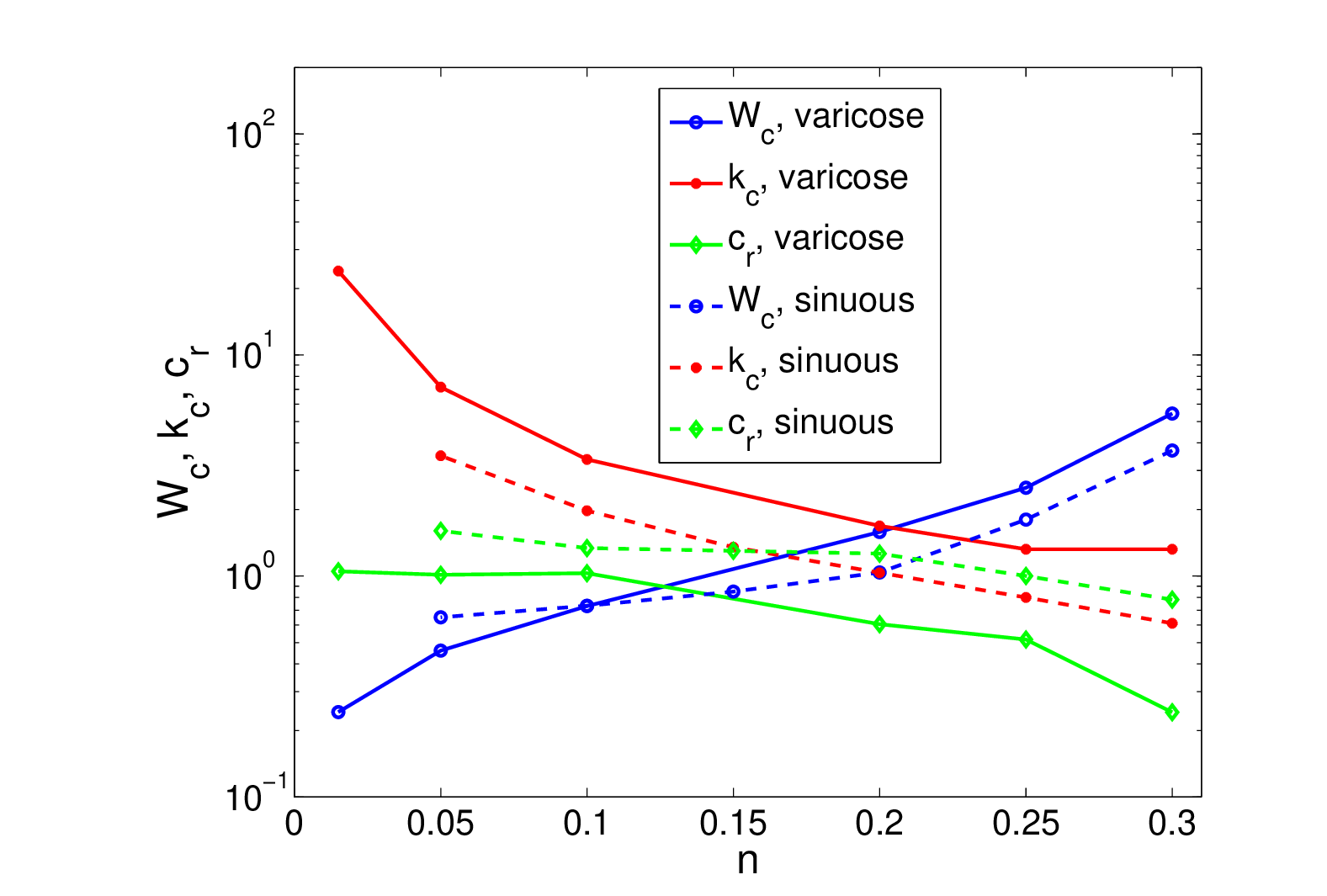}
\caption{\small Comparison of critical parameters for varicose and sinuous modes for $\beta=0$. The sinuous mode is more unstable than the varicose mode in agreement with the experiments of \cite{bodiguel-et-al-2015}.}
\label{fig:Wc_kc_vs_n}
\end{figure}

Finally, to understand the relative importance of varicose and sinuous modes in determining the stability of the system with change in $n$, we plot the variation of $W_c$ and $k_c$ with $n$ for both varicose and sinuous modes in figure~\ref{fig:Wc_kc_vs_n}. This shows that the sinuous mode is more unstable than the varicose mode for $n>0.1$, while the opposite is true for $n<0.1$. Furthermore, for both modes, $W_c$ decreases with decreasing $n$, but does not seem to exhibit a scaling similar to PCF as $n\rightarrow 0$. However,  and in contrast to PCF, $k_c$ for both sinuous and varicose modes increases rapidly with decreasing $n$; as discussed earlier, for PCF,  $k_c \rightarrow 0$ as $n \rightarrow 0$ (figure~\ref{fig:critical_beta0}) for $\beta=0$. This increase in $k_c$ with decreasing $n$ makes it numerically difficult to track the most unstable mode, as the pseudo-spectral method used requires more number of collocation points for higher values $k$. Thus, for varicose and sinuous modes, respectively for $n<0.025$ and $n<0.05$, we are unable to compute critical parameters very accurately. The phase speed data in figure~\ref{fig:Wc_kc_vs_n} reveals that as $n$ decreases $c_r \rightarrow 1$, i.e., the critical mode is a (varicose) center mode in conformity with PCF results (figure~\ref{fig:critical_beta0}). However, a comparison of figures~\ref{fig:critical_beta0} and \ref{fig:Wc_kc_vs_n} shows that unlike the PCF case, the switching from wall mode to center mode happens more gradually in the case of channel flow.  
To conclude, we find that sinuous mode is more unstable than the varicose mode for $n>0.1$ in broad agreement with the experimental findings of \cite{bodiguel-et-al-2015}. This aspect is discussed in more detail below in Sec.~\ref{sec:comparison-with-experiments}. The varicose and sinuous modes converge to the same wall mode at high $k$, and both modes are characterized by phase speed $c_r \rightarrow 0$ as $k \gg 1$.


\subsection{Tube flow} \label{sec:hpf-results}

\begin{figure}
\centering
\includegraphics[width=0.6\textwidth]{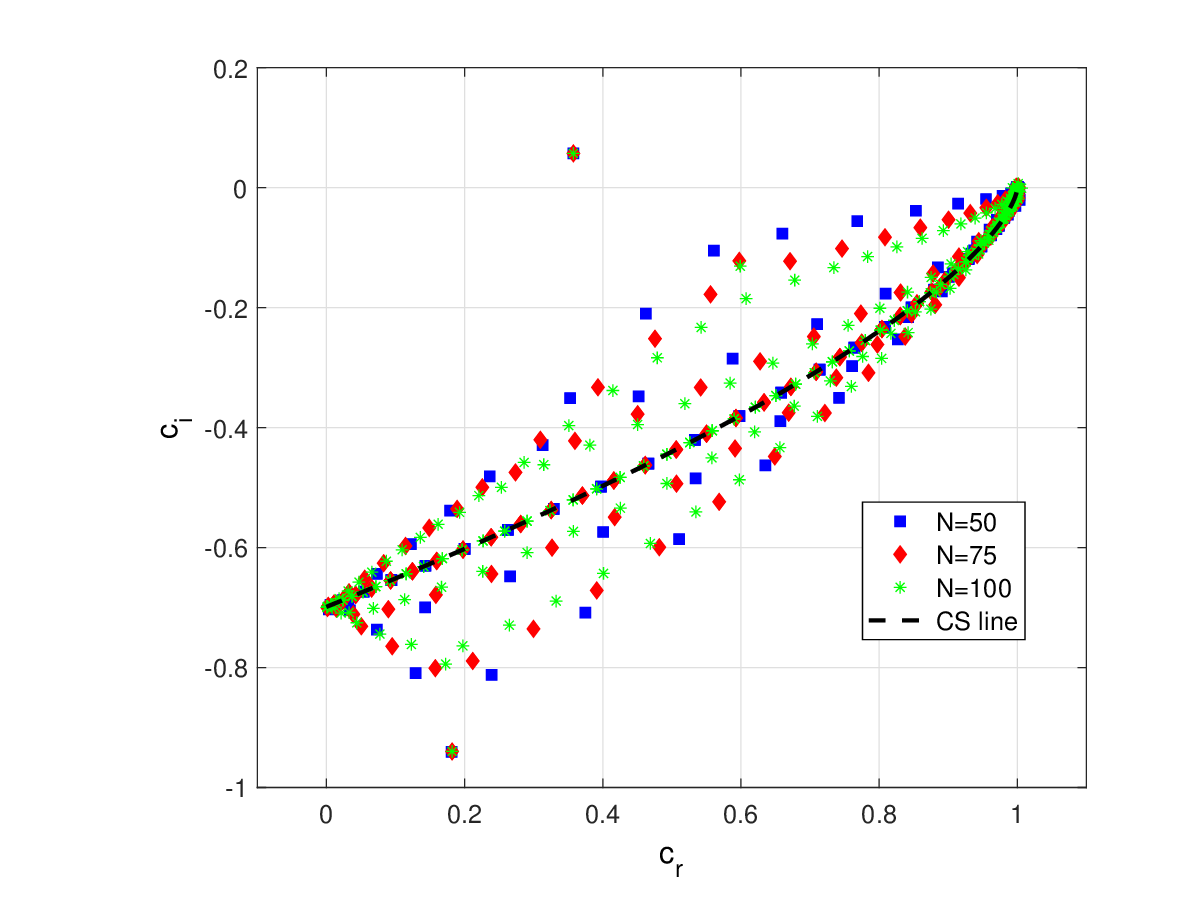}
\caption{\small  Continuous spectrum and discrete modes for tube flow at $W=3,k=2,n=0.2$ and $\beta=0$. The  most unstable eigenvalue is $c=0.3574466 +    0.05715674i$. The curved CS line is marked from eq. \ref{eq:hpf-continuous-spectra-c}}
\label{fig:cr_vs_ci_Re_0_W_3_k_2_n_0p2_hpf}
\end{figure}

\begin{figure*}
    \centering
    \begin{subfigure}[b]{0.5\textwidth}
        \centering
        \includegraphics[width=\textwidth]{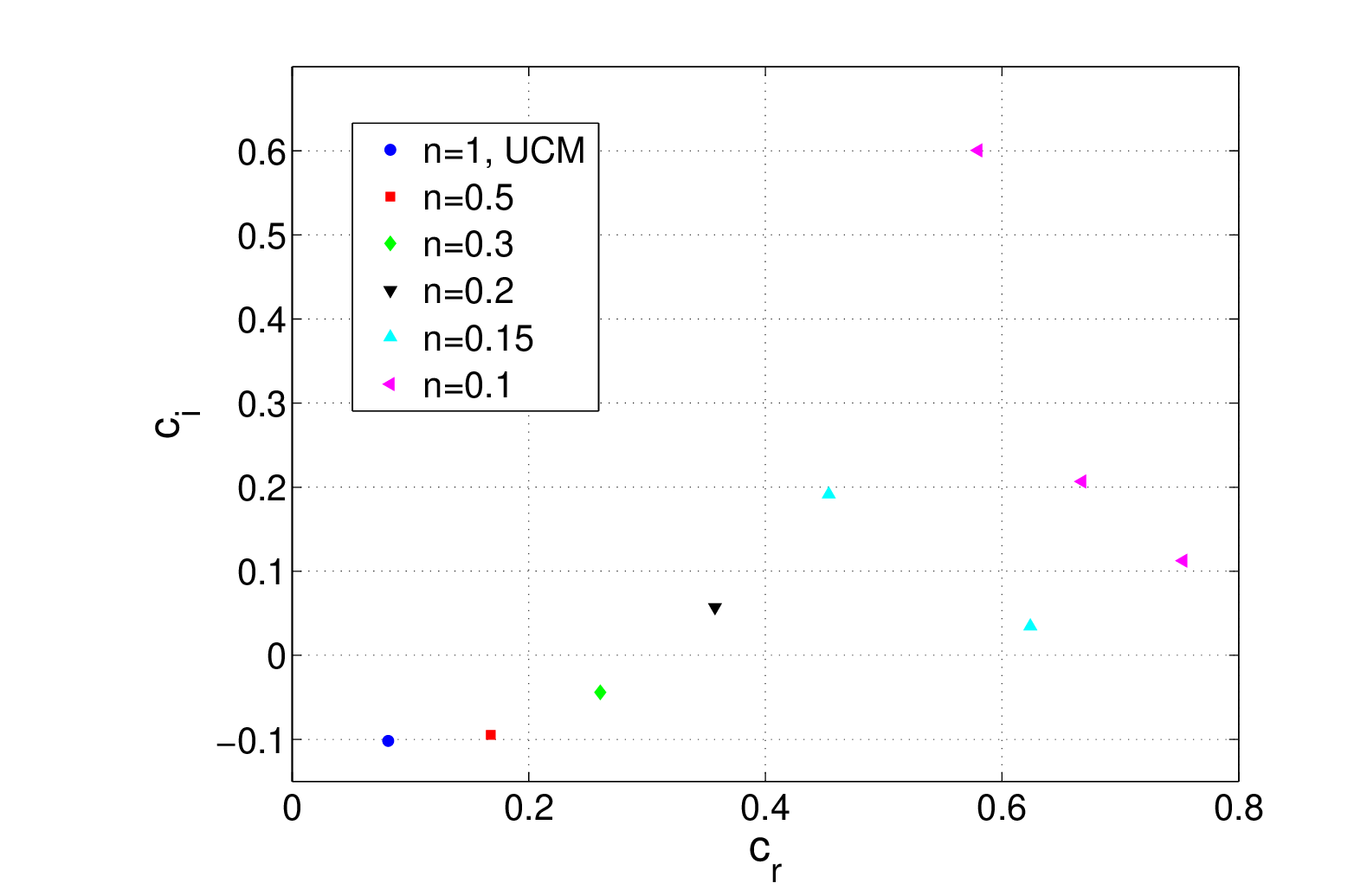}
        \caption{wall mode}
        \label{fig:cr_vs_ci_Re_0_W_3_k_2_hpf}
    \end{subfigure}%
    ~ 
    \begin{subfigure}[b]{0.5\textwidth}
        \centering
        \includegraphics[width=\textwidth]{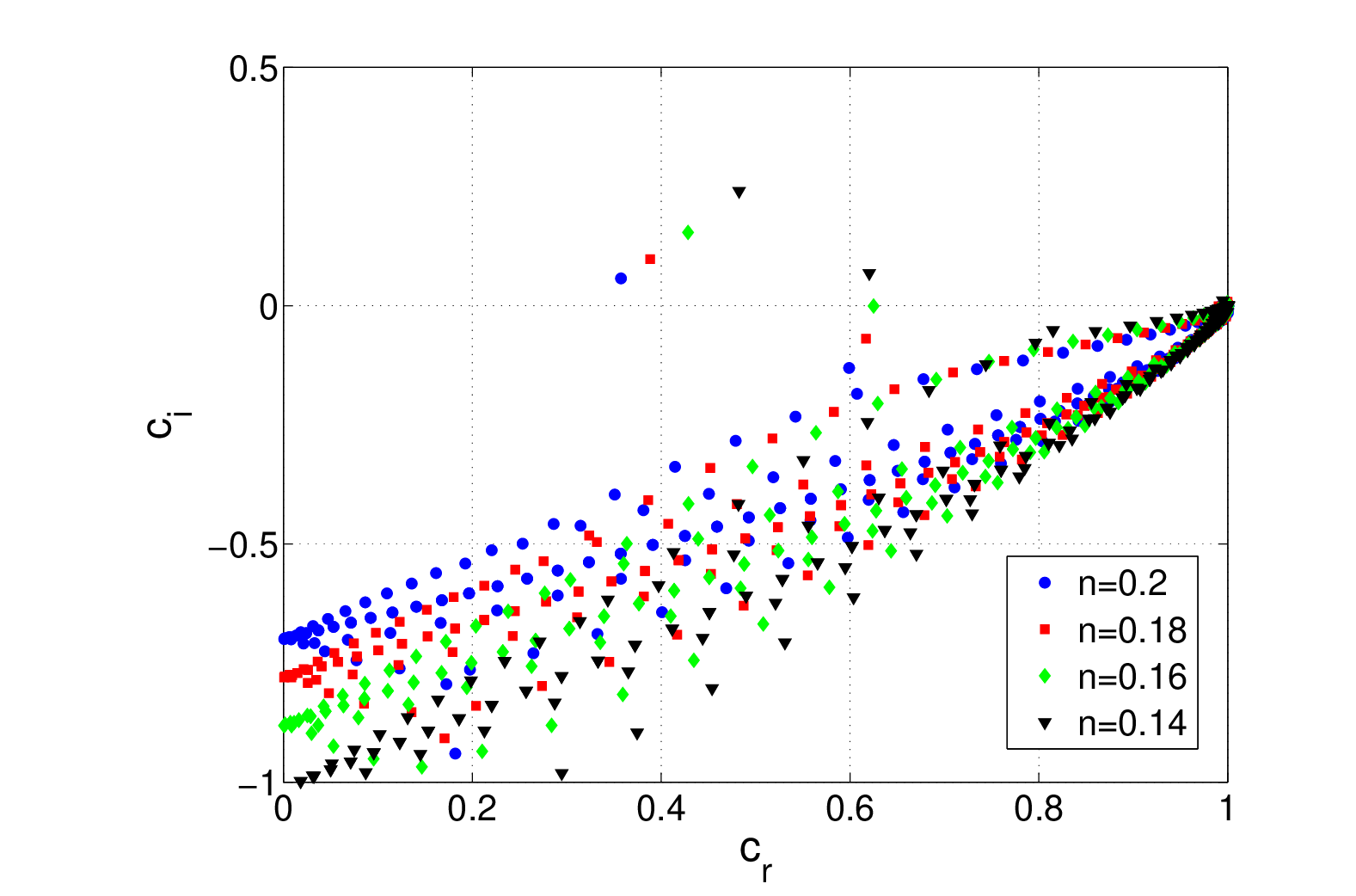}
        \caption{wall mode}
         \label{fig:cr_vs_ci_Re_0_W_3_k_2_hpf_origin}
    \end{subfigure}
    \caption{Panel~(a) shows the emergence and destabilisation of the wall mode for tube flow at $W=3,k=2$ and $\beta=0$.  A decrease in $n$ destabilises the UCM discrete mode and leads to creation of new unstable modes. Unstable modes with $c_r>0.6$ are the newly created modes. Panel~(b) shows that a new discrete mode emerges from the continuous spectrum with $c_r \sim 0.6$, which is destabilised as $n$ is decreased.}
\end{figure*}

\begin{figure*}
    \centering
    \begin{subfigure}[b]{0.5\textwidth}
        \centering
        \includegraphics[width=\textwidth]{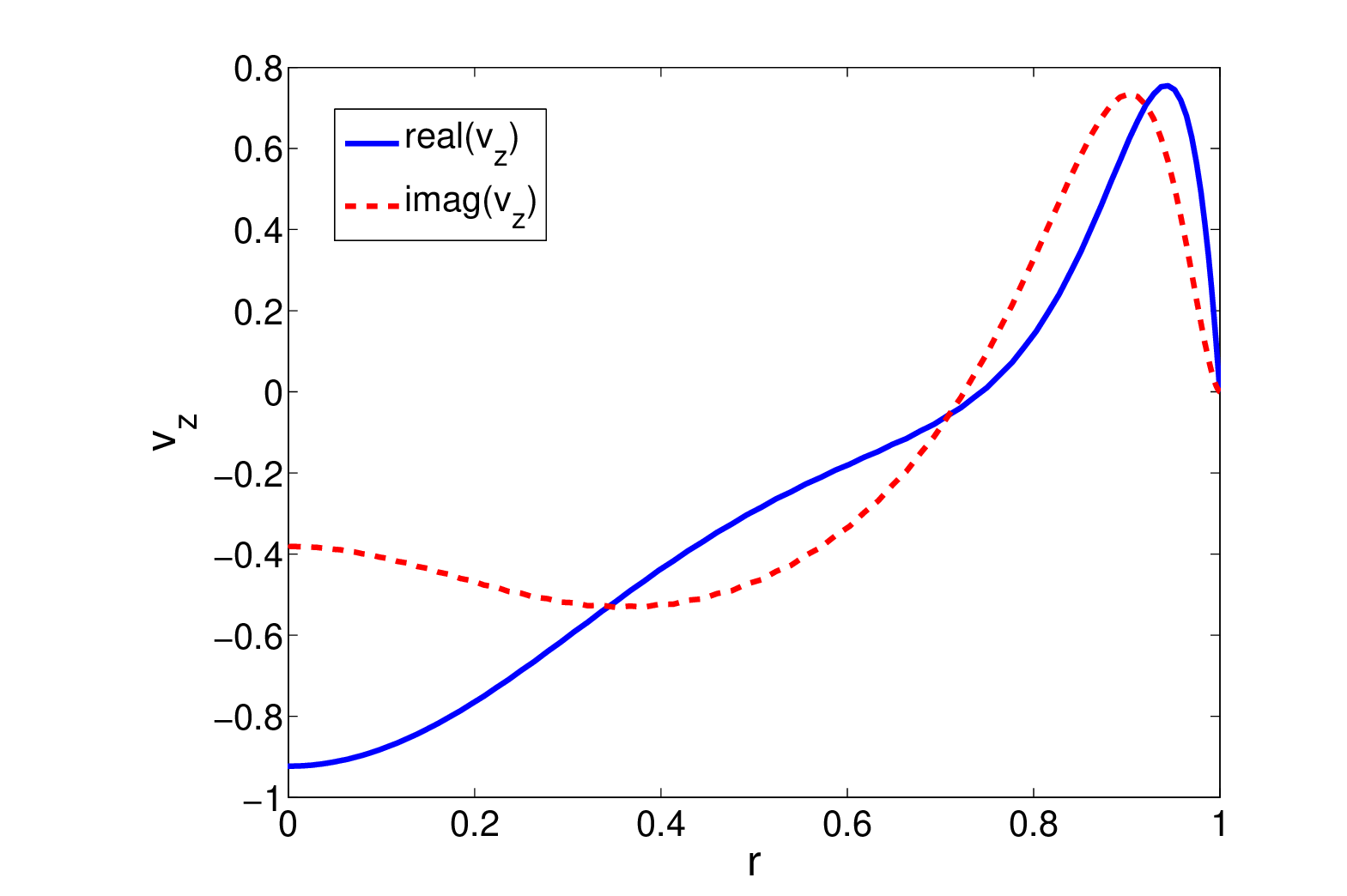}
        \caption{$v_z$-eigenfunction}
        \label{fig:vz_vs_r_W_3_k_2_n_0p2}
    \end{subfigure}%
    ~ 
    \begin{subfigure}[b]{0.5\textwidth}
        \centering
        \includegraphics[width=\textwidth]{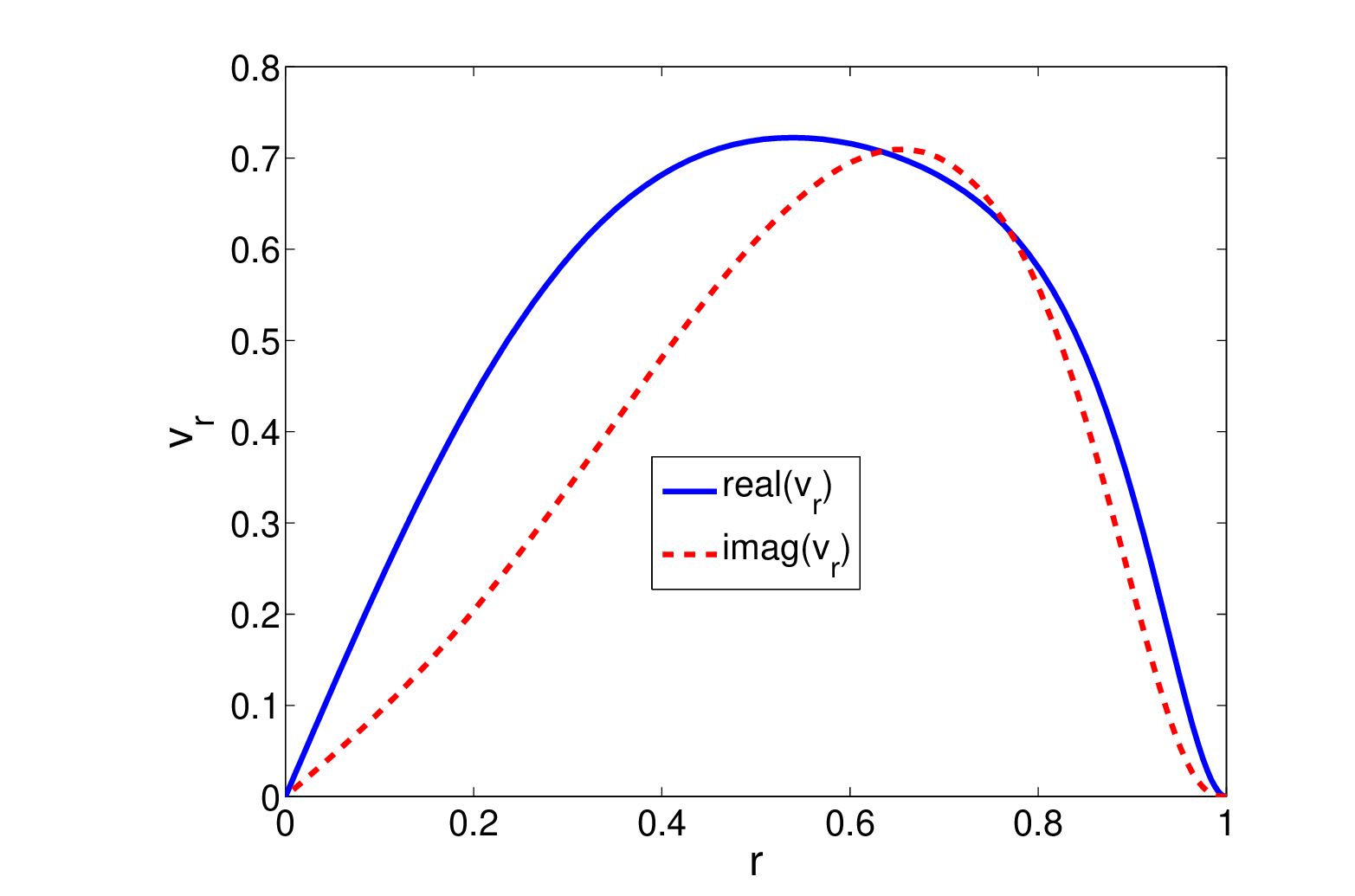}
        \caption{$v_r$-eigenfunction}
          \label{fig:vr_vs_r_W_3_k_2_n_0p2}
    \end{subfigure}
    \caption{\small Eigenfunctions for the most unstable mode ($c=0.3574466 +    0.0571568i$) for tube flow. Other parameters are $k=2, n=0.2, W=3$ and $\beta=0$. The smooth variation of eigenfunctions near $r = 0$ validates the regularization procedure used.}
\end{figure*}

The experiments of \cite{poole-2016,wen-et-al-2017,picaut-et-al-2017}  reported an instability in the flow of concentrated polymer solutions through tubes. Motivated by these experimental observations, and to ascertain the existence of the wall-mode instability similar to the one found in the previous sections for PCF and channel flows, we analyze the stability of tube flow using the WM model.
It is first useful to determine the theoretical location of continuous spectra which will aid in the identification of genuine (discrete) modes of instability. 
This is done by setting the coefficient of stresses in the stress Eqs.~(\ref{eq:hpf-Tauzz}-\ref{eq:hpf-Tautt}) to zero and then solving for $c$ to obtain
\begin{eqnarray}
c= 1-r^{\frac{1+n}{n}}-\frac{i}{kW} \left(\frac{n+1}{n} \right)^{1-n}r^{\frac{1-n}{n}} . \label{eq:hpf-continuous-spectra-c}
\end{eqnarray}
\noindent
Figure~\ref{fig:cr_vs_ci_Re_0_W_3_k_2_n_0p2_hpf} shows that the genuine discrete mode changes negligibly with increase in collocation points, unlike the ballooned-up continuous spectrum.
The discrete eigenvalue for a UCM fluid ($n=1$) is stable as shown in figure~\ref{fig:cr_vs_ci_Re_0_W_3_k_2_hpf}, but new unstable modes emerge as $n$ is decreased. Nevertheless, the continuation of the UCM discrete mode still remains the most unstable mode.
%
%
%
The new modes that appear with decreasing $n$ emerge from the continuous spectra as shown in figure~\ref{fig:cr_vs_ci_Re_0_W_3_k_2_hpf_origin}. From Eq.~(\ref{eq:hpf-continuous-spectra-c}), we expect that the eigenvalues originating from the continuous spectrum may not become unstable immediately, but require a further decrease in $n$  for their destabilisation, as shown in figure~\ref{fig:cr_vs_ci_Re_0_W_3_k_2_hpf_origin}. 
The tangential and normal velocity eigenfunction plots corresponding to the converged eigenvalue of figure~\ref{fig:cr_vs_ci_Re_0_W_3_k_2_n_0p2_hpf} are respectively plotted in figures~\ref{fig:vz_vs_r_W_3_k_2_n_0p2} and \ref{fig:vr_vs_r_W_3_k_2_n_0p2}.

A comparison of eigenfunctions for the varicose mode of channel flow shown in figures~\ref{fig:vx_vs_y_W_3_k_1_n_0p2_Var}-\ref{fig:vy_vs_y_W_3_k_1_n_0p2_Var} and the axisymmetric mode for tube flow  in figures~\ref{fig:vz_vs_r_W_3_k_2_n_0p2}-\ref{fig:vr_vs_r_W_3_k_2_n_0p2} shows a strong similarity between the disturbances in both flow geometries. This could be ascribed to the similarity in boundary conditions at the center shared by varicose and axisymmetric modes (Eqs.~\ref{eq:ppf-varicose} and \ref{eq:bc-r-zero}). Furthermore, there are several similarities between PCF, channel and tube flows in terms of the destabilization of the discrete stable modes from the UCM limit upon decrease in $n$, the origin and creation of new unstable modes, and the resemblance of the eigenfunction for the varicose and axisymmetric modes. This suggests that the instability predicted for PCF, channel and tube flows could be driven by a qualitatively similar mechanism. It is known from Sec.~\ref{sec:ppf-results} that, at high $k$, disturbances are confined near the wall and hence boundary conditions at the center should not affect the stability predictions to a large extent. In Sec.~\ref{sec:equivalence}, we analyse all the three flow geometries and demonstrate the existence of an identical wall mode instability at high $k$.

\begin{figure*}
    \centering
    \begin{subfigure}[b]{0.5\textwidth}
        \centering
        \includegraphics[width=\textwidth]{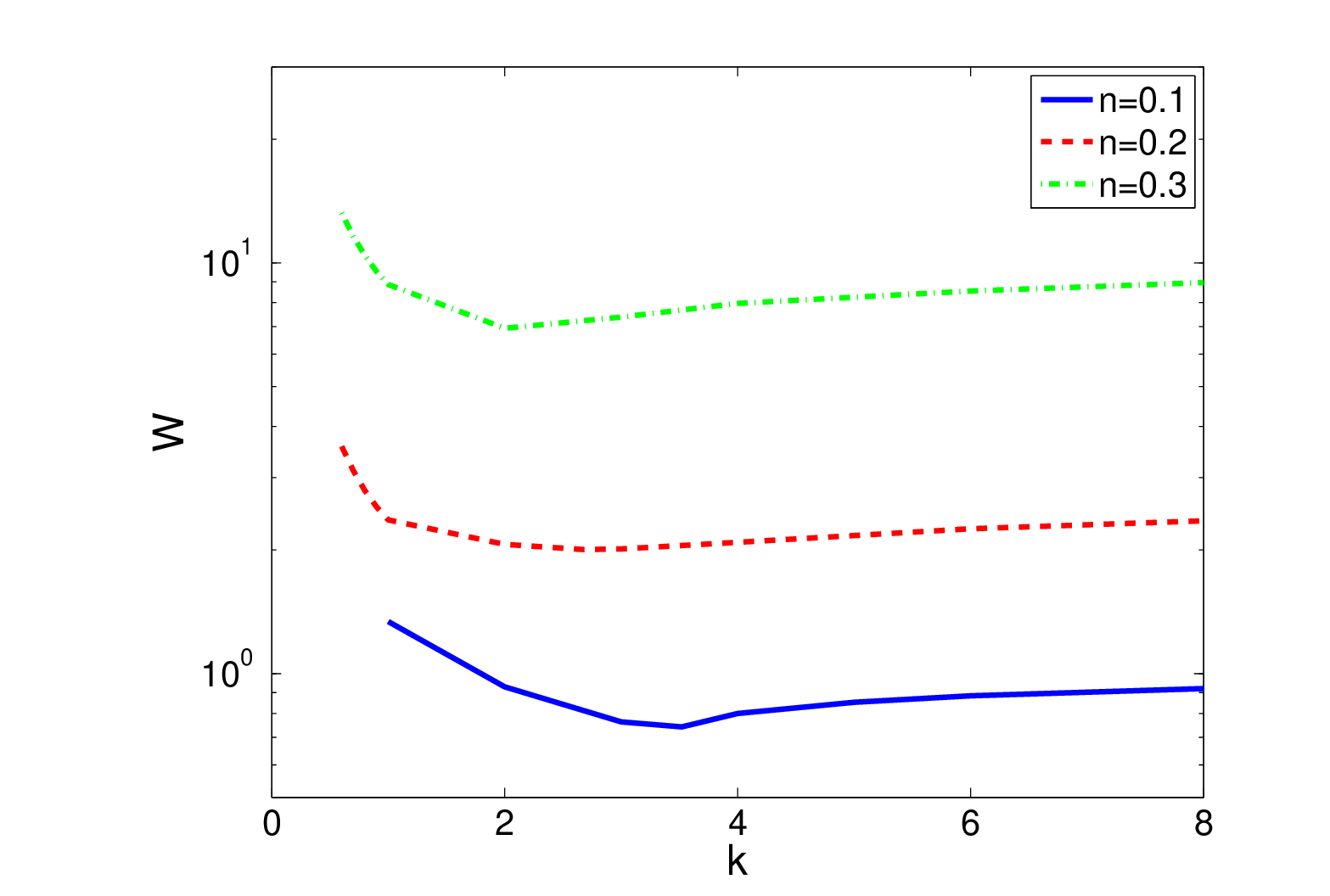}
        \caption{\small $W$--$k$ curves}
\label{fig:W_vs_k_Re_0_n_hpf}
    \end{subfigure}%
    ~ 
    \begin{subfigure}[b]{0.5\textwidth}
        \centering
        \includegraphics[width=\textwidth]{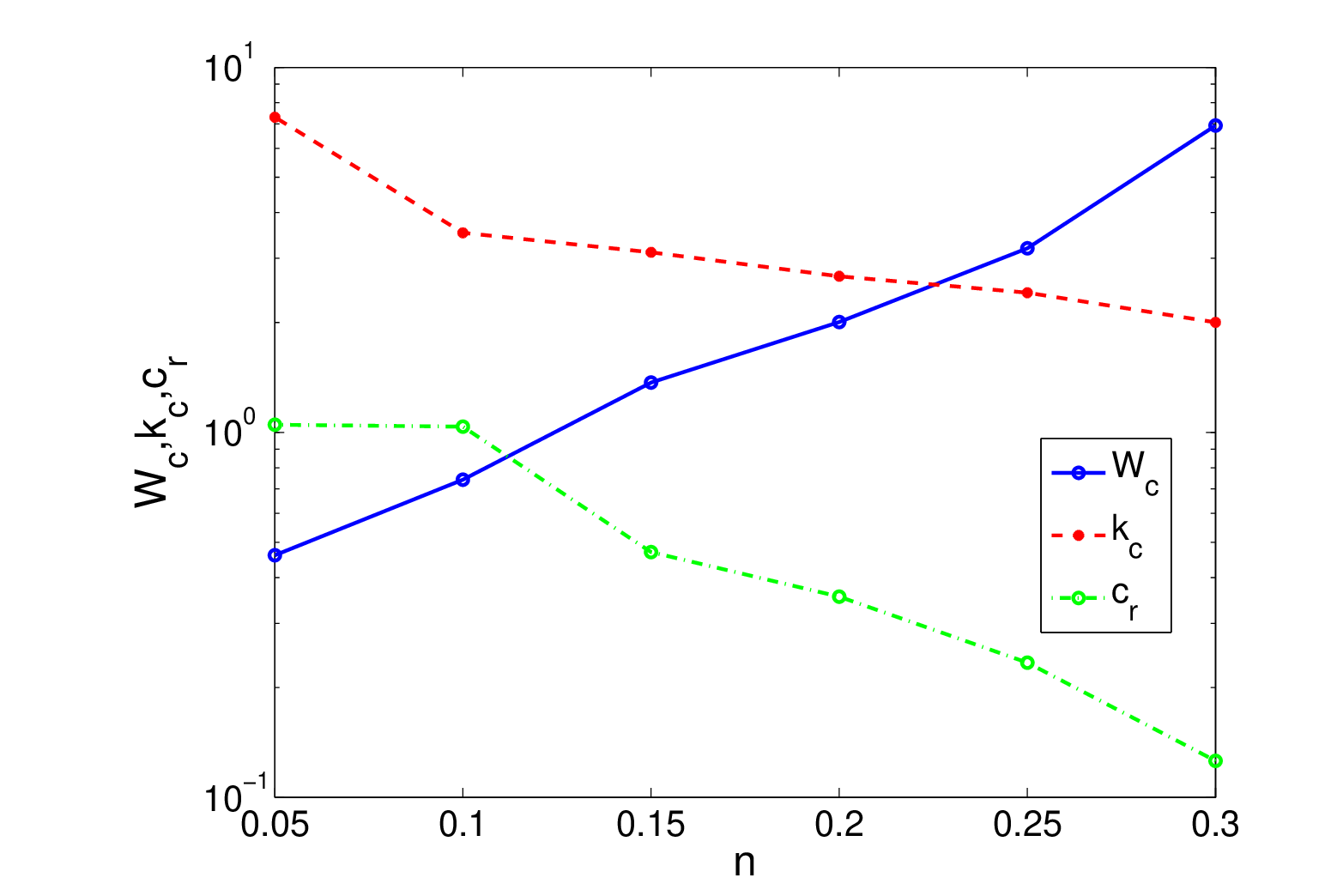}
        \caption{\small Critical parameters}
\label{fig:Wc_kc_vs_n_Re_0_n_hpf}
    \end{subfigure}
    \caption{\small Panel~(a) shows  neutral stability curves in $W$--$k$ plane for tube flow at $\beta=0$ showing a stabilising effect of increasing $n$.  Panel~(b) shows the variation of $W_c, k_c$ and $c_r$ of the most unstable mode with $n$.}
\end{figure*}

The neutral stability curves for tube flow are shown in figure~\ref{fig:W_vs_k_Re_0_n_hpf}. The structure of the neutral stability curves is very similar to those for the channel flow. But they are not identical due to the absence of the center mode for $n>0.1$ in tube flow. Thus, although all three geometries exhibit the wall mode instability (discussed in Sec.~\ref{sec:equivalence}), however for $0.1<k<5$, these flows exhibit different behaviour owing to the geometry-specific boundary conditions. The presence of a finite $W_c$ for very high $k$ indicates a  Hadamard instability similar to PCF and channel flow. The variation of the critical parameters with $n$ is plotted in figure~\ref{fig:Wc_kc_vs_n_Re_0_n_hpf}. Similar to the varicose mode in channel flow, $W_c$ increases, while $k_c$ and $c_r$ decrease with increasing $n$. A comparison of figures~\ref{fig:critical_beta}, \ref{fig:Wc_kc_vs_n} and \ref{fig:Wc_kc_vs_n_Re_0_n_hpf} shows that, unlike in the case of PCF, channel and tube flows do not have an abrupt crossover from center to wall modes.    

%
%

\section{Universality of the wall-mode instability} \label{sec:equivalence}

All the three canonical shear flows of a WM fluid exhibit a similar instability at high $k$, having modes with $c_r \rightarrow 0$ confined near the wall. This instability, thus, does not  depend on the flow geometry and $c_r$ corresponding to these modes indicate that the disturbances must be confined to the wall within a distance of $O(k^{-1})$ for high $k$, as noted by \cite{wilson-rallison-1999}.  In the following discussion, we demonstrate the universal nature of the wall mode instability for all the three geometries in the absence of solvent ($\beta=0$) and for constant shear modulus.
We introduce $\xi=k(\zeta-1)$ as the `inner' coordinate used to analyse the instability near the wall region. For planar flows, $\zeta=y$, while for the cylindrical coordinate system $\zeta=r$. The wall-normal coordinate considered for the channel and tube flows is in the interval $[0,1]$ with the wall at $\zeta=1$ being stationary and the channel or tube center at $\zeta=0$ where the base-flow velocity is maximum. To be consistent with this coordinate system, for the purposes of this discussion, even in the case of PCF, we take upper plate as stationary and the lower plate is moving. Consequently, in this section, we consider velocity profile of PCF as 
\begin{eqnarray}
\bar v_x=1-y.
\end{eqnarray}

We define the Weissenberg number based on the shear rate at the wall $\dot{\gamma}_w=|\frac{d \bar v_x}{dy}|_{y=1} $ as $W_w=W \dot{\gamma}_w^n $.  We use the complex frequency $\omega=\omega_r+i \omega_i=kc$, where $\omega_r$ and $\omega_i$ respectively denote the real and imaginary parts, in this section. Thus, the flow is unstable if $\omega_i>0$. The complex frequency of the perturbations scaled by shear rate at the wall is $\omega_w=\omega/\dot{\gamma}_w$. We restrict this discussion to $\beta=0$, consequently $\boldsymbol{\tilde \tau }=\boldsymbol{\tilde \tau^p}$ and the pressure and stresses are rescaled as $ (\tilde p^w,\boldsymbol{\tilde \tau}^w)=\frac{W \dot{\gamma}_w}{k} (\tilde p, \boldsymbol{\tilde \tau})$. By using these scalings and assuming creeping-flow limit, we arrive at the following continuity and momentum equations
\begin{eqnarray}
i \tilde v_1+D_\xi \tilde v_2=0, \label{eq:pf-coneq}\\
-i\tilde p^w+i\tilde \tau^w_{11}+D_\xi \tilde \tau^w_{12}=0,\\
-D_\xi \tilde p^w+i\tilde \tau^w_{12}+D_\xi \tilde \tau^w_{22}=0,
\end{eqnarray} 
where $D_\xi=\frac{d}{d \xi}$. In these equations and below, the subscripts $1=x, 2=y$ for Cartesian coordinate system while $1=z, 2=r$ for the cylindrical coordinate system.  Furthermore, for tube flow, the curvature terms in the continuity and momentum equations become subdominant and reduce to those corresponding to planar flows. This also means that the equation for $\tau_{\theta \theta}$ is decoupled from the momentum equations. The constitutive equations upon using the near-wall scaling become
\begin{eqnarray}
\nonumber
 2 i\tilde v_1 +2iW_w^2 \tilde v_1-2\tilde \tau^w_{12}-2 W_w D_\xi \tilde v_1+\underline{ 2 (n-1) W_w (i \tilde v_2+D_\xi \tilde v_1)}\\
 =\left[-i \xi -i \omega_w+\frac{1}{W_w} \right] \tilde \tau^w_{11}, \\
\nonumber
\underline{ i n \tilde v_2}+\underline{ n D_\xi \tilde v_1}-W_w D_\xi \tilde v_2-iW_w \tilde v_1-\tilde \tau^w_{22}+2iW_w^2 \tilde v_2 \\
=\left[-i \xi -i \omega_w+\frac{1}{W_w} \right] \tilde \tau^w_{12},\\
 2 D_\xi\tilde v_2 -2iW_w \tilde v_2=\left[-i \xi -i \omega_w+\frac{1}{W_w} \right] \tilde \tau^w_{22},
\end{eqnarray}
Interestingly, the non-trivial convective terms for channel and tube flows in the constitutive equations become subdominant at high wavenumbers. However, the terms introduced or modified (underlined terms) due to the WM model in comparison with UCM model, survive and destabilize the flow.

\begin{figure*}
    \centering
    \begin{subfigure}[b]{0.5\textwidth}
        \centering
        \includegraphics[width=\textwidth]{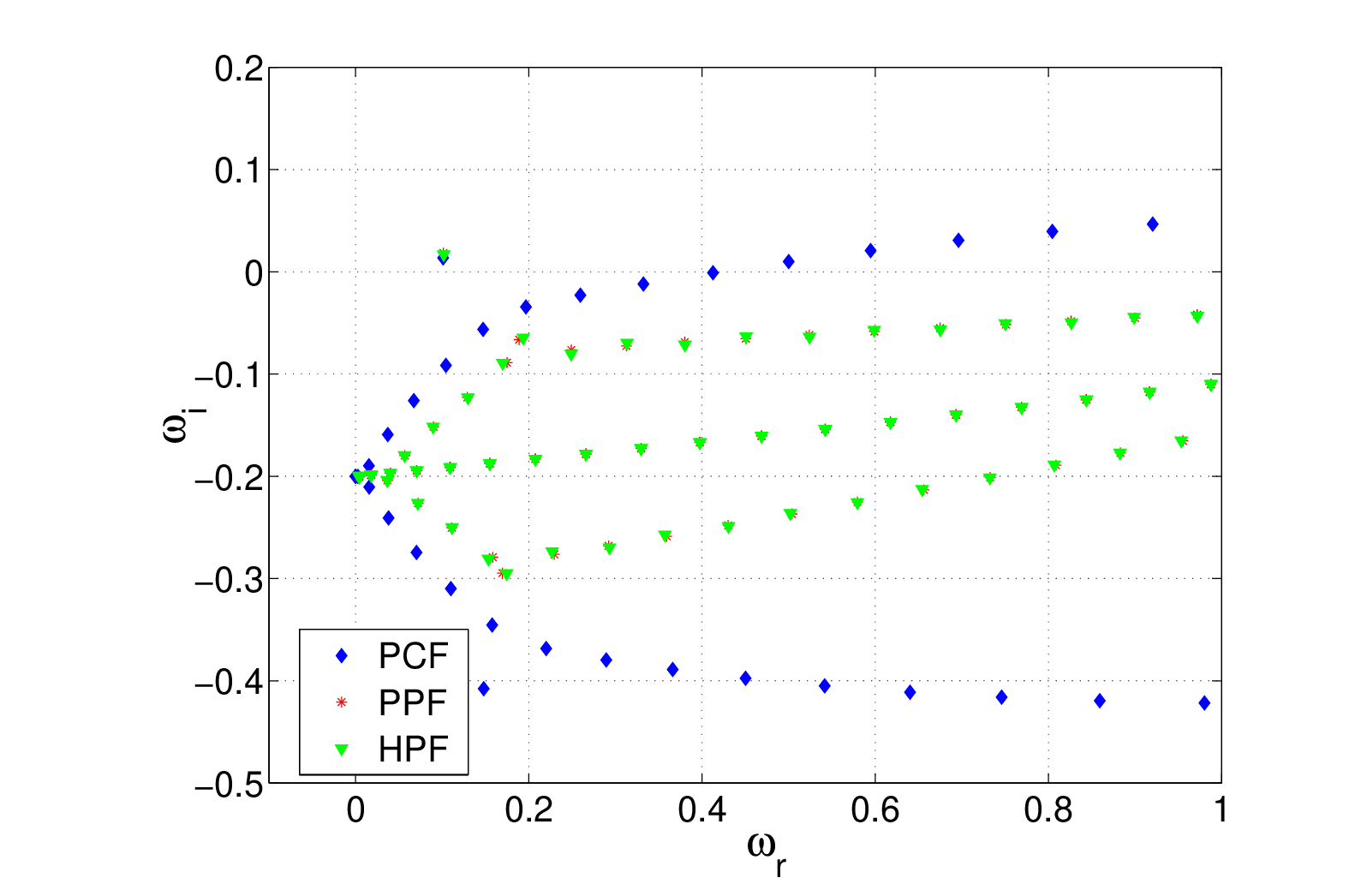}
        \caption{PCF, channel and tube flows}
        \label{fig:wr_vs_wi_Re_0_n_0p2_Ww_5_k_1_CPH}
    \end{subfigure}%
    ~ 
    \begin{subfigure}[b]{0.5\textwidth}
        \centering
        \includegraphics[width=\textwidth]{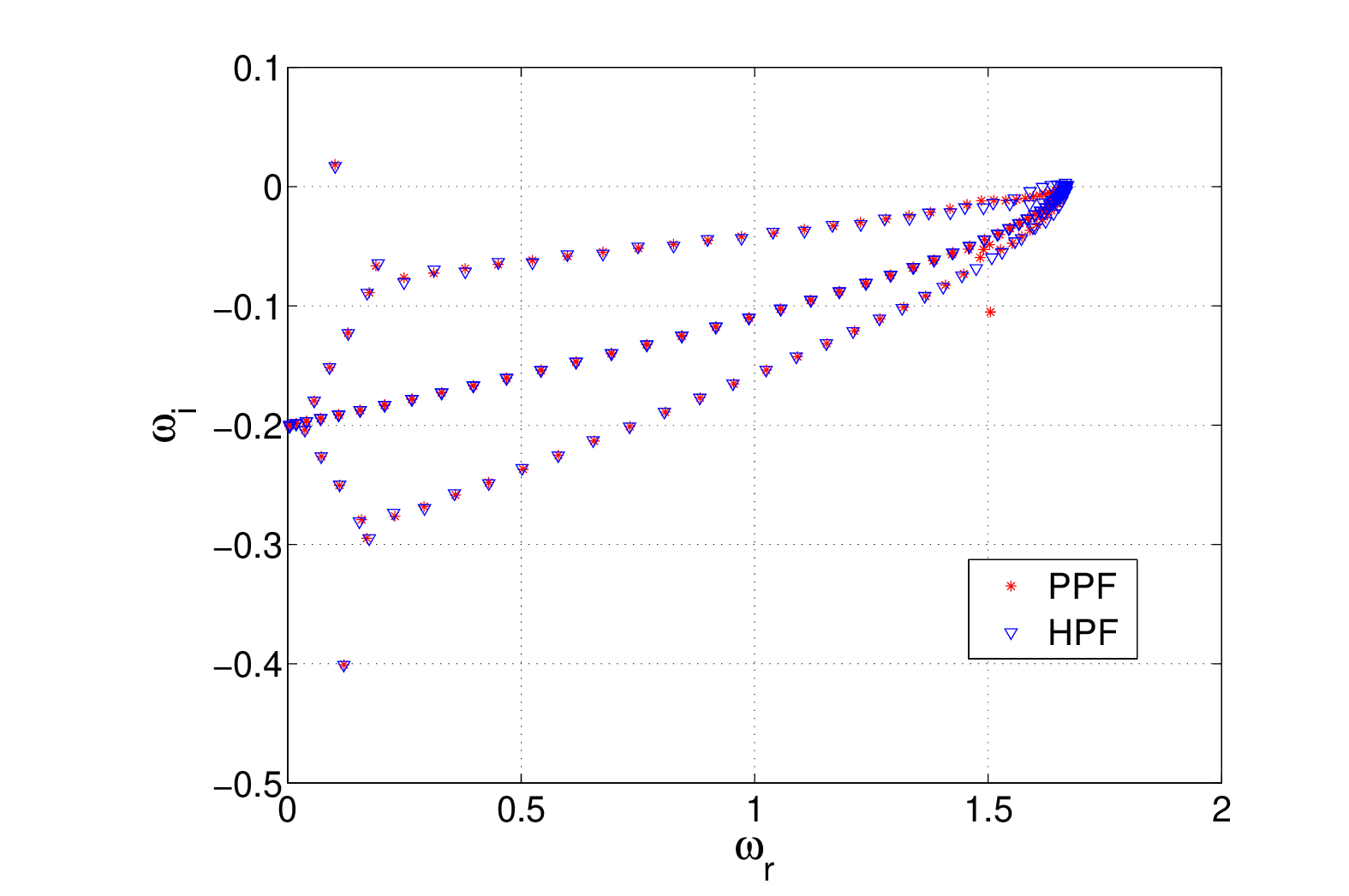}
        \caption{Channel and tube flows}
          \label{fig:wr_vs_wi_Re_0_n_0p2_Ww_5_k_1_PH}
    \end{subfigure}
    \caption{\small Comparison of spectra for PCF, channel and tube flows obtained by using $ W_w=5, n=0.2, k=10$. The discrete mode is identical for all the three geometries, while the ballooned-up continuous spectrum for PCF differs from channel and tube flows. The unstable eigenvalue is $\omega_w \sim 0.101+0.011i$. }
\end{figure*}

In the following discussion, we first demonstrate the similarity of the wall mode instability for PCF, channel and tube flow by using spectra and eigenfunctions obtained for equivalent parameters. Note that we will not solve the above set of equations, instead, as explained above, we will scale the Weissenberg number, stress components, eigenvalue and pressure for the three flows and compare the predicted eigenspectrum. As discussed in Sec.~\ref{sec:ppf-results}, sinuous and varicose modes show same eigenspectra and eigenfunctions at high $k$, consequently in this section we will consider only varicose modes for channel flow. In figure~\ref{fig:wr_vs_wi_Re_0_n_0p2_Ww_5_k_1_PH}, we compare the channel and tube flow spectra. The collapse of the eigenspectra (including that of the continuous spectrum balloon) for channel and tube flows demonstrates the similarity at high $k$. However, for PCF (figure~\ref{fig:wr_vs_wi_Re_0_n_0p2_Ww_5_k_1_CPH}), the continuous spectrum balloon is rather different from that of channel or tube flow, but  the discrete modes in all the three geometries show very good agreement. The agreement between PCF, channel and tube flows can be ascribed  to the presence of an identical instability leading to a collapse of the discrete unstable eigenvalue.

 \begin{table}
  \begin{center}
    \def~{\hphantom{0}}
    \begin{tabular}{lccc}
      Parameters  & PCF & channel flow  & tube flow \\[3pt]
      $n=0.25, k=5, W_w=7$ &  $0.07658 + 0.00180i$ & $0.07926 +0.00651i$ & $0.07921 +   0.00540i$ \\
      $n=0.25, k=10, W_w=7$ &  $0.07661 + 0.00182i$ & $0.07783 +   0.00418i$ & $0.07781 +   0.00357i$ \\
      $n=0.25, k=20, W_w=8$ &  $0.06899 +   0.00406i$ & $0.06948 +   0.00499i$ & $0.06946 +   0.00472i$ \\
      $n=0.2, k=15, W_w=5$ &  $0.10117 + 0.01358i$ & $0.10140 +     0.01681i$ & $0.10151 + 0.01583i$ \\
      $n=0.2, k=20, W_w=10$ &  $0.05883 +    0.01930i$ & $0.05898 +    0.02007i$ & $0.05898 +    0.01981i$ \\
    \end{tabular}
    \caption{\small Most unstable (or least stable) eigenvalues, $\omega_w$ (rounded off to fifth significant figure), for all three geometries for $\beta=0$.}
    \label{table:equivalence}
  \end{center}
\end{table}

\begin{figure*}
    \centering
    \begin{subfigure}[b]{0.5\textwidth}
        \centering
        \includegraphics[width=\textwidth]{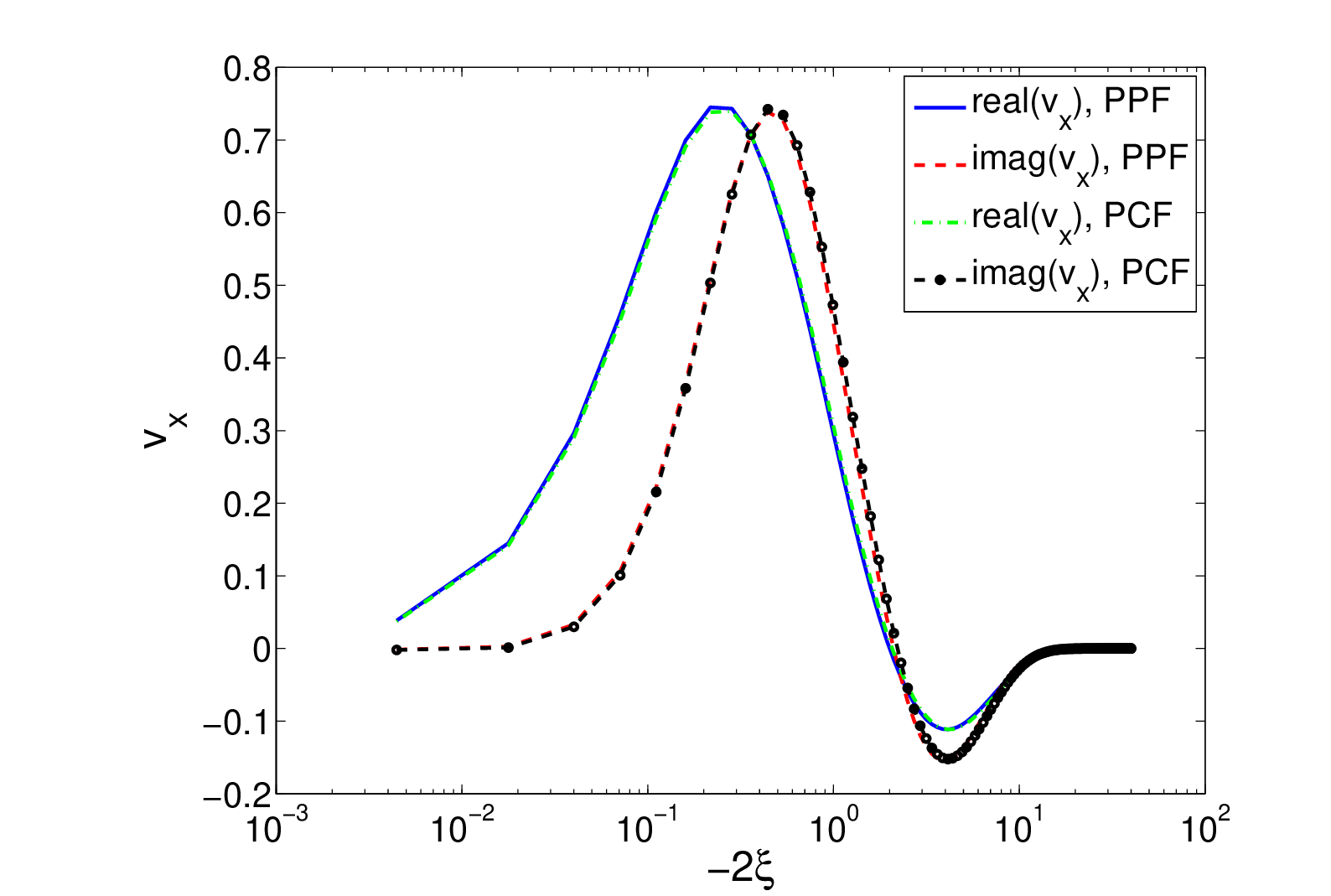}
        \caption{$v_x$-eigenfunction}
        \label{fig:vx_Re_0_n_0p25_Ww_7_k_20}
    \end{subfigure}%
    ~ 
    \begin{subfigure}[b]{0.5\textwidth}
        \centering
        \includegraphics[width=\textwidth]{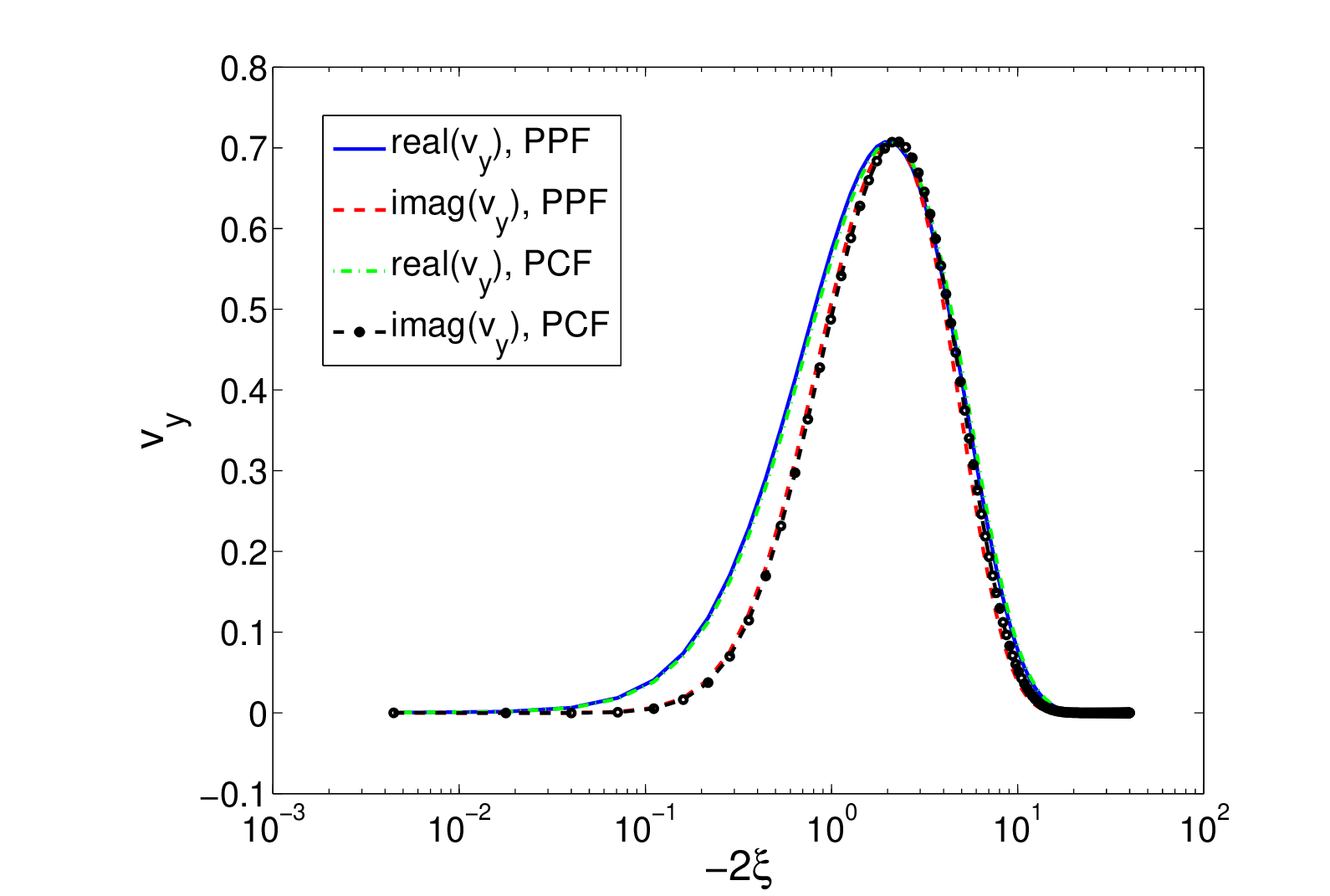}
        \caption{$v_y$-eigenfunction}
          \label{fig:vy_Re_0_n_0p25_Ww_7_k_20}
    \end{subfigure}
    \caption{\small Eigenfunctions for PCF and channel flow (varicose mode) for $W_w=7, n=0.25, k=20$. Real($f$) and imag($f$) respectively denote real and imaginary parts of the eigenfunction $f$. The agreement between the eigenfunctions demonstrates the identical nature of the wall mode instability for PCF and channel flows.}
\end{figure*}

\begin{figure*}
    \centering
    \begin{subfigure}[b]{0.5\textwidth}
        \centering
        \includegraphics[width=\textwidth]{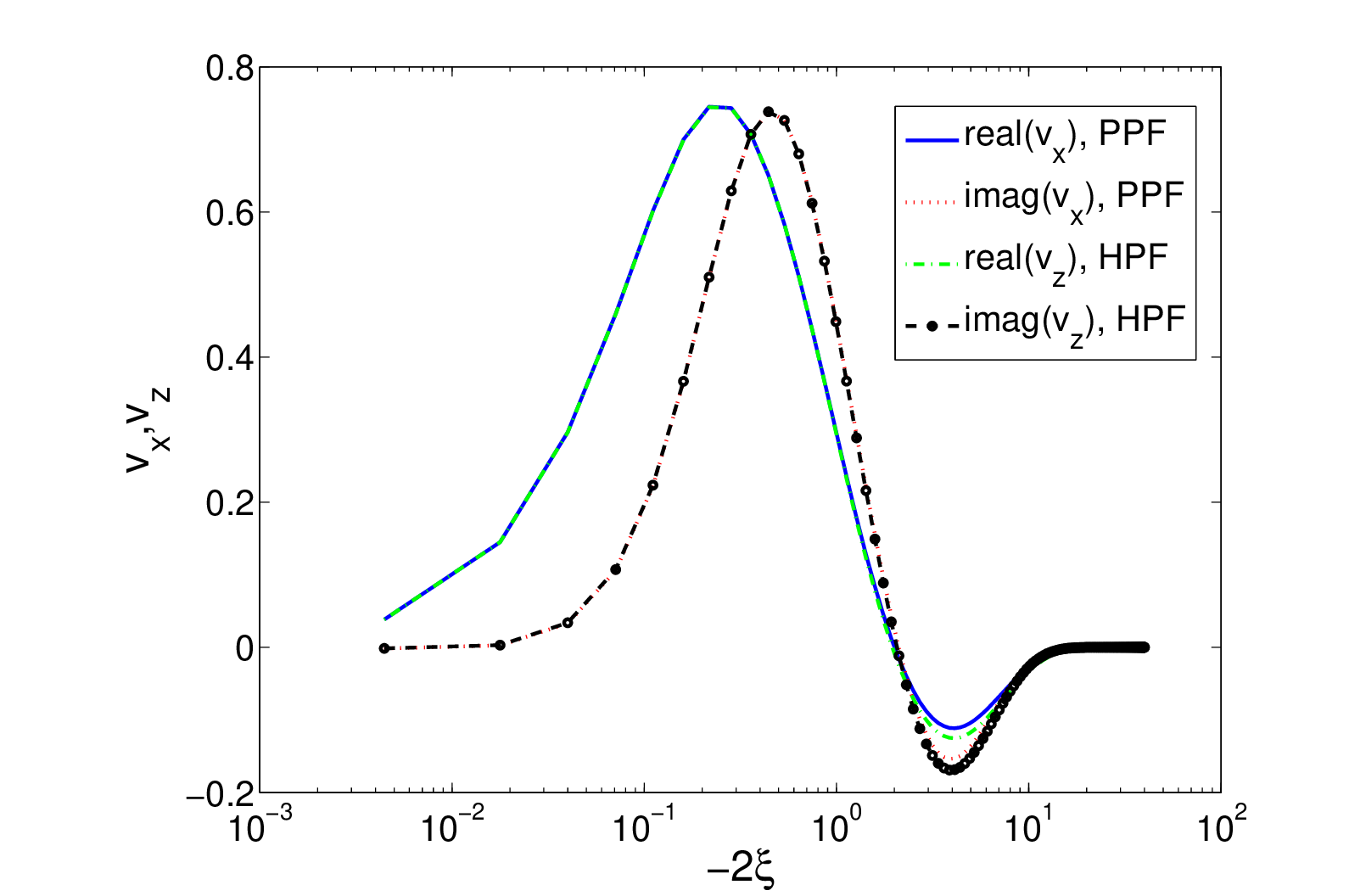}
        \caption{Stream-wise velocity eigenfunction}
        \label{fig:vx_vz_Re_0_n_0p25_Ww_7_k_20_HP}
    \end{subfigure}%
    ~ 
    \begin{subfigure}[b]{0.5\textwidth}
        \centering
        \includegraphics[width=\textwidth]{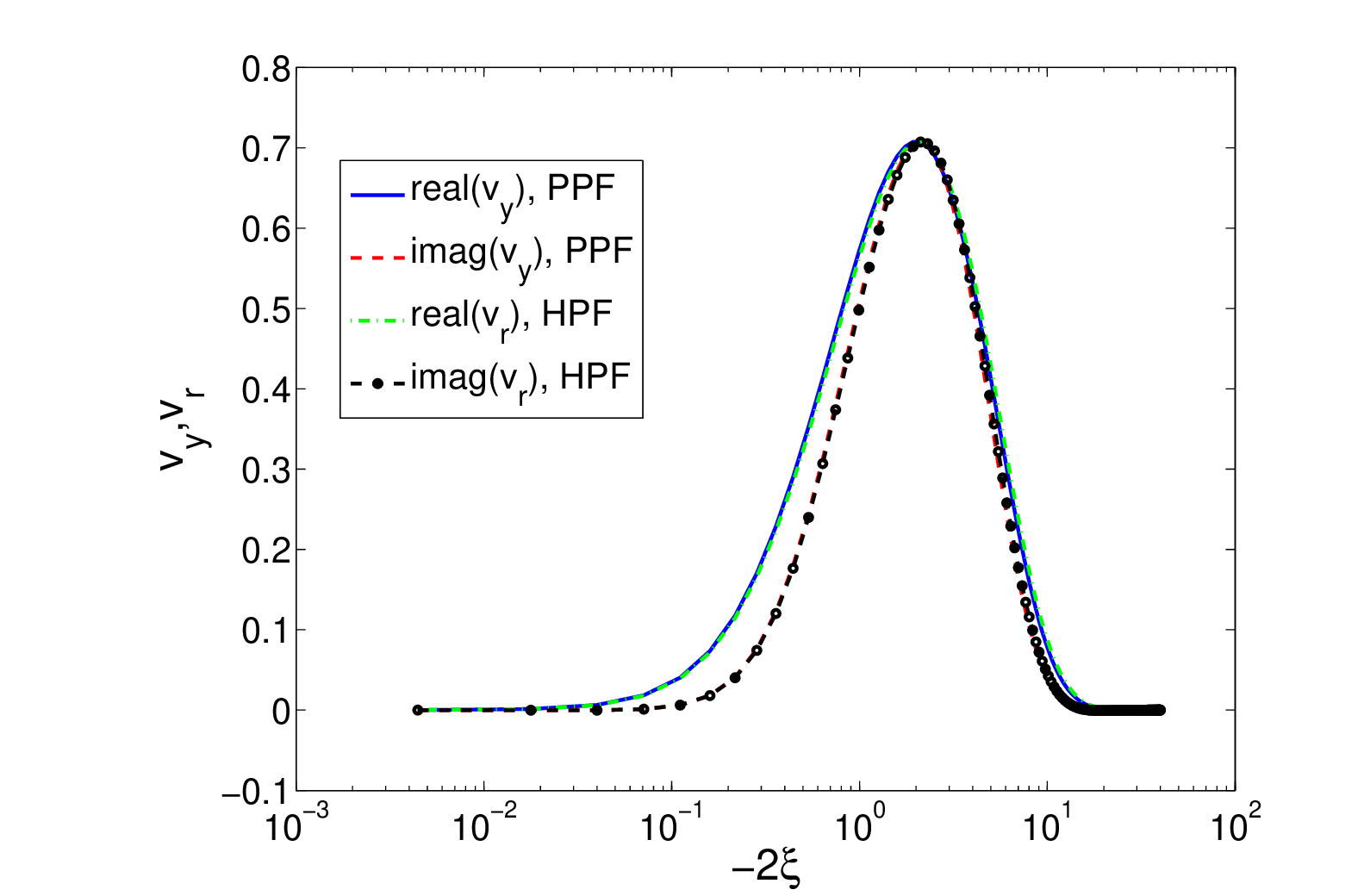}
        \caption{Normal velocity eigenfunction}
          \label{fig:vy_vr_Re_0_n_0p25_Ww_7_k_20_HP}
    \end{subfigure}
    \caption{\small Eigenfunctions for channel flow (varicose mode) and tube flow for $W_w=7, n=0.25, k=20$. The agreement between the eigenfunctions demonstrates the identical nature of the instability found in channel  and tube flows.}
\end{figure*}
Table \ref{table:equivalence} compares the most unstable (or least stable) discrete eigenvalue for all the three geometries at a given $W_w$. To further corroborate the universality, we plot the eigenfunctions for all the three geometries in terms of the inner  coordinate $\xi$. Here, the horizontal axis is $-2 \xi$ and not $\xi$ due to the simplicity this variable offers in plotting the results using the pseudo-spectral code. The $\log$ scale is used for the horizontal axis in the eigenfunction plots to magnify the area near the wall where the unstable disturbances are confined. As expected from the above analysis and shown in figures \ref{fig:vx_Re_0_n_0p25_Ww_7_k_20}-\ref{fig:vy_vr_Re_0_n_0p25_Ww_7_k_20_HP}, the eigenfunctions for all the three geometries collapse onto a single curve.   


\section{Comparison with experiments} \label{sec:comparison-with-experiments}

\subsection{\cite{bodiguel-et-al-2015}}

 \cite{bodiguel-et-al-2015} carried out experiments to characterize the instability in channel flow of semidilute polyacrylamide (PAA) solutions with dependence of shear stress ($\tau_B$) and Weissenberg number ($W_B$) on the shear-rate ($\dot{\gamma}_B$) given by,
\begin{eqnarray}
\tau_B = 3.73 \dot{\gamma}_B^{0.21}, \label{eq:shear-stress-bodiguel}\\
W_B = 3.63 \dot{\gamma}_B^{0.43},\label{eq:w-bodiguel}
\end{eqnarray} 
 \noindent
 where, subscript $B$ signifies quantities obtained from \cite{bodiguel-et-al-2015}. From the caption of figure~1 of \cite{bodiguel-et-al-2015}, $W_B$ is defined as the product of the relaxation time and the shear rate, similar to the definition used here. The difference in the power-law indices for ($\tau_B$) and ($W_B$) is due to the shear-softening of the solution. Here, the shear-softening implies decrease in the shear-modulus of the fluid with an increasing shear rate.  \cite{bodiguel-et-al-2015}characterized the transition in terms of the shear stress at the wall. Thus, in order to estimate the corresponding critical Weissenberg number, Eq. (\ref{eq:shear-stress-bodiguel}) is first solved for $\dot{\gamma}_B$, and the result is substituted in Eq. (\ref{eq:w-bodiguel}) to obtain
 \begin{eqnarray}
W_B = 3.63 \left( \frac{\tau_{B}^w}{3.73} \right)^\frac{0.43}{0.21},\label{eq:w-bodiguel-2}
\end{eqnarray} 
 \noindent
 here $\tau_{B}^w$ is the shear stress acting at the wall. where the superscript $w$ signifies that the quantities are obtained at the wall. \cite{bodiguel-et-al-2015} used particle image velocimetry (PIV) to track the trajectories of tracer particles in channel flow, and observed well-aligned trajectories for $\tau_{B}^w < 4.7$ Pa, and a time-periodic variation of normal velocity for $\tau_{B}^w > 4.7$ Pa. Substituting this critical stress in Eq. (\ref{eq:w-bodiguel-2}), we obtain $W_B^w \approx 5.8$. The relation between the Weissenberg numbers at the wall ($W_B^w$) and in the bulk ($W_B$) is
 \begin{eqnarray}
 W_B=W_B^w \frac{n}{1+n}. 
\end{eqnarray}   
 \noindent
 For $W_B^w \approx 5.8$ and $n=0.21$, the above relation gives $W_B \approx 1$ for the onset of instability in experiments.  Thus, the critical Weissenberg number for the onset of instability from the experiments of \cite{bodiguel-et-al-2015} is $\approx 1$. From the theoretical predictions shown in figure~\ref{fig:Wc_kc_vs_n}, for $n=0.21$, $W_c \approx 1$ sinuous modes while  $W_c \approx 1.8$ for varicose modes. The lower $W_c$ predicted from the stability analysis  for sinuous modes is in good agreement with the threshold $W$ inferred from the experiments of \cite{bodiguel-et-al-2015}.  The authors also observed sinuous modes at the onset of instability, as reported from the supplementary video that accompanied their paper, and as also noted by \cite{wilson-loridan-2015}.
 
 We estimate the Reynolds number, $Re=\rho V_m R/\eta^*_p$, for the experiments of \cite{bodiguel-et-al-2015} as follows. The shear stress at the wall in their experiments was estimated by using expression
 \begin{eqnarray}
 \tau_{B}^w=\frac{\delta P_c }{L} \frac{R}{2}=-\frac{dp}{dx} \frac{R}{2}, \label{eq:tw-bodiguel}
\end{eqnarray} 
\noindent
here $\frac{\delta P_c }{L}=\frac{dp}{dx}$ is the effective applied pressure gradient after eliminating pressure drop loss at the ends. In the above expression, $\delta P_c$ and $L$ are respectively pressure difference between the ends of the channel and length of the channel. The maximum velocity ($V_m$) for the channel flow of a power-law fluid is
\begin{eqnarray}
V_m=\frac{n}{1+n} R^{\frac{1+n}{n}} \left( - \frac{1}{K}\frac{dp}{dx} \right)^{1/n}.
\end{eqnarray} 
\noindent
Solving (\ref{eq:tw-bodiguel}) for $\frac{dp}{dx}$ and substituting in the above equation we obtain 
\begin{eqnarray}
V_m=\frac{n}{1+n} R^{\frac{1+n}{n}} \left(  \frac{2}{K}\frac{\tau_{bw}}{R} \right)^{1/n}.
\end{eqnarray} 
\noindent
The channel half height in their experiments was $R=76 \mu m$ and from (\ref{eq:shear-stress-bodiguel}), $n=0.21$ and $K=3.73 Pa s^{0.21}$. Substituting these values in the above expression for maximum velocity, we obtain $V_m \sim 0.001 m/s$ which corresponds to $Re \sim 10^{-4}$. Thus, fluid inertia can be safely neglected for the experimental regime of \cite{bodiguel-et-al-2015}, and it is reasonable to assume the creeping-flow limit in the analysis. It must be noted that the experimental fluid of \cite{bodiguel-et-al-2015} showed shear softening and substantial wall-slip. But the excellent agreement between our theoretical predictions and the experimental transition indicates that shear-softening and wall slip are perhaps not dominant effects required to predict this instability.  
 

\subsection{\cite{poole-2016}}

\cite{poole-2016} experimentally studied the transition of a strongly shear-thinning viscoelastic fluid through channels and tubes of significantly larger diameters ($\sim 25-100$ mm) unlike the microfluidic studies of \cite{bodiguel-et-al-2015}. The larger channel/tube dimensions in \cite{poole-2016} reduced wall slip and thus experimental and analytical  velocity profiles  were in quantitative agreement. The experimental fluid of \cite{poole-2016} was PAA dissolved in water and the concentration was higher than the critical overlap concentration and thus solution was highly concentrated. The power-law index of the viscosity for the channel and tube flows was $0.19$. The experimental fluid was shear-softening thus the power-law indices of the relaxation time for the channel and tube flows were $0.36$ and $0.32$, respectively.   

The Reynolds and Weissenberg numbers of \cite{poole-2016} were defined using the average velocity of the fluid. The relation between the dimensionless numbers defined in the present work and that of \cite{poole-2016} are
\begin{eqnarray}
W=W_P \left(\frac{1+2n}{1+n}\right)^n,\label{eq:w-relation}\\
Re=Re_P \left(\frac{1+2n}{1+n}\right)^{(2-n)}, \label{eq:re-relation}
\end{eqnarray}
\noindent
where, $W_P$ and $Re_P$ are the Weissenberg and Reynolds numbers defined by using average velocity and subscript $P$ indicates they are dimensionless numbers from \cite{poole-2016}. 

By using above equation for $W$, the critical Weissenberg number for the channel flow is $~9.2$. From figure~\ref{fig:Wc_kc_vs_n}, our linear stability analysis predicts $W \sim 1.2$. Thus, there is some discrepancy between theory and observations, and the agreement is not as good as the one obtained with the experiments of \cite{bodiguel-et-al-2015}. For tube flow, experiments of \cite{poole-2016} predicted $W_c \sim 6.1$ and from figure~\ref{fig:Wc_kc_vs_n_Re_0_n_hpf}, $W_c \sim 2$. 

The discrepancy between the theoretical predictions of the present work and the experimental observations of \cite{poole-2016} could be attributed to the following reasons. In the present work, we have neglected fluid inertia, but due to the larger diameter of tube,   the role of inertia is not negligible in \cite{poole-2016}'s experiments. The Reynolds number in the experiments of \cite{poole-2016} was of $O(10^2)$ based on the average velocity and shear-rate dependent viscosity. From relation (\ref{eq:re-relation}), we estimate the Reynolds number defined in the present work is also of $O(10^2)$. The experiments of Bodiguel et al., however, involved much smaller channel dimensions (gap width $\sim 150 \mu m$), thus inertia was negligible, which explains the better agreement between theory and experiments. Hence, it could be speculated that consideration of finite inertia could lead to better agreement between the theory and the experiments of \cite{poole-2016}.

Another factor the present theoretical work neglects is the consideration of the shear-softening which was experimentally observed by both \cite{bodiguel-et-al-2015} and \cite{poole-2016}. However, an analysis involving the shear-rate dependence of the relaxation modulus leads to a singularity in the constitutive equations at the channel or tube center. The above singularity can only be removed by using a numerical patch at the center by means of using a Frobenius series solution.
Also, in the present model for the shear-thinning viscoelastic fluid for channel and tube flows, we have ignored the contribution of the solvent because of the lack of an analytical base state velocity profile when both shear-thinning  and Newtonian solvent are present in the model.  To this end, \cite{castillo-wilson-2017} used an inelastic shear-thinning solvent having power-law index equal to that of the polymer and predicted strong stabilizing effect of the addition of the inelastic shear-thinning solvent. However, in the experiments of \cite{bodiguel-et-al-2015} and \cite{poole-2016}, the solvent is water, which is a Newtonian fluid. Thus, a realistic consideration of the solvent in the theoretical analysis could also result in a better agreement between theory and experiments.


\subsection{\cite{picaut-et-al-2017}}
\label{subsec:picaut}

\cite{picaut-et-al-2017} carried out experiments on the flow of entangled polymer solutions extruded through capillaries. The solution used was obtained by dissolving sodium alginate (a biopolymer) in deionized water and 500 ppm of sodium azide.  The power-law index of solutions used was reported to be $n \sim 0.29$. The  capillaries used were with radii $100, 150$ and $250 \mu m$, thus the dimensions of the geometry  studied were similar to the channel experiments of \cite{bodiguel-et-al-2015}. The rheological data of \cite{picaut-et-al-2017} shows that the shear modulus of the fluid is independent of the shear rate unlike in the case of \cite{bodiguel-et-al-2015,poole-2016}. Thus, the assumption of constant shear modulus employed in the present work is appropriate  for the system of \cite{picaut-et-al-2017}. The authors provide the critical average bulk shear rate for the instability, from which we can estimate the Reynolds number and critical Weissenberg number to be compared with the predictions of the present study. The relation between  the average shear-rate ($V_{avg}/R$) and maximum shear-rate ($V_m/R$) for the tube flow is
\begin{eqnarray}
\frac{V_m}{R}=\frac{V_{avg}}{R} \left(\frac{1+3n}{1+n}\right),
\end{eqnarray}

We estimated $Re \sim O(10^{-4})$ in the experiments of  \cite{picaut-et-al-2017}  based on the zero-shear viscosity and the critical maximum shear-rate obtained from their table~1. Thus, the creeping-flow limit assumption of the present study is justified for the experimental regime probed by \cite{picaut-et-al-2017}. Next, the Weissenberg number based on the average shear-rate in their experiments  was
\begin{eqnarray}
W_{Pi}=\frac{1}{\dot{\gamma}_0^n} \left( \frac{V_{avg}}{R}  \right)^n=\frac{1}{\dot{\gamma}_0^n} \left(\frac{1+n}{1+3n}\right)^n \left( \frac{V_m}{R}  \right)^n,
\end{eqnarray}
where subscript $Pi$ signifies quantity defined in the study of \cite{picaut-et-al-2017}. The parameter $\dot{\gamma}_0$ is the shear-rate at the onset of shear-thinning in the viscosity vs shear-rate curve whose values are provided in their table~1. Comparison of the above equation with $W$ defined in Sec.~\ref{sec:pf-problem-formulation} we obtain
\begin{eqnarray}
W_{Pi}=\lambda^*_{sc} \left(\frac{1+n}{1+3n}\right)^n \left( \frac{V_m}{R}  \right)=W \left(\frac{1+n}{1+3n}\right)^n \,.
\end{eqnarray}
Solving above equation for $W$, we obtain
\begin{eqnarray}
W=W_{Pi} \left(\frac{1+3n}{1+n}\right)^n,
\end{eqnarray}
 From above relation, for all three solutions $W_c \sim 3$ in the experiments of \cite{picaut-et-al-2017}. From figure~\ref{fig:Wc_kc_vs_n_Re_0_n_hpf}, for $n=0.29$, the theoretical estimate shows $W_c \sim 6$. Thus, there is a reasonable agreement between our theoretical prediction and experimental observations of \cite{picaut-et-al-2017}.
They further observed a helical instability which implies an instability due to the non-axisymmetric disturbances.  However, our present analysis is restricted to axisymmetric disturbances. It is possible that consideration of non-axisymmetric disturbances could improve the agreement between the theoretical predictions and experimental observations.

\section{Conclusions} \label{sec:conclusions}

A comprehensive linear stability analysis of canonical rectilinear shear flows, viz., plane Couette flow and pressure-driven flows through channels and tubes, of a strongly shear-thinning viscoelastic (White--Metzner) fluid is presented in the creeping-flow limit. We demonstrate the existence of an identical short-wave, `wall mode' instability in all three flow configurations.  The stability analysis for PCF reveals that the instability in the WM fluid is governed by a continuation of the stable GL \citep{gorodtsov-leonov-1967} modes in a UCM fluid to $n \leq 0.3$.  This illustrates that shear-thinning has a destabilizing effect on the stable elastic discrete GL modes. With decrease in $n$, more discrete modes emerge from the continuous spectrum which are eventually destabilized for sufficiently small $n$. Our numerical results further showed the presence of both  center ($c_r \sim 0.5$) and wall ($c_r \sim 0,1$) modes for PCF. However, these are not completely distinct, and the center modes smoothly cross over to wall modes at high $k$.
 
Similar to PCF, even for pressure-driven channel and tube flows, the discrete elastic modes in the UCM limit are destabilized with decreasing $n$ and new modes originate from the continuous spectrum.  This trend appears to be generic to shear thinning WM fluids. In prior literature, only varicose modes were analysed for channel flow due to an apparent singularity at the center of channel. In the present work, however, we showed that the governing equations for the disturbances can be regularised for channel and tube flows,  and that the pseudospectral code can be used to determine the eigenspectrum. Results for channel flow show that sinuous modes are more unstable than varicose modes, in agreement with experimental observations \citep{bodiguel-et-al-2015}. The results for tube flow (in terms of the critical $W_c$ required for the instability) are in qualitative agreement with experimental observations of \cite{poole-2016,picaut-et-al-2017}.

An asymptotic analysis at high wavenumbers showed that an identical wall mode instability exists for PCF, channel and tube flows of a WM fluid. The localization of disturbances near the wall in this limit further shows
that the instability predicted in the present work is likely to be present in any wall-bounded shear flow of a shear-thinning viscoelastic fluid.
Although all three geometries exhibit an identical wall mode instability, we find that for $k<1$, PCF exhibits center mode instability while channel and tube flows exhibit center mode instability for $0.5<k<5$. Furthermore, the center mode for channel and tube flows exhibit different behaviour with variation in $n$ owing  to  flow-specific boundary conditions. The boundary condition dependent behaviour of the instability makes the sinuous mode more unstable than the varicose mode for channel flow, in agreement with experimental observations. 
%

\appendix
\section{Origin of wall and center modes} \label{appA}

\subsection{Origin of discrete wall modes: $n \in [0.4, 1]$}\label{appA1}

\begin{figure}
\centering
\includegraphics[width=0.6\textwidth]{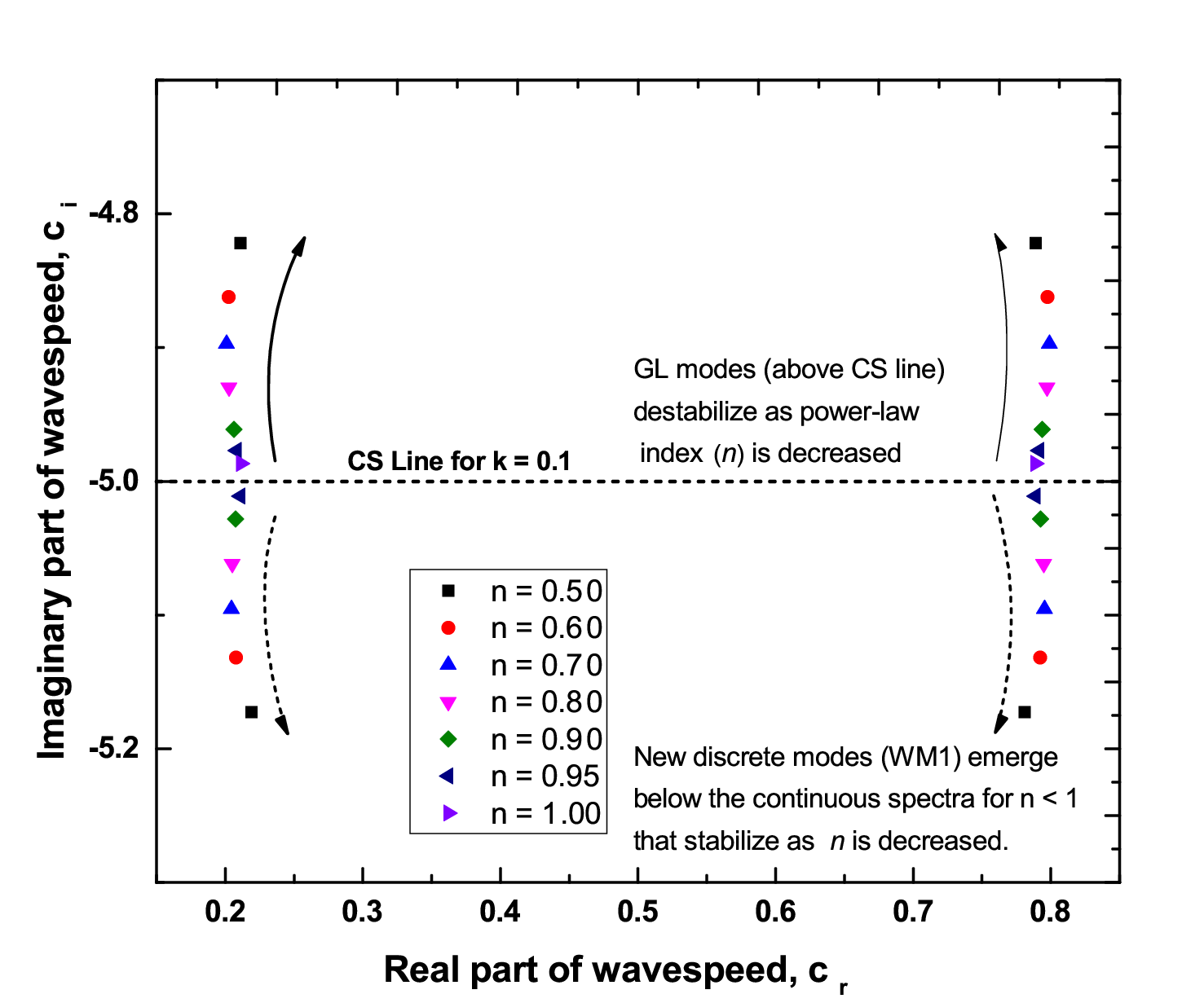}
\caption{Spectra for $W=2$, $Re=0$, $\beta=0$, $k= 0.1$ and at different $n$ showing the emergence of discrete wall modes below the CS line, and the trajectory of GL modes with decreasing $n$.}
\label{fig:k0p5n0p5to1}
\end{figure}

\begin{figure}
\centering
\includegraphics[width=0.6\textwidth]{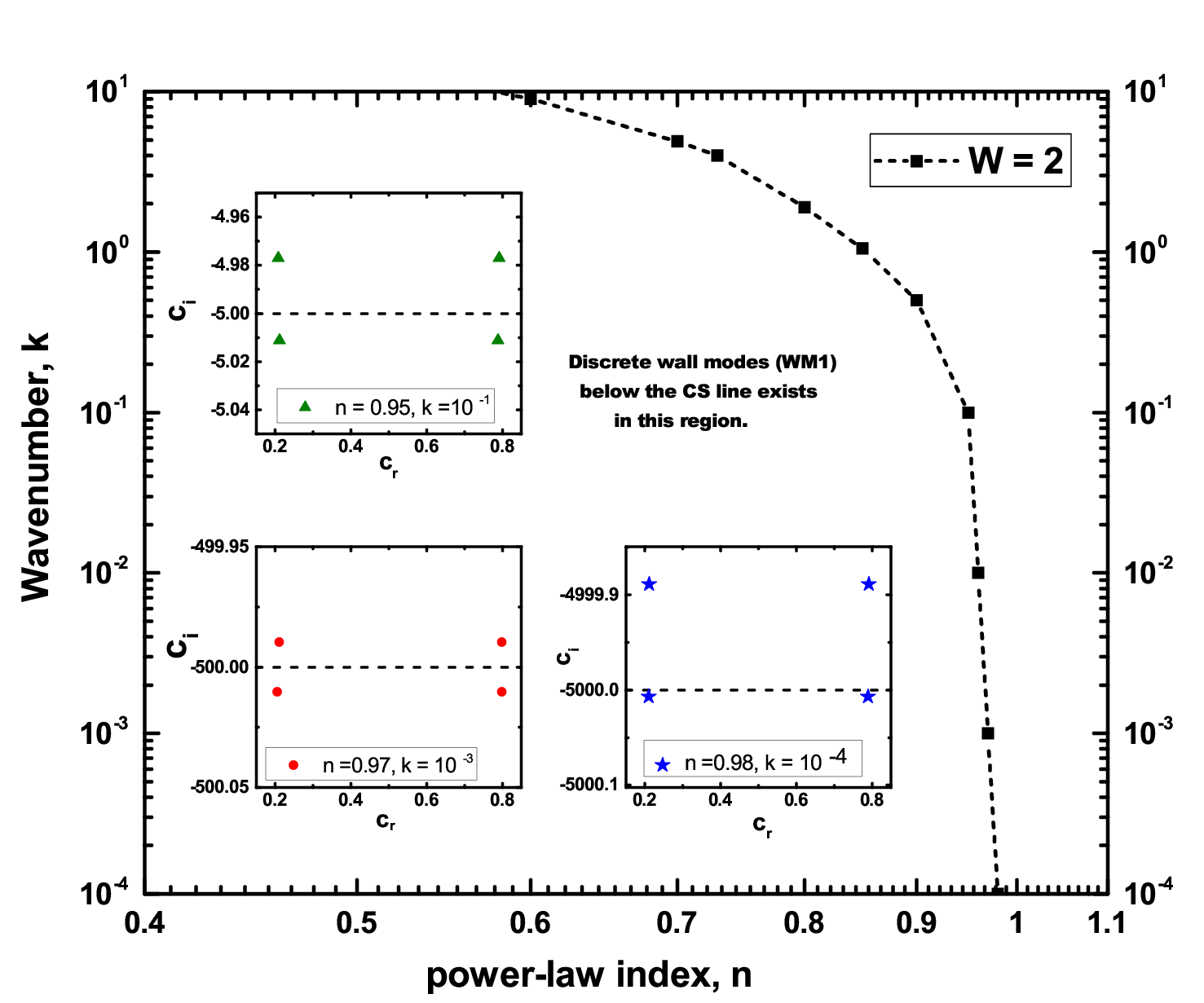}
\caption{Variation of the wavenumber at which the discrete wall modes (WM1) emerge below the CS line as a function of $n$. Data for $Re=0$, $W=2$, $\beta=0.0$.}
 \label{fig:kvsn_highn}
\end{figure}

\begin{figure}
\centering
\includegraphics[width=0.6\textwidth]{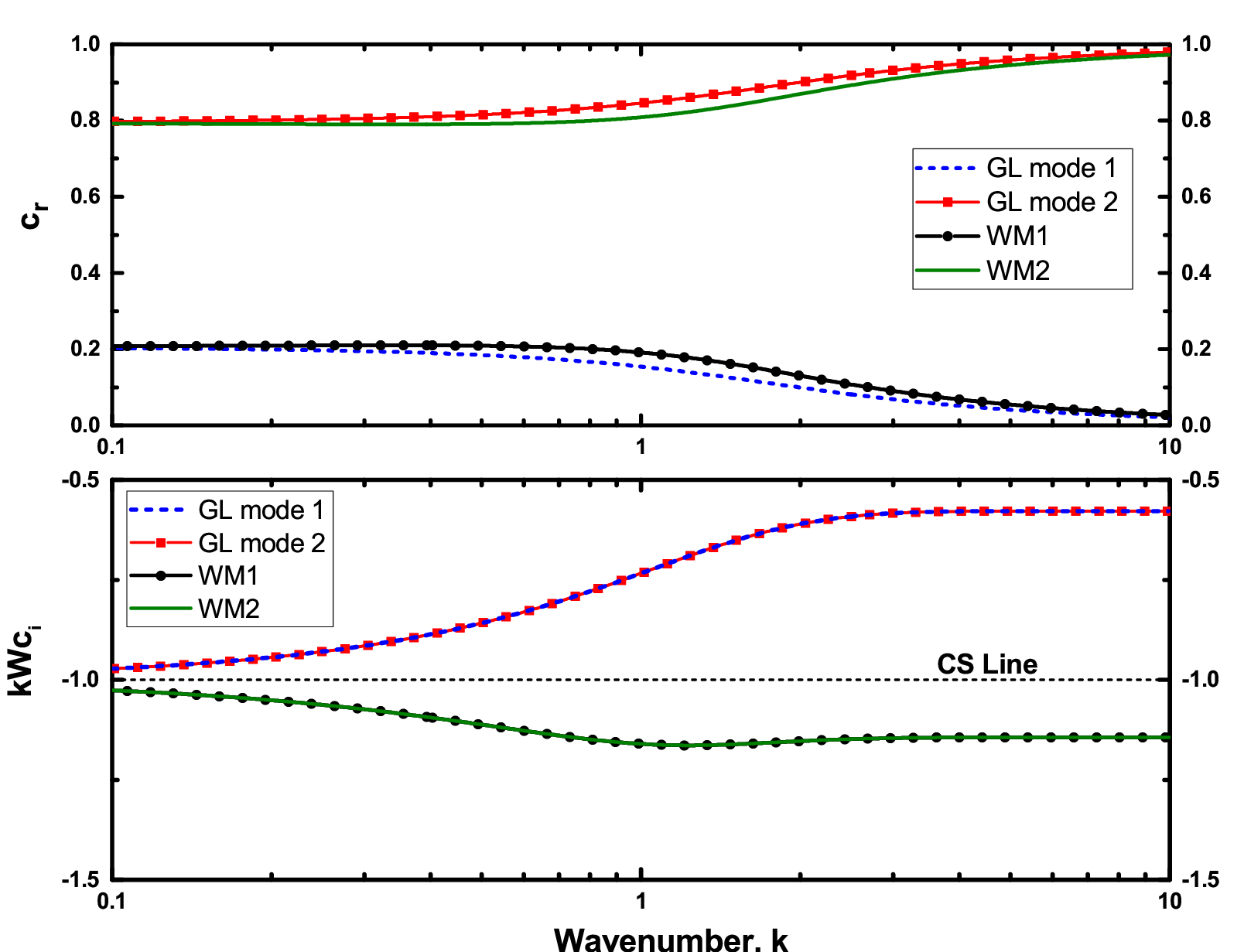}
\caption{The effect of variation of wavenumber for a fixed $n$: Panel~(1) of the figure shows the evolution of real part of wavespeed $c_r$ with wavenumber for $W=2$, $Re=0$, $\beta=0$, $n = 0.6$.  Panel~(2) of the figure shows the variation of the scaled growth rate $kWc_{i}$ with wavenumber ($k$) for the same set of parameters. The scaled CS Line is shown in the lower panel for visual reference. We observe that the GL and wall modes above and below the CS line exhibit a smooth variation as $k$ is decreased in this parameter regime.}
\label{fig:n0p6_kwci_shoot}
\end{figure}

\begin{figure*}
    \begin{subfigure}[b]{0.5\textwidth}
        \includegraphics[width=\textwidth]{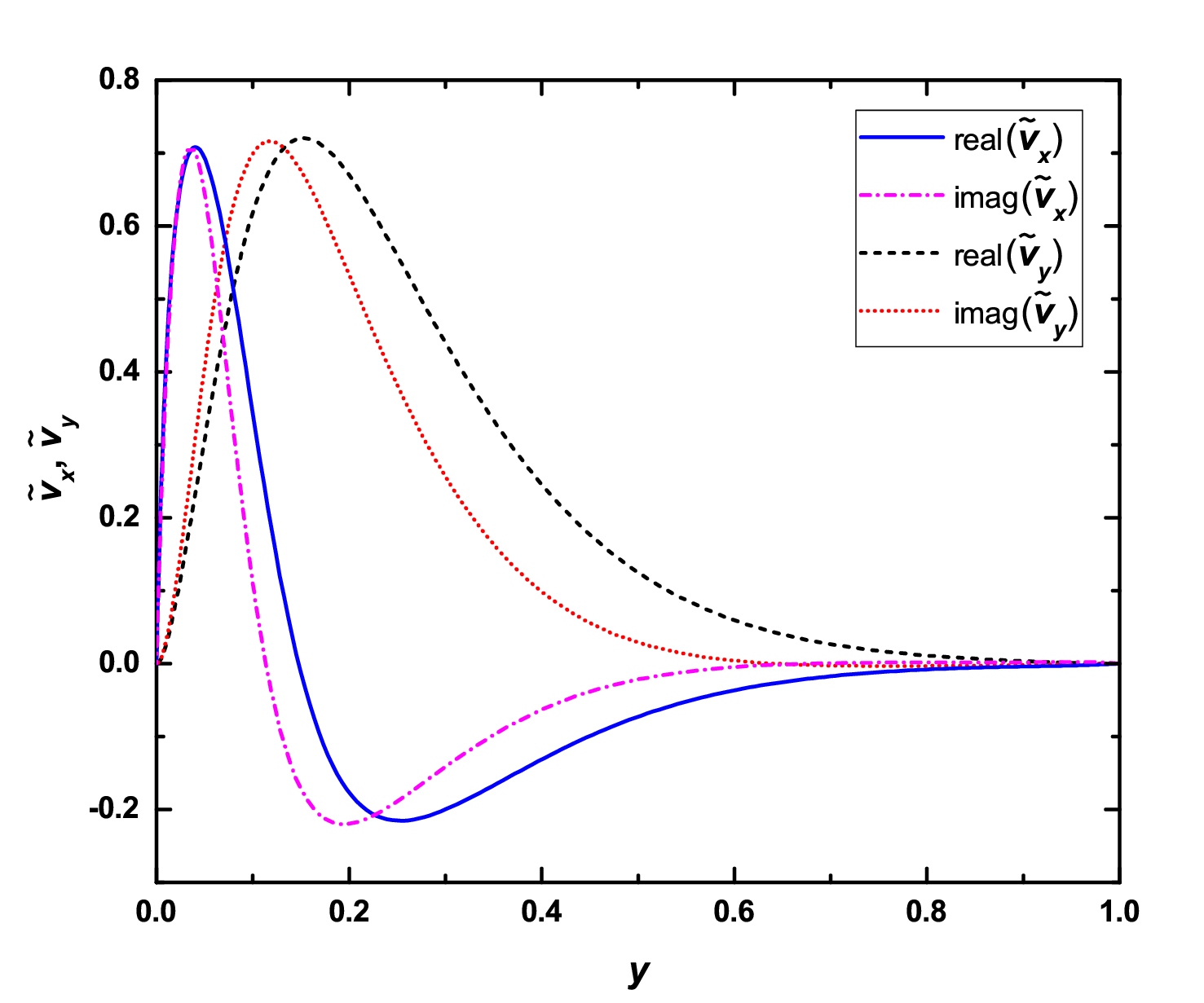}
   \caption{GL mode $c_r \rightarrow 0$ }
   \label{fig:GL1}
\end{subfigure}\hspace{1em}
  \begin{subfigure}[b]{0.5\textwidth}
         \includegraphics[width=\textwidth]{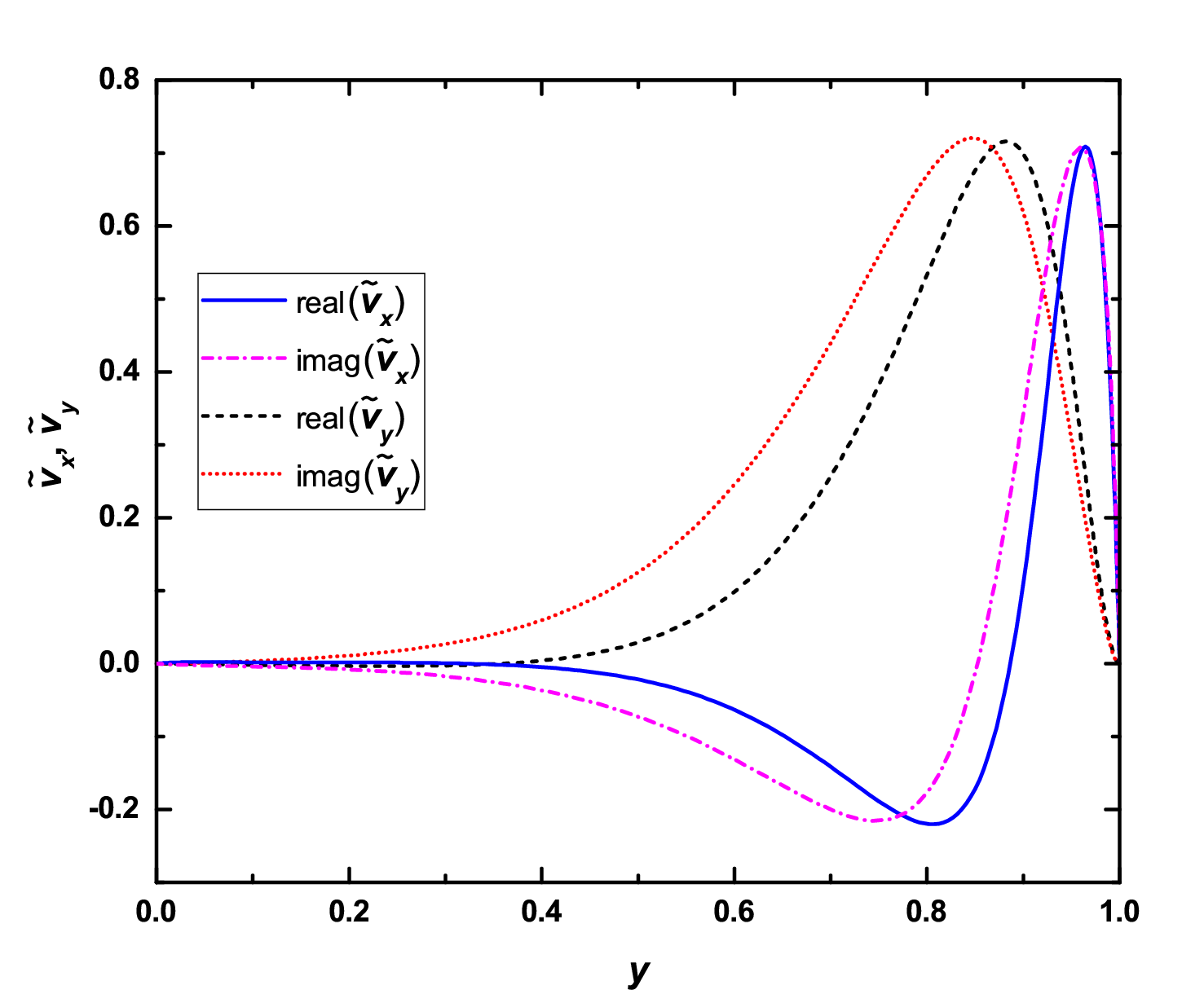}
   \caption{GL mode $c_r \rightarrow 1$}
    \label{fig:GL2}
\end{subfigure}
 \begin{subfigure}[b]{0.5\textwidth}
         \includegraphics[width=\textwidth]{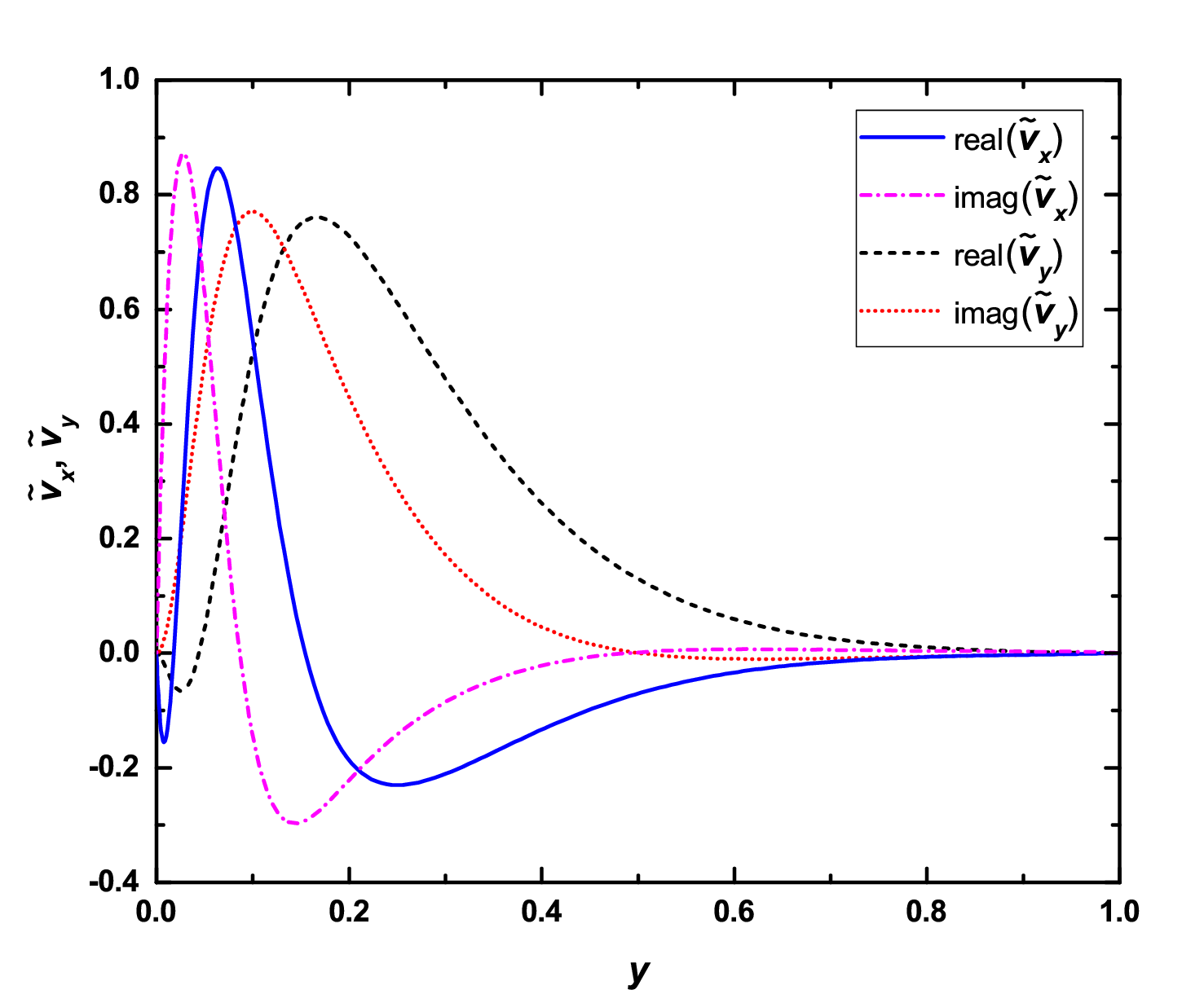}
   \caption{WM1 $c_r \rightarrow 0$}
    \label{fig:WM1}
\end{subfigure}\hspace{1em}
 \begin{subfigure}[b]{0.5\textwidth}
            \includegraphics[width=\textwidth]{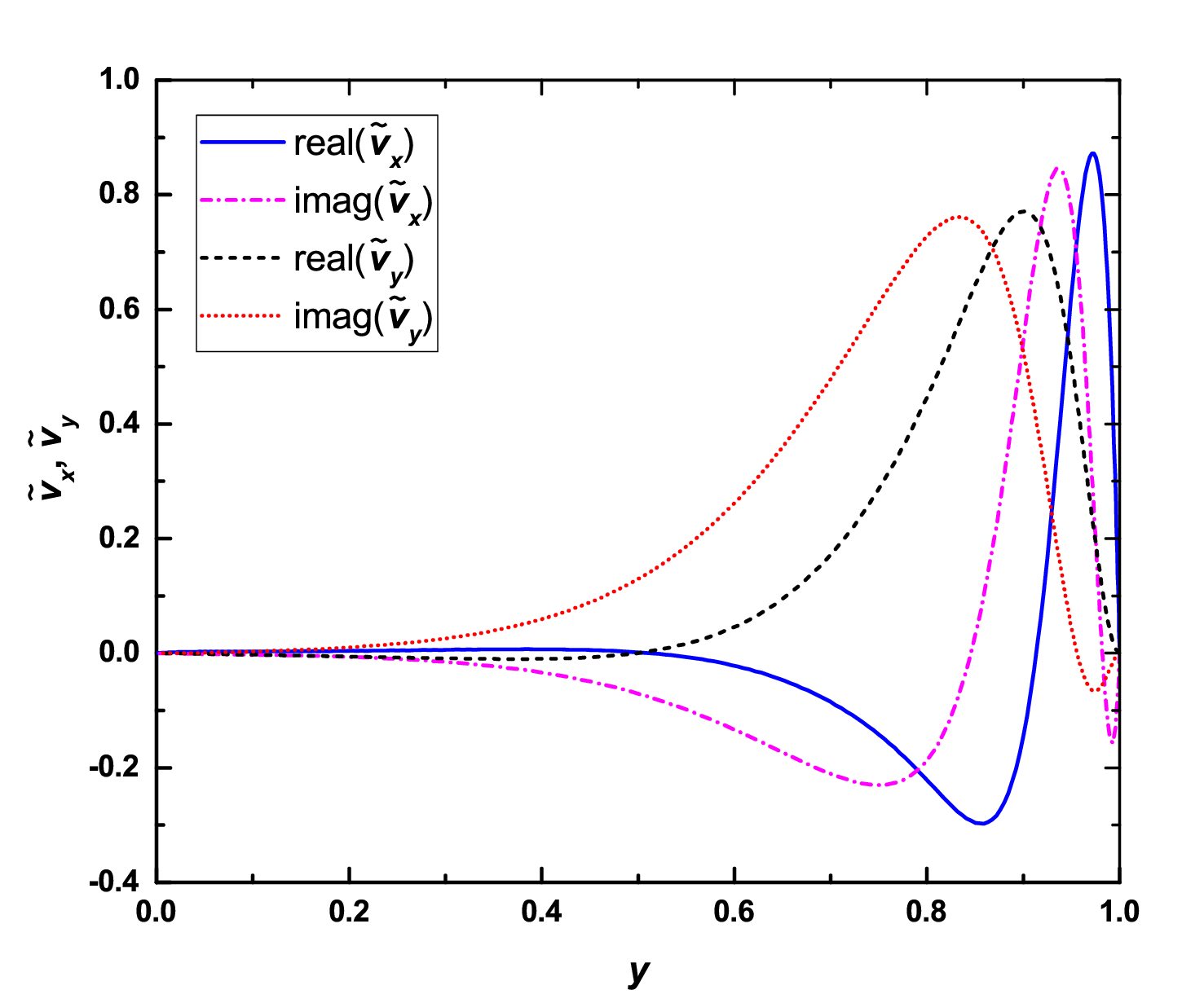}
   \caption{WM1 $c_r \rightarrow 1$}
    \label{fig:WM2}
\end{subfigure}
\caption{Representative eigenfunctions for the GL and wall modes: Data for $n = 0.6$, $k =10$, $W =2$, $\beta = 0$ and $Re =0$. In panel (a) we show the eigenfunctions for the GL mode $c= 0.02056255 - 0.02889299i$ and panel~(b) represents the eigenfunctions for the GL mode $c =0.97943745 - 0.02889299i$. Panels (c) and (d) demonstrate the eigenfunctions for the WM1 present below the CS line given by $c = 0.97276311 - 0.05720071i
$ and $c = 0.02723689- 0.0572007i$.}
 \label{fig:Eigenfunctions}
\end{figure*}

 As the power-law index $n$ is decreased from $n = 1$ (i.e. UCM limit) to $0.5$ in figure \ref{fig:k0p5n0p5to1}, the two GL modes (corresponding to $n=1$, $W\neq 0$ and $\beta = 0$) above the CS becomes less stable. Also, a pair of new discrete wall modes (abbreviated as WM1 henceforth) emerge below the polymer continuous spectrum. These modes  appear in the spectrum as a pair (one with $c_r \rightarrow 0$ and another with $c_r \rightarrow 1$), with the same decay rates ($c_i$) owing to the (anti)symmetry of the base plane Couette flow. These modes, however, become more stable as $n$ is decreased further.
The value of $n$ at which new discrete modes emerge depends on the wavenumber $k$. Figure \ref{fig:kvsn_highn} shows that as $n$ is increased towards unity, the $k$ at which the new discrete wall modes (WM1) appear below the CS also decreases, and $k$ appears to decrease rapidly to zero as $n \rightarrow 1$. The filtered spectra for $n = 0.95$, $n = 0.97$ and $n = 0.98$ corresponding to the wavenumber at which the discrete modes (WM1) emerge below the polymer continuous spectra, are shown in the insets of figure \ref{fig:kvsn_highn}. 
Figure\,\ref{fig:n0p6_kwci_shoot} shows the variation of $c_r$ and the scaled growth rate $kWc_i$ with $k$ for both the GL and WM1 modes, for a fixed power-law index ($n =0.6$) and Weissenberg number ($W=2$). The scaled growth rate  $kWc_i$ fixes the location of the polymer continuous spectrum at $k W c_i = -1$ for $\beta = 0$ and $\bar{\eta_{p}} = 1$, as shown in the figure. We observe that for both the modes (GL and WM1), the growth rate plateaus off in the high wavenumber regime ($3<k<10$). As the wavenumber is decreased from higher to lower values, the GL modes become more stable. In contrast, WM1 becomes slightly less stable in lower wavenumber regime. However, the real part of wavespeed ($c_r$) does not exhibit a significant change with $k$. Also, no new discrete modes emerge from the CS for any $k$ in this power-law index regime where $n \in [0.4, 1]$. We only observe a pair of discrete GL modes above the CS line and a pair of discrete wall modes (WM1) below the CS line for $n\in [0.4, 1]$ within the wavenumber regimes under consideration. 
We next plot the real and imaginary parts of the eigenfuctions $\tilde{v}_x$ and $\tilde{v}_y$ for the GL and wall modes (WM1), in high wavenumber regime ($k =10$) in figure \ref{fig:Eigenfunctions}. We observe that for both the wall and GL modes, the normalised (such that maximum absolute value is unity) eigenfunctions $\tilde{v}_x$ and $\tilde{v}_y$ (shown in subfigures (a) to (d)), are more localized near the two walls at $y =0$ and $y =1$, in the given parametric regime. However, as $k$ is decreased these eigenfunctions spread out in entire $y$-domain (not shown here). 

\subsection {Origin of center modes: $n \in [0.3, 0.4]$ }\label{appA2}
\begin{figure}
\centering
\includegraphics[width=0.6\textwidth]{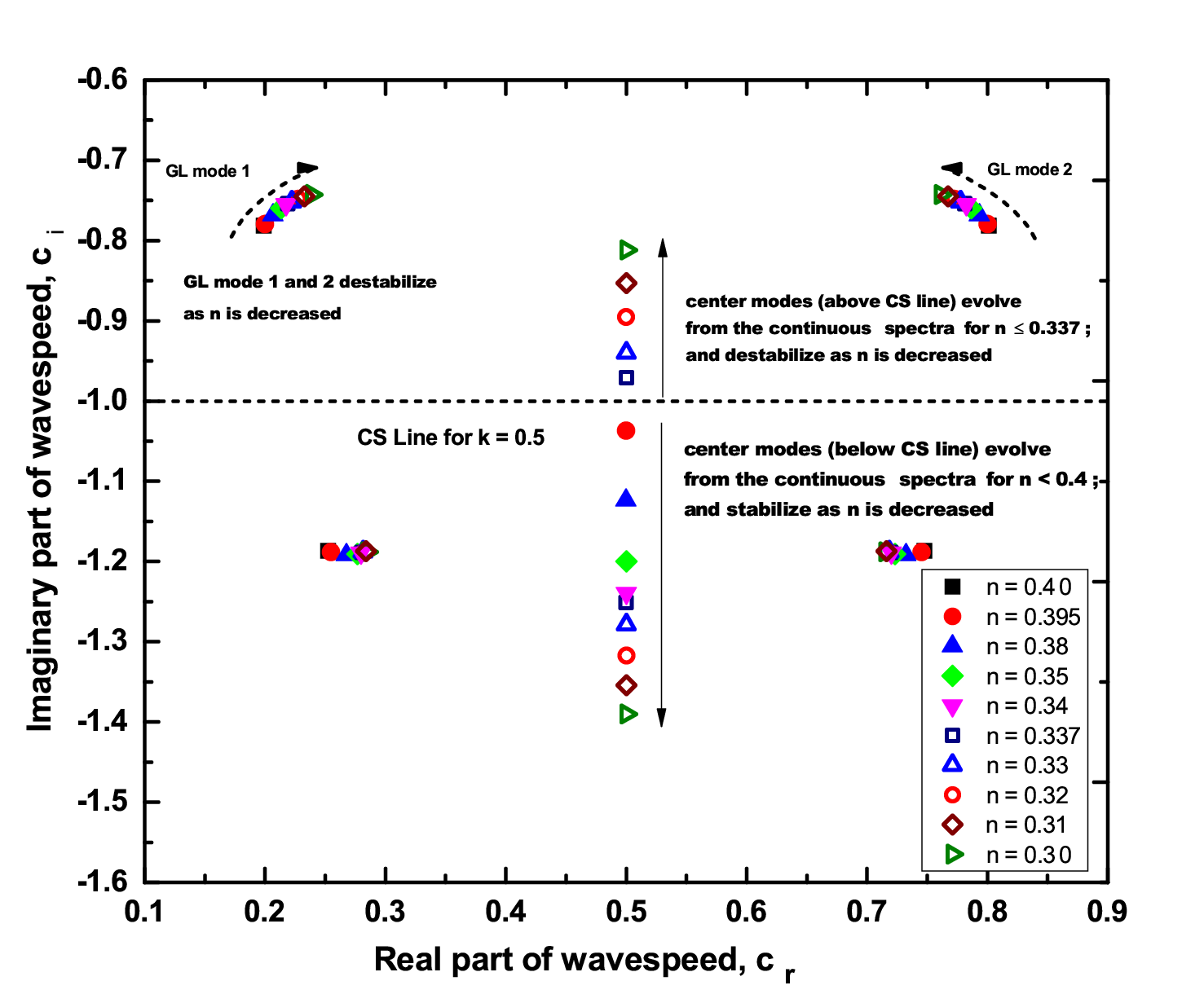}
\caption{Filtered eigenvalue spectrum demonstrating the emergence and subsequent trajectory of center modes for $W=2$, $Re=0$, $\beta=0$, $k=0.5$ and at different $n$. As the power-law index is decreased from $n = 0.4$ to $n = 0.33$, the center modes ($c_r = 0.5$) emerge both below (marked by solid symbols) and above (marked by open symbols) the CS line.}
\label{fig:n0p3to4}
\end{figure}
In figure \ref{fig:n0p3to4}, we demonstrate the effect of variation of $n$ on the spectra 
 for $k = 0.5$, $W=2$, and $\beta=0$. As $n$ is decreased from $0.4$ to $0.3$, we observe that new discrete `center modes' that are characterized by $c_r = 0.5$ emerge from the polymer continuous spectrum in low-wavenumber regime. These modes (marked by solid symbols) initially emerge below the continuous spectra, which further stabilize as  $n$ is decreased. However, as the power-law index is further decreased below a critical $n$ ($n \approx 0.337$ for fixed $W, k$ and $\beta$), additional center modes begin to emerge \textit{above} the polymer continuous spectra. These center modes (marked by open symbols) become less stable as the power-law index is decreased further. The location of the polymer continuous spectrum for a given parameter set is depicted using the analytical expression (Eq.\,\ref{CSline}) in figure \ref{fig:n0p3to4}. The GL modes from the UCM limit (denoted as GL mode $1$ and $2$ in the figure) become slightly less stable as $n$ is decreased from $0.4$ to $0.3$. However, there is no significant impact of variation of $n$ on the wall modes in this range of $n$. In the high-wavenumber regime, we only observe the GL and wall modes (WM1) (similar to the high-$k$ behavior for $n>0.4$).

\begin{figure}
\centering
        \includegraphics[width=0.6\textwidth]{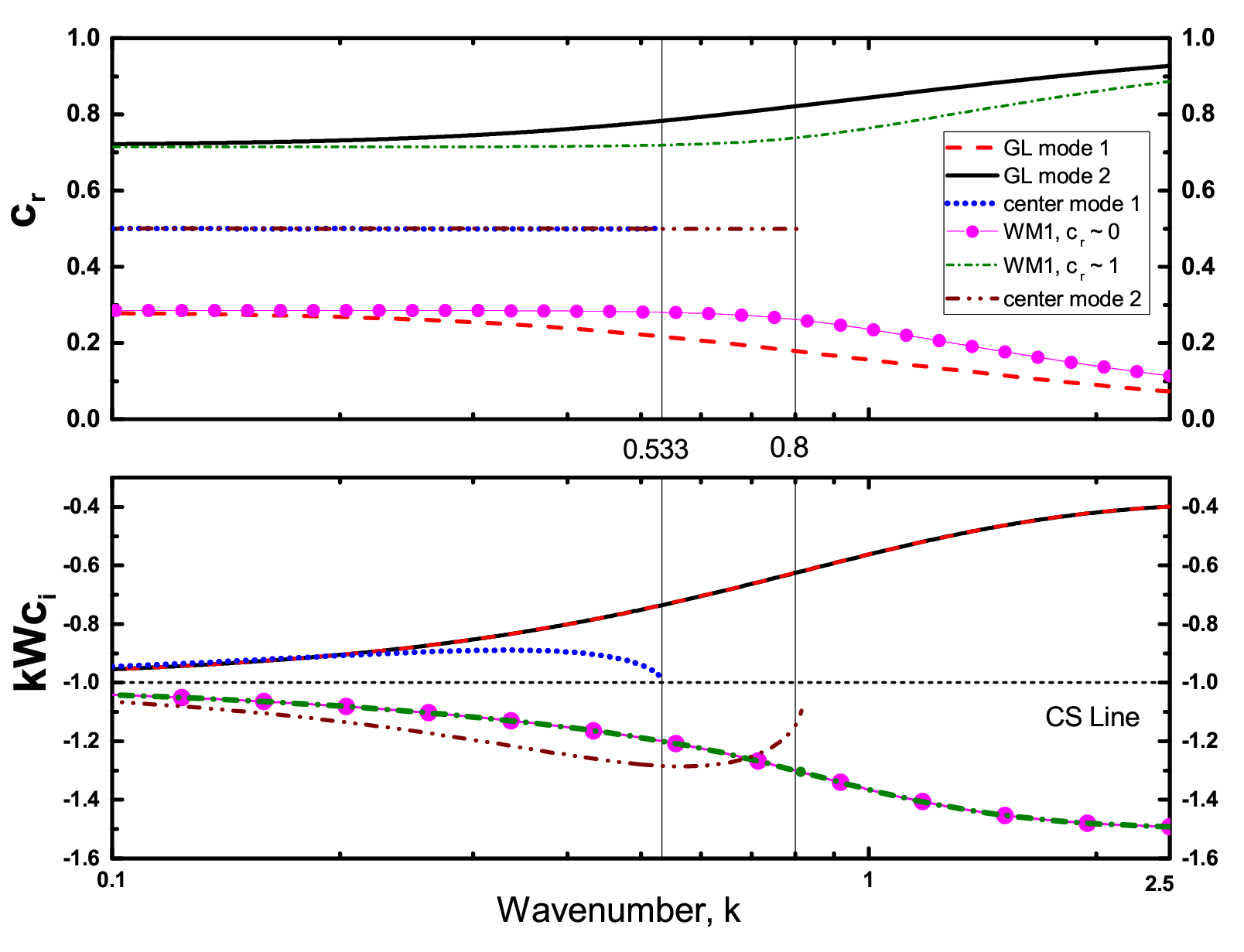}
\caption{Variation of the phase speed and scaled growth rate with wavenumber $k$. Data for $W=2$, $Re=0$, $n=0.33$ and $\beta=0$.}
\label{fig:n0p33_shoot_kwci}
\end{figure}

\begin{figure}
\centering
    \includegraphics[width=0.6\textwidth]{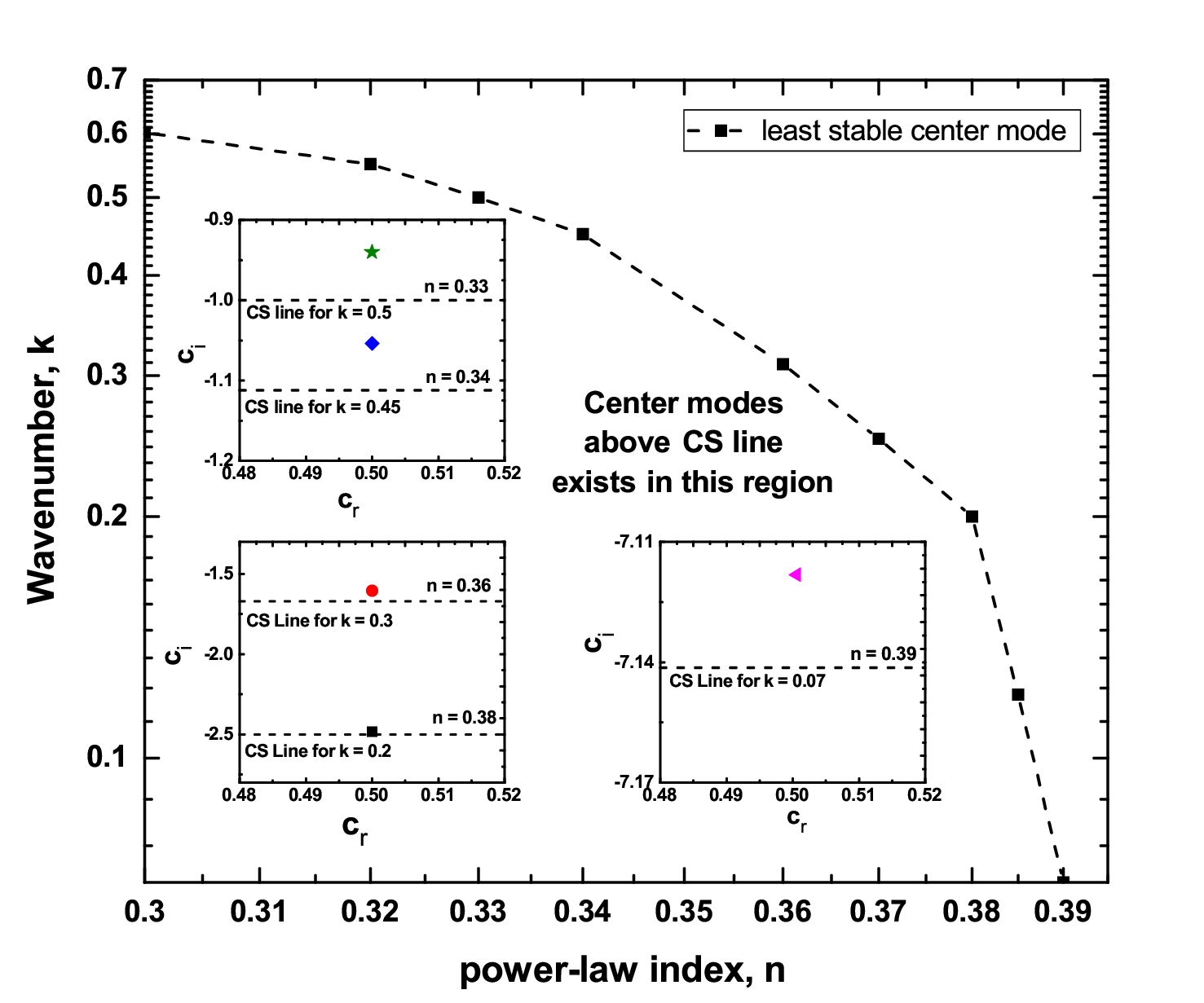}
\caption{Variation of the wavenumber at which center modes emerge above the CS with $n$. Data for $W=2$, $Re=0$, $\beta=0$ $n$ and $k$ is as shown in the figure. Eigenvalue spectrum in the inset of the figure shows the center modes above the CS at different $n$. As the wavenumber is decreased, the center modes can be observed above the CS  for $n \rightarrow 0.4$. However, we do not observe centermodes for $n>0.4$ for $W=2$.}
\label{fig:kvsn}
\end{figure}

\begin{figure}
\centering
    \includegraphics[width=0.6\textwidth]{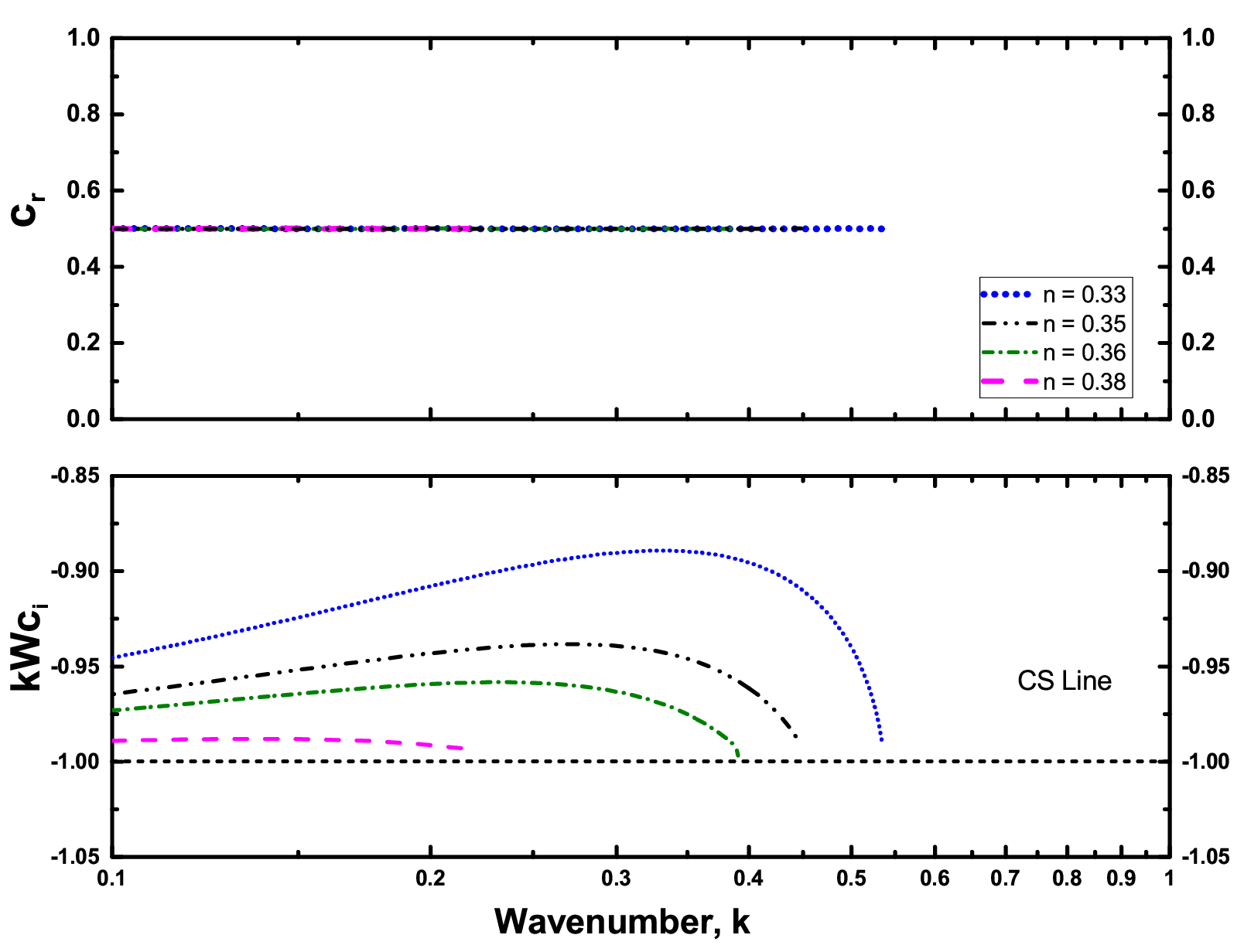}
\caption{Variation of real part of wavespeed ($c_r$) and scaled growth rate $kWc_i$ with $k$ for $W=2$, $Re=0$, $\beta=0$ and $n\in[0.33,0.4]$ as shown in figure; this figure shows the emergence and evolution of the center modes above the CS line with $k$. }
\label{fig:centermodes_kwci_shoot}
\end{figure}

To elucidate the emergence of the modes better, we next show the variation of $c_r$ and $kW c_i$ with $k$ at a fixed power-law index ($n = 0.33$) on the GL, wall (WM1) and center modes in figure \ref{fig:n0p33_shoot_kwci}. As the wavenumber is decreased, the scaled decay rate, $|kWc_i|$, for the GL modes smoothly decreases with $k$. However, the wall modes (WM1) become slightly less stable as $k$ is decreased in this power-law index regime, but remain below the CS. While the variation of the growth rates for the GL modes and WM1 is similar to what was observed for $n> 0.4$, a new center mode (with $c_r = 0.5$ ) emerges from the polymer continuous spectrum below the CS line at $k \sim 0.8 $ and exists for all wavenumbers below it. As the wavenumber is decreased further, another center mode emerges above the CS line at $k \sim 0.533$ and exists for all $k<0.533$. 
The critical values of the power-law index $n$ and the wavenumber $k$ at which the center modes emerge above the CS, for a fixed $W$, is shown in figure \ref{fig:kvsn}. In the inset of the figure \ref{fig:kvsn}, we have plotted the zoomed-in region regions of the spectra showing the emergence of the center modes from the polymer continuous spectra. We notice that the center mode above the CS line exists for power-law index $n \rightarrow 0.4$ as the wavenumber $k$ decreases to $0.01$. However, we do not observe any center modes for $n\geq 0.4$ for given values of $Re$, $\beta$ and $W$. While this behavior for center modes is similar to what was observed for WM1 for $n\in[0.4, 1]$ (Fig.~\ref{fig:kvsn_highn}), interestingly, the least stable center modes exist for $k<1$ in this power-law index regime ($n \in[0.33, 0.4]$). 
In figure \ref{fig:centermodes_kwci_shoot}, we show the variation of the scaled growth rates with $k$ for $n \in [0.33,0.4]$. For $n = 0.38$, we observe that the center modes emerge from the CS balloon at $k \sim 0.22$ and persist in the lower-wavenumber regime. However, as the power-law index is decreased, these modes emerge from the polymer continuous spectra at higher-wavenumbers ($k \sim 0.56$ for $n = 0.33$).

\subsection{Evolution of center and wall modes: $n \in [0.15, 0.33]$}\label{appA3}
\begin{figure*}
\centering
    \begin{subfigure}[b]{0.47\textwidth}
        \includegraphics[width=\textwidth]{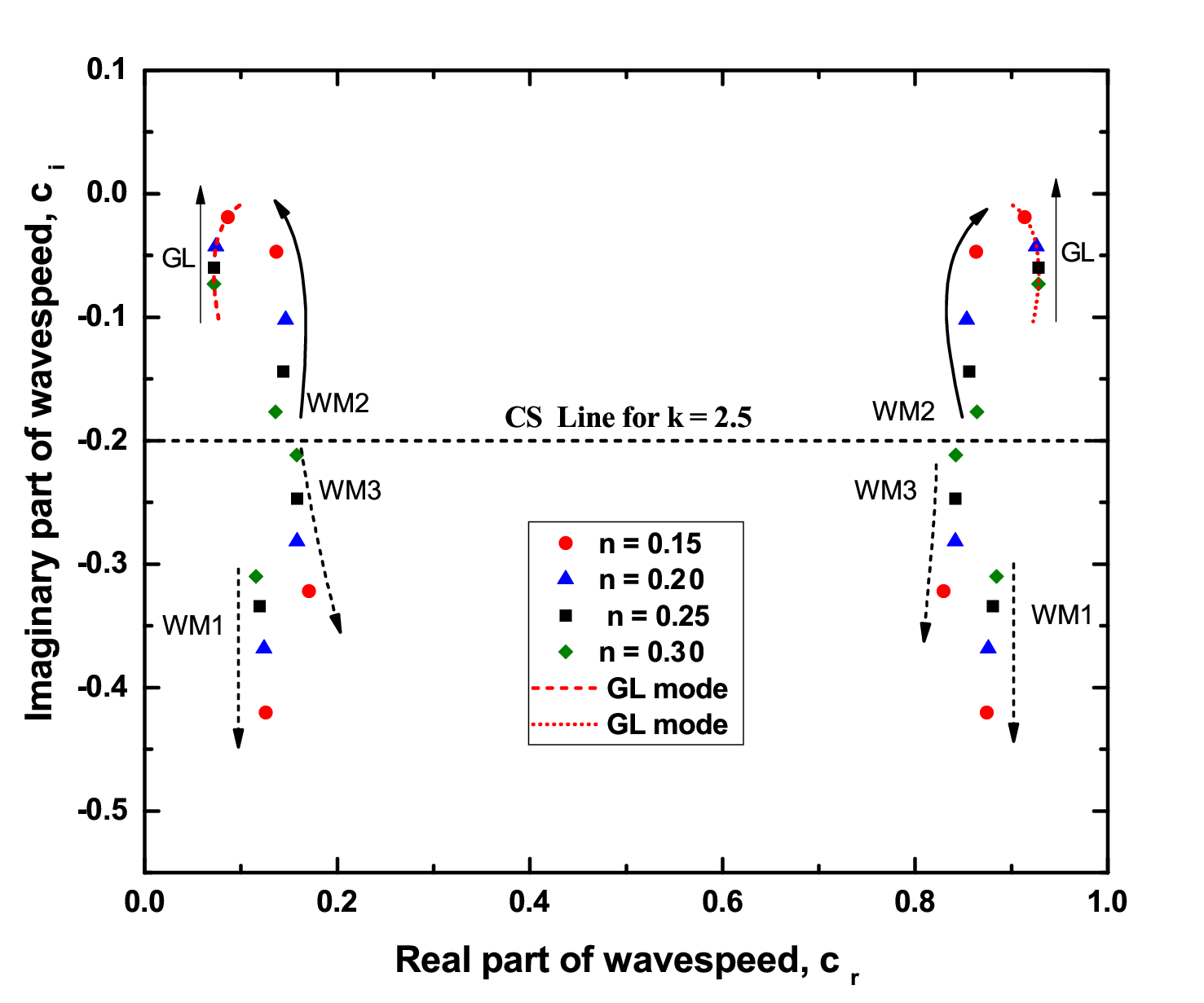}
\caption{$k = 2.5$}
\label{fig:n0p3t0p15}
\end{subfigure}\hspace{1em}
\begin{subfigure}[b]{0.47\textwidth}
        \includegraphics[width=\textwidth]{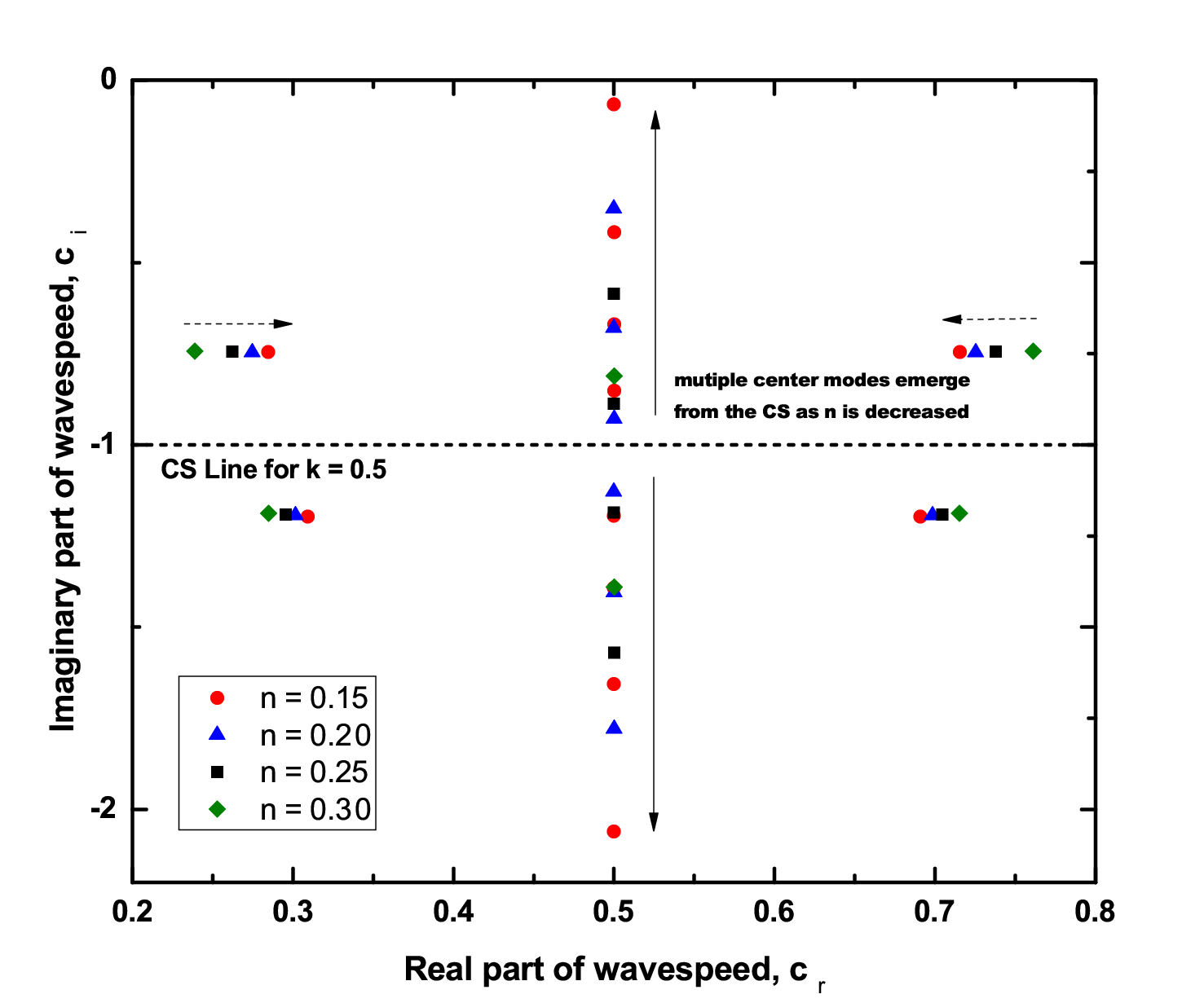}
\caption{$k = 0.5$}
\label{fig:k0p5_n0p3to15}
\end{subfigure}
\caption{Eigenvalue spectrum for $W=2$, $Re=0$, $\beta=0$, $k$ and different $n$ to illustrate the emergence of new wall and center modes.}
\label{fig:high_lowk}
\end{figure*}

In Fig.~\ref{fig:high_lowk} we show the filtered spectrum at fixed wavenumbers in the high ($k =2.5$) as well as low-$k$ ($ k =  0.5$) regimes for decreasing power-law index to show the trajectory of the discrete modes discussed above, and  to demonstrate the emergence of new discrete modes for $n \in [0.15,0.3]$ at a  fixed Weissenberg number. In the high wavenumber regime, we observe that the GL modes become less stable while the wall modes below the CS line (WM1) stabilize with a decrease in the power-law index as shown in figure \ref{fig:n0p3t0p15}. We further note a new pair of discrete wall modes emerge out of the polymer continuous spectrum for $n \leq 0.3$ above and below the CS line, henceforth referred as WM2 (above CS line) and WM3 (below CS line). Similar to the GL modes, the wall mode above the CS line (WM2) becomes less stable while the wall mode below CS line (WM3) becomes more stable as the power-law index is decreased.
In the low-$k$ regime, for $n\in[0.15, 0.3]$ we find multiple discrete center modes above and below the CS line that emerge from the polymer continuous spectrum as shown in Fig.\,\ref{fig:k0p5_n0p3to15}. The center modes above (below) the CS line become less (more) stable as $n$ is decreased. More center modes (many above and one below the CS) originate from the polymer continuous spectra as $n$ is decreased in this regime. However, the $c_i$ for the GL and wall modes are not affected by the change in $n$ for lower-$k$ values.
\begin{figure}
\centering
    \includegraphics[width=0.6\textwidth]{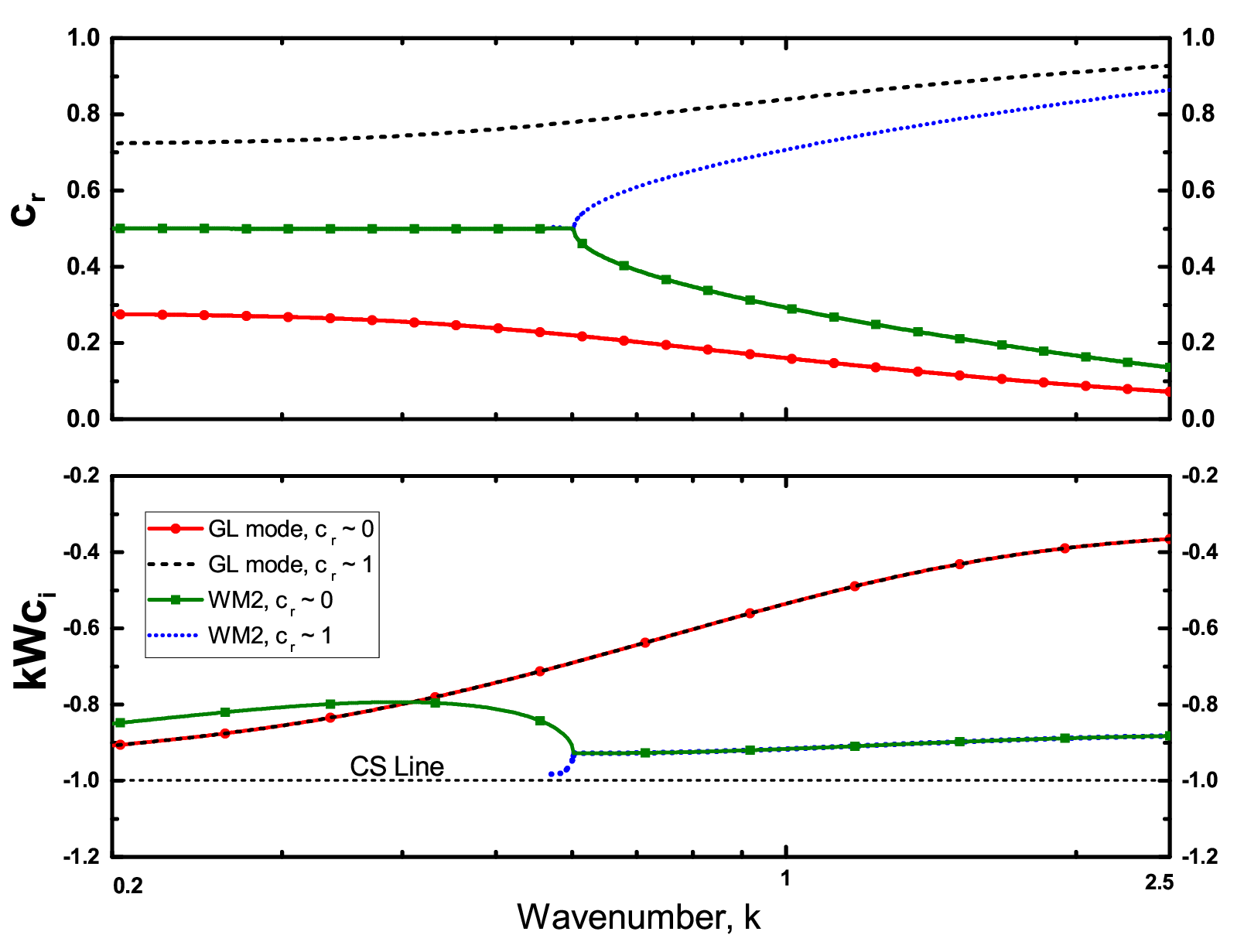}
\caption{Variation of real and imaginary parts of wavespeeds ($c_r$ and $c_i$) with wavenumber, $k$. Data for $W=2$, $Re=0$, \textbf{$n=0.3$} and $\beta=0$.}
\label{fig:n0p3_kwci_shoot}
\end{figure}
\begin{figure}
\centering
    \includegraphics[width=0.6\textwidth]{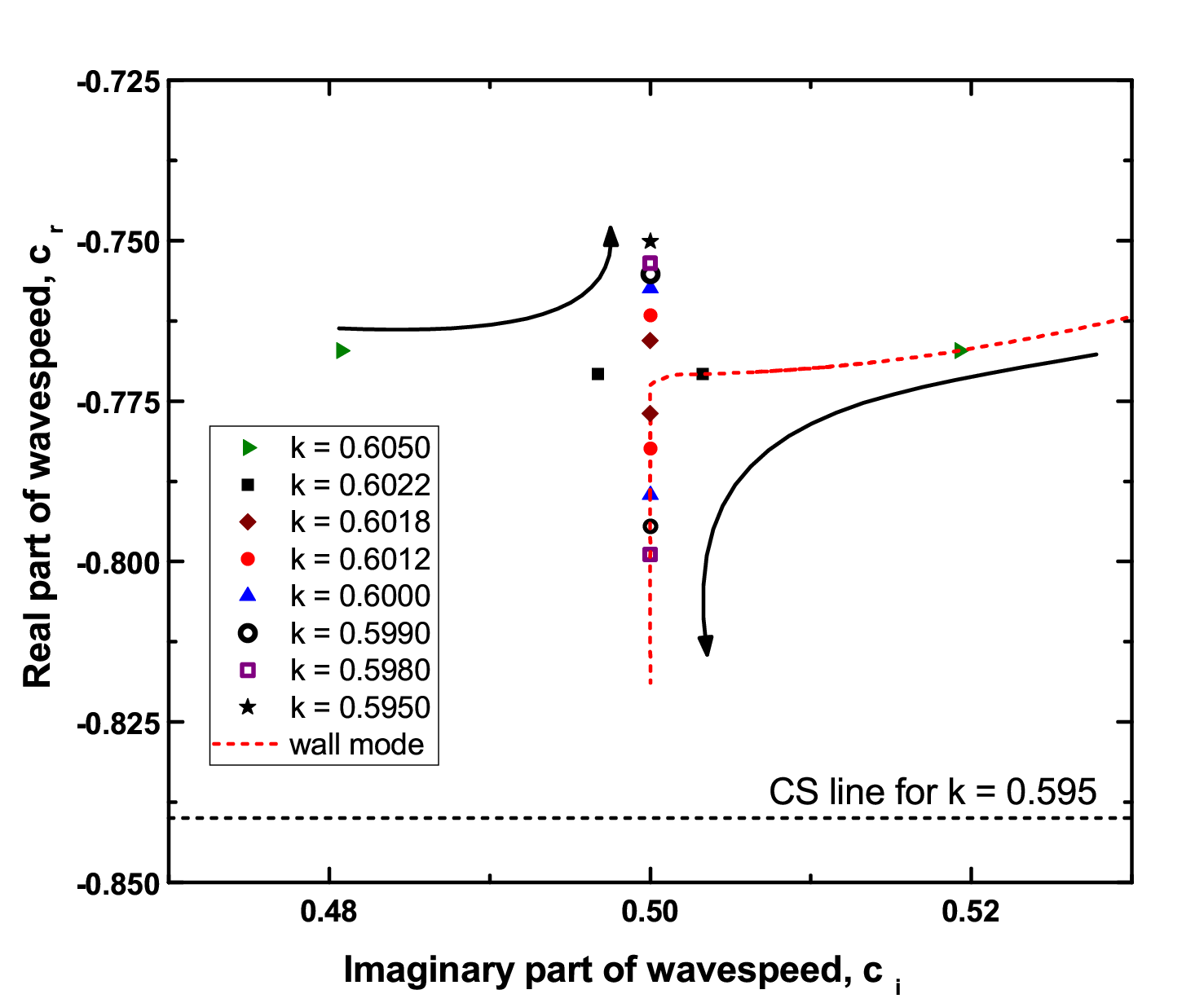}
\caption{Spectrum showing the variation of real and imaginary parts of wavespeeds ($c_r$ and $c_i$) at different $k$  for $W=2$, $Re=0$, $n=0.3$ and $\beta=0$, to demonstrate the transition of wall modes to center modes for $n=0.3$. }
\label{fig:wallmerge}
\end{figure}

We next show, in figure \ref{fig:n0p3_kwci_shoot}, the effect of wavenumber on the scaled growth rate ($kWc_i$) at fixed power-law index ($n = 0.3$) for the GL and wall modes above the CS line (WM2).
As the power-law index is decreased to $n = 0.3$, we observe an interesting variation in the growth rate ($kWc_i$) with wavenumber. While the growth rates for GL modes decrease with decrease in wavenumber, the wall modes (indicated as WM2 ) with different $c_r$ but same growth rate in high-$k$ regime morph into two center modes with same $c_r$ ($c_r \sim 0.5$) and different growth rates at $k \sim 0.6$. While one center mode (the continuation of WM2, $c_r \sim 0$ ) continues to persist in lower $k$ regime, the other center mode (continuation of WM2, $c_r \sim 1$) merges into the polymer continuous spectrum at $k\sim 0.56$ as indicated in Fig.~\ref{fig:n0p3_kwci_shoot}. The filtered spectrum for the wavenumbers near the transition of wall modes to center modes is shown in  Fig.\,\ref{fig:wallmerge} to mark the transition of the two wall modes into two center modes. Also, the center mode (i.e. the continuation of WM2, $c_r \sim 1$) moves towards the CS line as $k$ is decreased and merges into the continuous spectrum balloon at $k \sim 0.56$. 

\begin{figure}
\centering
    \includegraphics[width=0.6\textwidth]{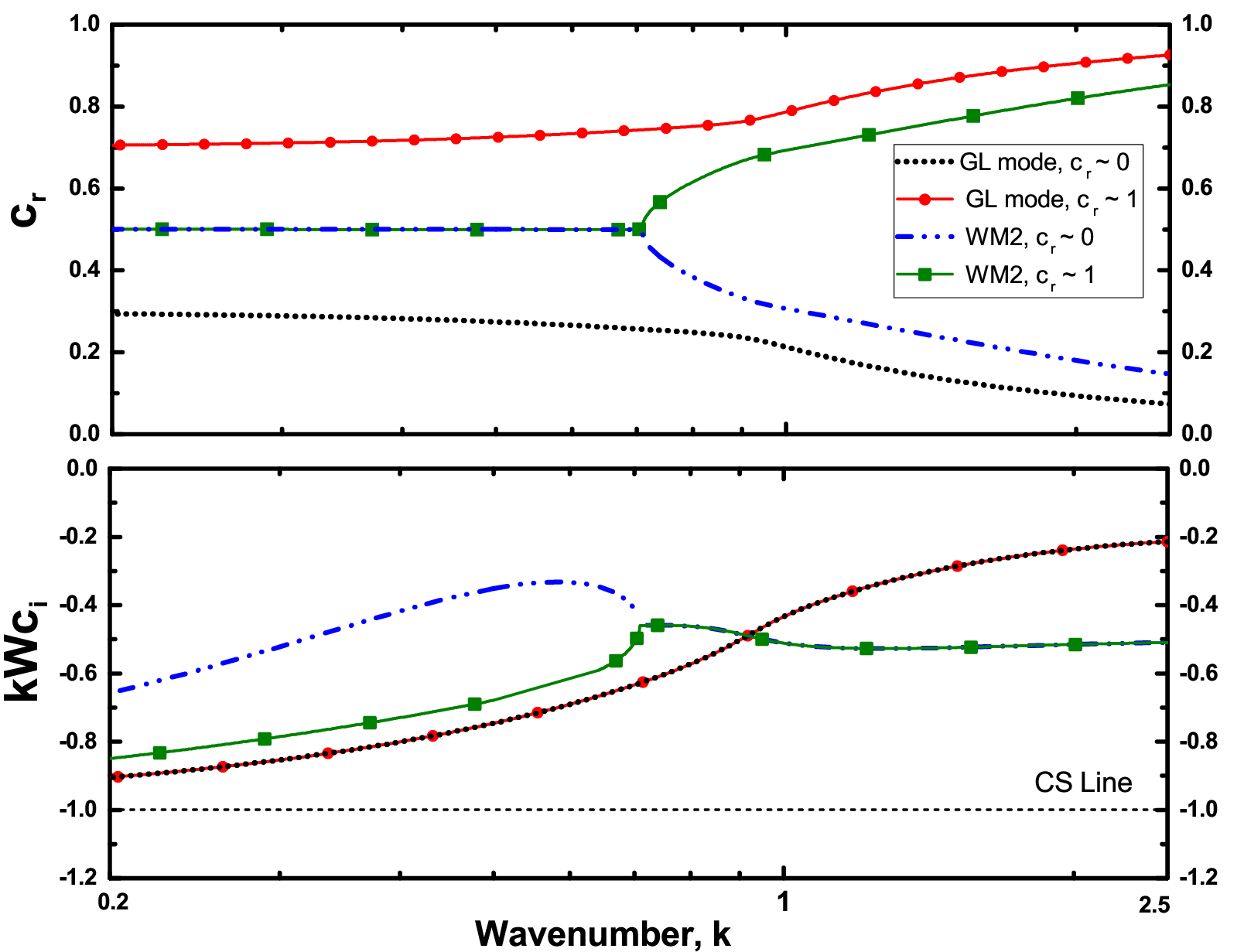}
\caption{Variation of real part of wavespeed ($c_r$) and scale growth rate $kWc_i$ with wavenumber, $k$. Data for $W=2$, $Re=0$, $n=0.2$, and $\beta=0$.}
\label{fig:n0p2}
\end{figure}
\begin{figure*}
\begin{subfigure}[b]{0.5\textwidth}
        \includegraphics[width=\textwidth]{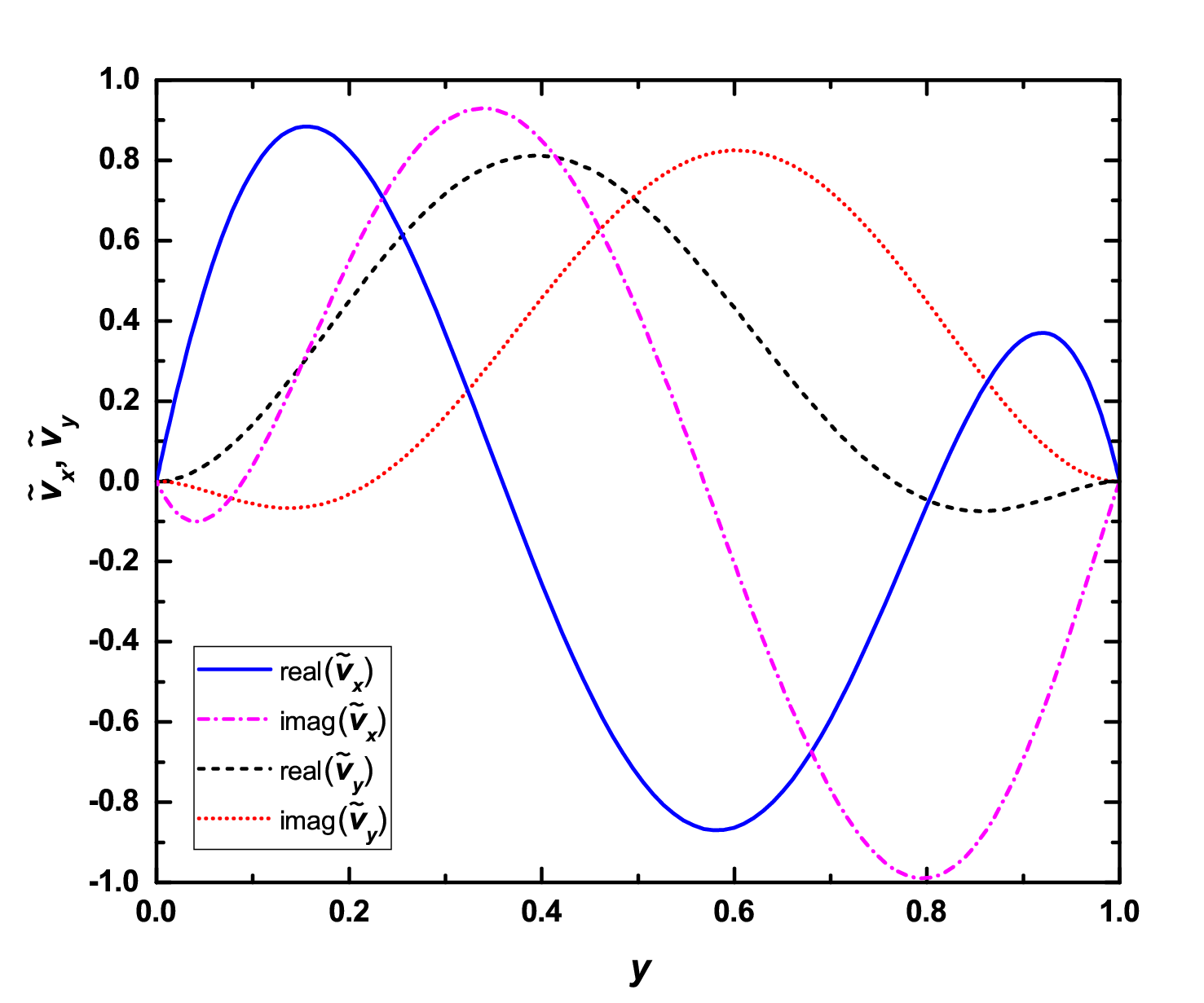}
\caption{Least stable center mode}
\label{fig:eigenfunctions_centremodes}
\end{subfigure}\hspace{1em}
\begin{subfigure}[b]{0.5\textwidth}
        \includegraphics[width=\textwidth]{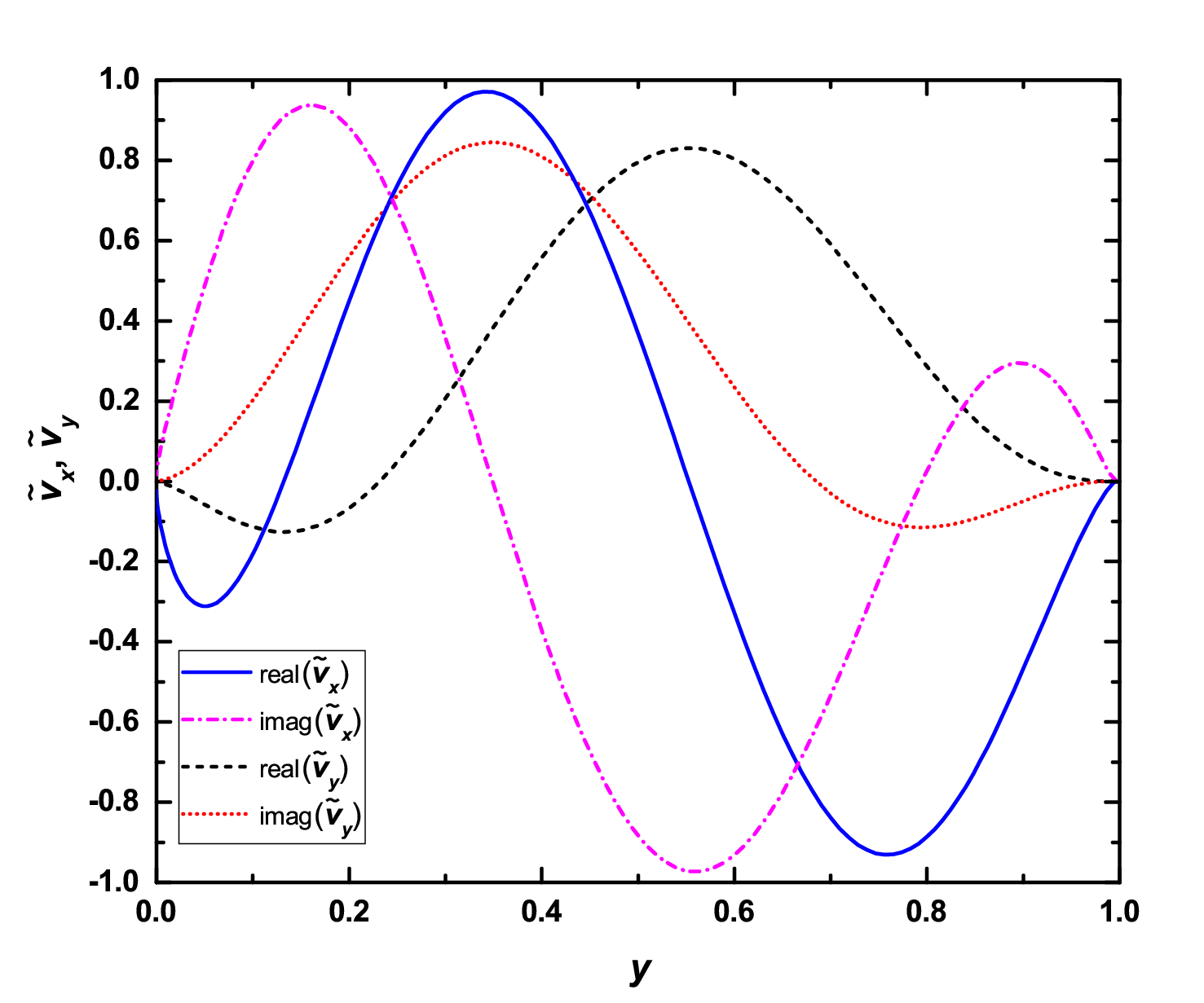}
\caption{Most stable center mode}
\label{fig:eigenfunctions_wallmodes}
\end{subfigure}
\caption{Real and imaginary parts of the eigenfunctions $\tilde v_{x}, \tilde v_{y}$ corresponding to the least and most stable center mode (above and below CS line respectively) for $W=2$, $Re=0$, $\beta=0$, $k=0.5$ and $n=0.2$. }
\label{fig:eigenfunctions_center}
\end{figure*}

As the power-law index is decreased further to $n = 0.2$, we once again observe, similar to the trend for $n = 0.3$, that the GL modes smoothly varies with $k$ on continuation to lower wavenumbers as shown in Fig.\,\ref{fig:n0p2}. Whereas, the two wall modes above the CS line (different $c_r$, same $c_i$ values in high-$k$ regime) morph into two center modes (same $c_r$ but different $c_i$) on numerical continuation to lower wavenumbers. In contrast to the scenario for $n =0.3$ (Fig.\,\ref{fig:n0p3_kwci_shoot}), however, both the center modes for $n= 0.2$ continue to exist in the lower wavenumber regime. 
Note that for $n = 0.2$ and $W = 2$, the GL modes are less stable than wall modes for $k>1$, whereas, for $k<1$, the center modes (i.e. the continuation of wall modes) are less stable than the GL modes.\\
We next plot the real and imaginary parts of the normalized eigenfunctions $\tilde v_{x}, \tilde v_{y}$ with $y$, representing the least stable center mode ($c_r = 0.5$) above the CS line and most stable center mode below the CS line for fixed values of the parameters $W, Re, \beta, n$ and $k = 0.5$ in figure \ref{fig:eigenfunctions_center}. The eigenfunctions for the center modes are spread over the entire $y$-domain as compared to the much-localized oscillations near the walls for the GL and wall modes shown earlier in figure \ref{fig:Eigenfunctions}.

\subsection{Origin of GL modes: $n>1$}\label{appA4}
\begin{figure*}
 \begin{subfigure}[b]{0.5\textwidth}
        \includegraphics[width=\textwidth]{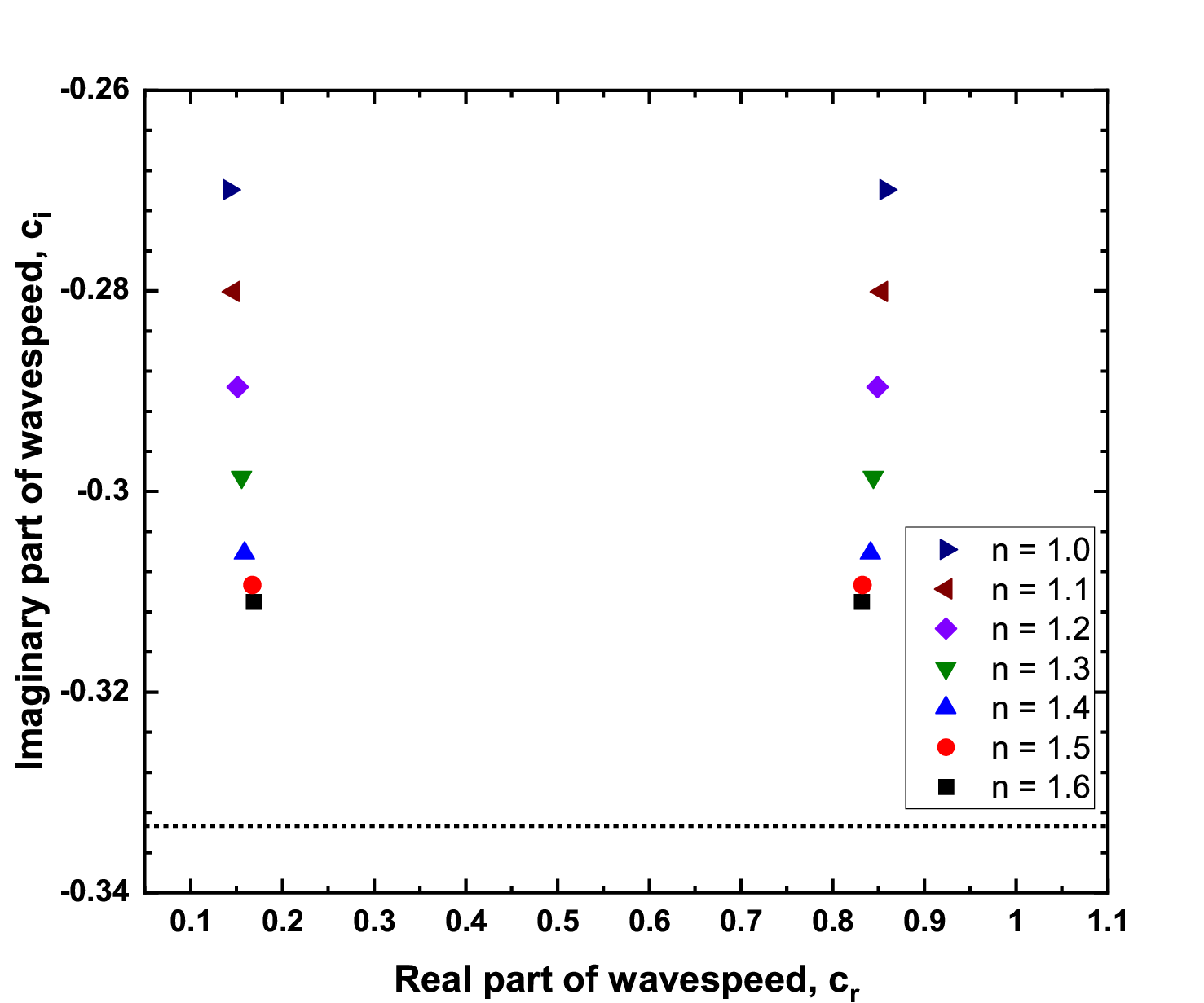}
   \caption{}
    \label{fig:G_critical_vs_W}
\end{subfigure}\hspace{1em}
\begin{subfigure}[b]{0.5\textwidth}
        \includegraphics[width=\textwidth]{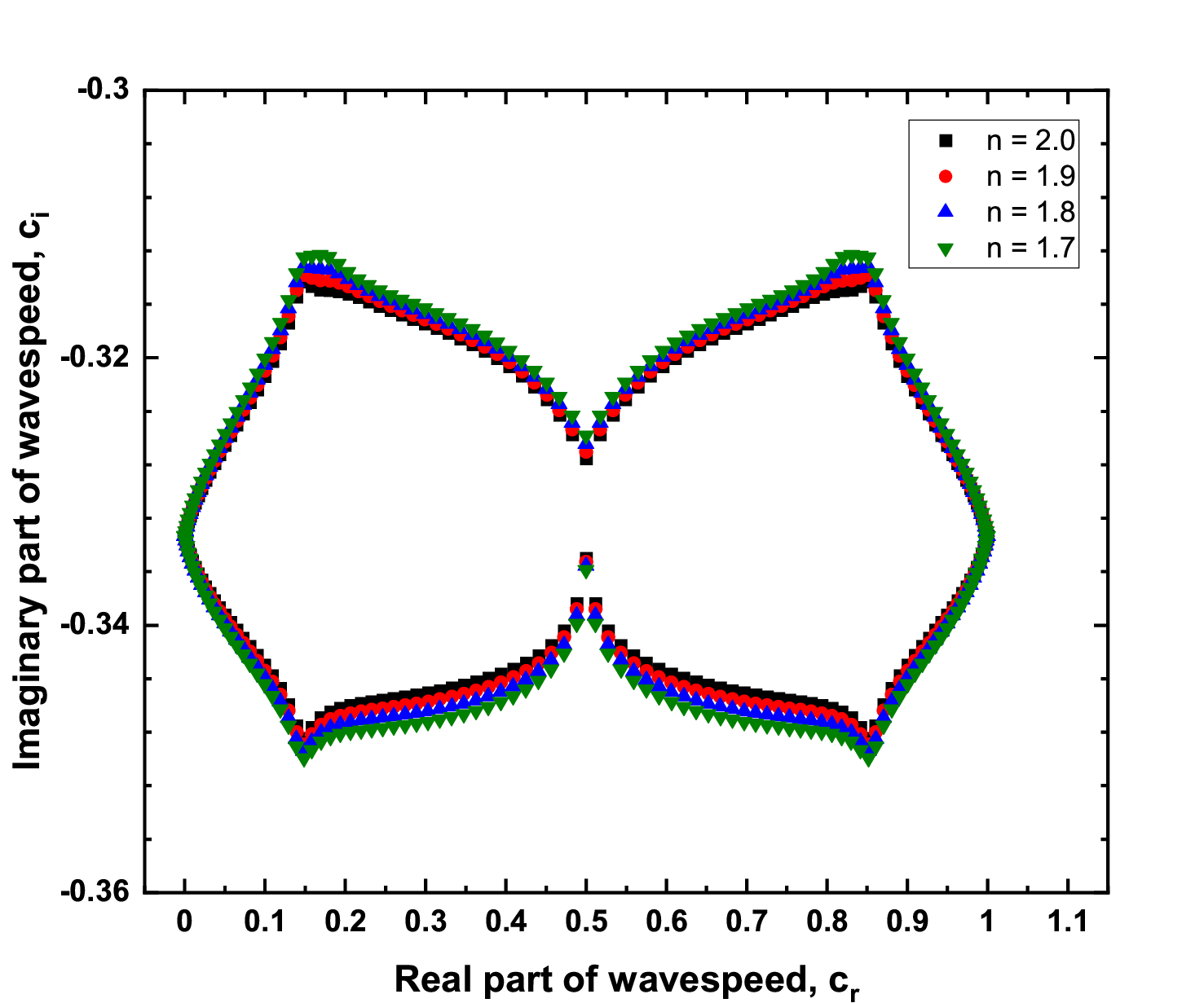}
   \caption{}
   \label{fig:Gvsk_for_all_W}
\end{subfigure}
\caption{The origin of GL modes: Data for $W=2$, $Re=0$, $\beta=0$, $k =1.5$ and at different $n$. In panel~(a) we show the origin of GL modes from the continuous spectra as $n$ is increased. Whereas in panel~(2) we show that for $n >1.6$ we do not observe any discrete modes.}
\label{fig:GLmodes}
 \end{figure*}

In this subsection, we explore the origin of the well-known GL modes present in the UCM limit ($W\neq 0$, $n =1$, $\beta = 0$). In the previous discussion, we demonstrated the origin and evolution of wall and center modes above and below the CS line from the polymer continuous spectra for  $n < 1$. In  Fig.\,\ref{fig:GLmodes}, we show the even the GL modes present in the UCM fluid ($n = 1$) also emerge from the polymer continuous spectra at $n > 1$. In panel~(a), we show that as the power-law index is increased from $n = 1$ to the higher values, the GL modes stabilize and move towards the theoretical CS line (CS balloon). However, as the power-law index is increased beyond $n = 1.6$, we do not observe any discrete GL modes for the given set of parameters as shown in Fig.\,\ref{fig:Gvsk_for_all_W}. Thus, the well-known GL modes in plane Couette flow of the UCM fluid, also merge into/emerge from the polymer continuous spectra for the power-law index $n>1$.

\bibliographystyle{jfm}
\bibliography{revision1}

\end{document}